\documentclass[12pt]{article}
\usepackage{amsmath,amsthm}
\usepackage{eucal}
\usepackage{amsfonts}
\usepackage{pb-diagram,lamsarrow,pb-lams}
\newcommand{\ar}{\renewcommand{\arraystretch}{1}} % 1.0 % 0.65
\gdef\C{\Bbb C}

\gdef\dS{\Bbb S}
\gdef\R{\Bbb R}

\DeclareMathOperator{\spin}{{\bf Spin}}

\DeclareMathOperator{\Sym}{Sym}
\DeclareMathOperator{\arc}{arccos}
\DeclareMathOperator{\diver}{div}
\DeclareMathOperator{\rot}{rot}

\newcommand{\tg}{\tan}
\newcommand{\ch}{\cosh}
\newcommand{\sh}{\sinh}
\newcommand{\tnh}{\tanh}
\newcommand{\ctg}{\cot}

\newcommand{\re}{\mbox{\rm Re}\,}
\newcommand{\im}{\mbox{\rm Im}\,}

\newcommand{\cA}{\mathcal{A}}

\newcommand{\cL}{\mathcal{L}}

\newcommand{\sA}{{\sf A}}
\newcommand{\sB}{{\sf B}}
\newcommand{\sI}{{\sf I}}

\newcommand{\sL}{\Lambda}
\newcommand{\sT}{T}

\newcommand{\sY}{{\sf Y}}
\newcommand{\sX}{{\sf X}}

\newcommand{\bj}{{\bf j}}

\newcommand{\bx}{{\bf x}}
\newcommand{\by}{{\bf y}}
\newcommand{\bz}{{\bf z}}
\newcommand{\bB}{{\bf B}}

\newcommand{\bF}{{\bf F}}
\newcommand{\bE}{{\bf E}}

\newcommand{\fM}{\mathfrak{M}}

\newcommand{\fG}{\mathfrak{G}}
\newcommand{\fT}{\mathfrak{M}}
\newcommand{\fg}{\mathfrak{g}}

\newcommand{\hypergeom}[5]{\mbox{$
_#1 F_#2 \left.
\!\!
\left(
\!\!\!\!
\begin{array}{c}
\multicolumn{1}{c}{\begin{array}{c}
#3
\end{array}}\\[1mm]
\multicolumn{1}{c}{\begin{array}{c}
#4
   \end{array}}\end{array}
\!\!\!\!
\right|\displaystyle{#5}\right)
$}
}
\newcommand{\cl}{C\kern -0.2em \ell}

\newcommand{\e}{\mbox{\bf e}}

\newcommand{\ld}{\left[}
\newcommand{\rd}{\right]}

\begin{document}
\title{General Solutions of Relativistic Wave Equations}
\author{V.~V. Varlamov}
\date{}
\maketitle
\begin{abstract}
General solutions of relativistic wave equations are studied in terms
of the functions on the Lorentz group. A close relationship between
hypersperical functions and matrix elements of irreducible representations
of the Lorentz group is established. A generalization of the
Gel'fand--Yaglom formalism for higher--spin equations is given.
It is shown that a two--dimensional complex sphere is associated with
the each point of Minkowski spacetime. The separation of variables in
a general relativistically invariant system is obtained via the
hypersperical functions defined on the surface of the two--dimensional
complex sphere. In virtue of this the wave functions are represented
in the form of series on the hyperspherical functions.
Such a description allows to consider all the physical fields on
an equal footing. General solutions of the Dirac and Weyl equations,
and also the Maxwell equations in the Majorana--Oppenheimer form,
are given in terms of the functions on the Lorentz group.
\end{abstract}
{\bf PACS 1998:} 03.65.Pm, 02.30.Gp, 02.10.Tq
\section{Introduction}
Traditionally, equations of motion play a basic role in physics.
Relativistic wave equations (RWE) play the same role in quantum field
theory. The wave function (main object of quantum theory) is a solution
of RWE. For that reason all textbooks on quantum field theory began with
a brief introduction to RWE. As a rule, solutions of the Dirac equations
are represented by plane--wave solutions (see for example \cite{Sch61}, and
also many other textbooks). However, the plane--wave solutions are
strongly degenerate. These solutions do not contain parameters of the
Lorentz group and weakly reflect a relativistic nature of the wave
function described by the Dirac equations.

In the present paper general solutions of RWE are given in the form of
series on hyperspherical functions. In turn, matrix elements of 
irreducible representations of the
Lorentz group are expressed via the hyperspherical functions
(for more details see recent paper \cite{Var022}). By this reason
the wave function hardly depends on the parameters of the Lorentz group.
Moreover, it allows to consider solutions of RWE as the functions on
the Lorentz group.

A starting point of the research is an isomorphism
$SL(2,\C)\sim\mbox{\sf complex}(SU(2))$, that is, the group $SL(2,\C)$
is a complexification of $SU(2)$. The well--known Van der Waerden
representation for the Lorentz group \cite{Wa32} is a direct consequence
of this isomorphism.
\begin{sloppypar}
The following important point of the present research is a generalization
of the Gel'fand--Yaglom formalism. In accordance with the fundamental
isomorphism $SL(2,\C)\sim\mbox{\sf complex}(SU(2))$, a general
relativistically invariant system is obtained in the result of
complexification of three--dimensional Gel'fand--Yaglom equations.
Further, a correspondence of the obtained invariant system with
four--dimensional Gel'fand--Yaglom equations is established via the
mapping of complexified equations into a six--dimensional bivector
space $\R^6$ associated with the each point of the Minkowski
spacetime $\R^{1,3}$.\end{sloppypar}

The following step is a separation of variables in the relativistically
invariant system. At this point, all the variables are parameters of the
Lorentz group. The main tool for separation of variables is a
two--dimensional complex sphere (this sphere was firstly considered
by Smorodinsky and Huszar at the study of helicity states \cite{SH70}).
Hyperspherical functions, defined on the surface of the
two--dimensional complex sphere, allow to separate variables in the
general relativistically invariant system.

Group theoretical description of RWE allows to present all the physical
fields on an equal footing. Namely, all these fields are the functions
on the Lorentz group. For example, in the sections 4--6 solutions of the
Dirac, Weyl and Maxwell equations, which considered as the particular
cases of the general relativistically invariant system, are represented
via the hyperspherical series in terms of the functions on the
Lorentz group.
\section{Lorentz group and hyperspherical functions}
\subsection{Van der Waerden representation of the Lorentz group}
Let $\mathfrak{g}\rightarrow T_{\mathfrak{g}}$ be an arbitrary linear
representation of the proper orthochronous Lorentz group $\fG_+$ and let
$\sA_i(t)=T_{a_i(t)}$ be an infinitesimal operator corresponded the rotation
$a_i(t)\in\fG_+$. Analogously, we have $\sB_i(t)=T_{b_i(t)}$, where
$b_i(t)\in\fG_+$ is a hyperbolic rotation. The operators $\sA_i$ and
$\sB_i$ satisfy the following commutation 
relations\footnote{Denoting $\sI^{23}=\sA_1$, $\sI^{31}=\sA_2$,
$\sI^{12}=\sA_3$, and $\sI^{01}=\sB_1$, $\sI^{02}=\sB_2$, $\sI^{03}=\sB_3$
we can write the relations (\ref{Com1}) in a more compact form:
\[
\ld\sI^{\mu\nu},\sI^{\lambda\rho}\rd=\delta_{\mu\rho}\sI^{\lambda\nu}+
\delta_{\nu\lambda}\sI^{\mu\rho}-\delta_{\nu\rho}\sI^{\mu\lambda}-
\delta_{\mu\lambda}\sI^{\nu\rho}.
\]}:
\begin{equation}\label{Com1}
\left.\begin{array}{lll}
\ld\sA_1,\sA_2\rd=\sA_3, & \ld\sA_2,\sA_3\rd=\sA_1, &
\ld\sA_3,\sA_1\rd=\sA_2,\\[0.1cm]
\ld\sB_1,\sB_2\rd=-\sA_3, & \ld\sB_2,\sB_3\rd=-\sA_1, &
\ld\sB_3,\sB_1\rd=-\sA_2,\\[0.1cm]
\ld\sA_1,\sB_1\rd=0, & \ld\sA_2,\sB_2\rd=0, &
\ld\sA_3,\sB_3\rd=0,\\[0.1cm]
\ld\sA_1,\sB_2\rd=\sB_3, & \ld\sA_1,\sB_3\rd=-\sB_2, & \\[0.1cm]
\ld\sA_2,\sB_3\rd=\sB_1, & \ld\sA_2,\sB_1\rd=-\sB_3, & \\[0.1cm]
\ld\sA_3,\sB_1\rd=\sB_2, & \ld\sA_3,\sB_2\rd=-\sB_1. &
\end{array}\right\}
\end{equation}
Let us consider the operators
\begin{gather}
\sX_k=\frac{1}{2}(\sA_k+i\sB_k),\quad\sY_k=\frac{1}{2}(\sA_k-i\sB_k),
\label{SL25}\\
(k=1,2,3).\nonumber
\end{gather}
Using the relations (\ref{Com1}) we find that
\begin{gather}
\ld\sX_1,\sX_2\rd=\sX_3,\quad\ld\sX_2,\sX_3\rd=\sX_1,\quad
\ld\sX_2,\sX_1\rd=\sX_2,\nonumber\\
\ld\sY_1,\sY_2\rd=\sY_3,\quad\ld\sY_2,\sY_3\rd=\sY_1,\quad
\ld\sY_3,\sY_1\rd=\sY_2,\nonumber\\
\ld\sX_k,\sY_l\rd=0,\quad(k,l=1,2,3).\label{Com2}
\end{gather}
Further, taking
\begin{equation}\label{SL26}
\left.\begin{array}{cc}
\sX_+=\sX_1+i\sX_2, & \sX_-=\sX_1-i\sX_2,\\[0.1cm]
\sY_+=\sY_1+i\sY_2, & \sY_-=\sY_1-i\sY_2
\end{array}\right\}
\end{equation}
we see that in virtue of commutativity of the relations (\ref{Com2}) a
space of an irreducible finite--dimensional representation of the group
$\fG_+$ can be stretched on the totality of $(2l+1)(2\dot{l}+1)$ basis
vectors $\mid l,m;\dot{l},\dot{m}\rangle$, where $l,m,\dot{l},\dot{m}$
are integer or half--integer numbers, $-l\leq m\leq l$,
$-\dot{l}\leq \dot{m}\leq \dot{l}$. Therefore,
\begin{eqnarray}
&&\sX_-\mid l,m;\dot{l},\dot{m}\rangle=
\sqrt{(\dot{l}+\dot{m})(\dot{l}-\dot{m}+1)}\mid l,m;\dot{l},\dot{m}-1
\rangle\;\;(\dot{m}>-\dot{l}),\nonumber\\
&&\sX_+\mid l,m;\dot{l},\dot{m}\rangle=
\sqrt{(\dot{l}-\dot{m})(\dot{l}+\dot{m}+1)}\mid l,m;\dot{l},\dot{m}+1
\rangle\;\;(\dot{m}<\dot{l}),\nonumber\\
&&\sX_3\mid l,m;\dot{l},\dot{m}\rangle=
\dot{m}\mid l,m;\dot{l},\dot{m}\rangle,\nonumber\\
&&\sY_-\mid l,m;\dot{l},\dot{m}\rangle=
\sqrt{(l+m)(l-m+1)}\mid l,m-1,\dot{l},\dot{m}\rangle
\;\;(m>-l),\nonumber\\
&&\sY_+\mid l,m;\dot{l},\dot{m}\rangle=
\sqrt{(l-m)(l+m+1)}\mid l,m+1;\dot{l},\dot{m}\rangle
\;\;(m<l),\nonumber\\
&&\sY_3\mid l,m;\dot{l},\dot{m}\rangle=
m\mid l,m;\dot{l},\dot{m}\rangle.\label{Waerden}
\end{eqnarray}
From the relations (\ref{Com2}) it follows that each of the sets of 
infinitisimal operators $\sX$ and $\sY$ generates the group $SU(2)$ and these
two groups commute with each other. Thus, from the relations (\ref{Com2})
and (\ref{Waerden}) it follows that the group $\fG_+$, in essence,
is equivalent to the group $SU(2)\otimes SU(2)$. In contrast to the
Gel'fand--Naimark representation for the Lorentz group \cite{GMS,Nai58}
which does not find a broad application in physics,
a representation (\ref{Waerden}) is a most useful in theoretical physics
(see, for example, \cite{AB,Sch61,RF,Ryd85}). This representation for the
Lorentz group was firstly given by Van der Waerden in his brilliant book
\cite{Wa32}.

As known, a double covering of the proper orthochronous Lorentz group $\fG_+$,
the group $SL(2,\C)$, is isomorphic to the Clifford--Lipschitz group
$\spin_+(1,3)$, which, in its turn, is fully defined within a
biquaternion algebra $\C_2$, since
\[\ar
\spin_+(1,3)\simeq\left\{\begin{pmatrix} \alpha & \beta \\ \gamma & \delta
\end{pmatrix}\in\C_2:\;\;\det\begin{pmatrix}\alpha & \beta \\ \gamma & \delta
\end{pmatrix}=1\right\}=SL(2,\C).
\]\begin{sloppypar}\noindent
Thus, a fundamental representation of the group $\fG_+$ is realized in a
spinspace $\dS_2$. The spinspace $\dS_2$ is a complexification of the
minimal left ideal of the algebra $\C_2$: $\dS_2=\C\otimes I_{2,0}=\C\otimes
\cl_{2,0}e_{20}$ or $\dS_2=\C\otimes I_{1,1}=\C\otimes\cl_{1,1}e_{11}$
($\C\otimes I_{0,2}=\C\otimes\cl_{0,2}e_{02}$), where $\cl_{p,q}$
($p+q=2$) is a real subalgebra of $\C_2$, $I_{p,q}$ is the minimal left ideal
of the algebra $\cl_{p,q}$, $e_{pq}$ is a primitive idempotent.
\end{sloppypar}
Further, let $\overset{\ast}{\C}_2$ be the biquaternion algebra, in which
all the coefficients are complex conjugate to the coefficients of the
algebra $\C_2$. The algebra $\overset{\ast}{\C}_2$ is obtained from 
$\C_2$ under action of the automorphism $\cA\rightarrow\cA^\star$
(involution) or the antiautomorphism $\cA\rightarrow\widetilde{\cA}$ (reversal),
where $\cA\in\C_2$ (see \cite{Var99,Var00}). Let us compose a tensor
product of $k$ algebras $\C_2$ and $r$ algebras $\overset{\ast}{\C}_2$:
\begin{equation}\label{Ten}
\C_2\otimes\C_2\otimes\cdots\otimes\C_2\otimes\overset{\ast}{\C}_2\otimes
\overset{\ast}{\C}_2\otimes\cdots\otimes\overset{\ast}{\C}_2\simeq
\C_{2k}\otimes\overset{\ast}{\C}_{2r}.
\end{equation}
The tensor product (\ref{Ten}) induces a spinspace
\begin{equation}\label{Spin}
\dS_2\otimes\dS_2\otimes\cdots\otimes\dS_2\otimes\dot{\dS}_2\otimes
\dot{\dS}_2\otimes\cdots\otimes\dot{\dS}_2=\dS_{2^{k+r}}
\end{equation}
with `vectors' (spintensors) of the form
\begin{equation}\label{Vect}
\xi^{\alpha_1\alpha_2\cdots\alpha_k\dot{\alpha}_1\dot{\alpha}_2\cdots
\dot{\alpha}_r}=\sum\xi^{\alpha_1}\otimes\xi^{\alpha_2}\otimes\cdots\otimes
\xi^{\alpha_k}\otimes\xi^{\dot{\alpha}_1}\otimes\xi^{\dot{\alpha}_2}\otimes
\cdots\otimes\xi^{\dot{\alpha}_r}.
\end{equation}\begin{sloppypar}\noindent
The full representation space $\dS_{2^{k+r}}$ contains both symmetric and
antisymmetric spintensors (\ref{Vect}). Usually, at the definition of
irreducible finite--dimensional representations of the Lorentz group
physicists confined to a subspace of symmetric spintensors
$\Sym(k,r)\subset\dS_{2^{k+r}}$. Dimension of $\Sym(k,r)$ is equal to
$(k+1)(r+1)$ or $(2l+1)(2l^\prime+1)$ at $l=\frac{k}{2}$,
$l^\prime=\frac{r}{2}$. It is easy to see that the space $\Sym(k,r)$ is a
space of Van der Waerden representation (\ref{Waerden}). The space
$\Sym(k,r)$ can be considered as a space of polynomials\end{sloppypar}
\begin{gather}
p(z_0,z_1,\bar{z}_0,\bar{z}_1)=\sum_{\substack{(\alpha_1,\ldots,\alpha_k)\\
(\dot{\alpha}_1,\ldots,\dot{\alpha}_r)}}\frac{1}{k!\,r!}
a^{\alpha_1\cdots\alpha_k\dot{\alpha}_1\cdots\dot{\alpha}_r}
z_{\alpha_1}\cdots z_{\alpha_k}\bar{z}_{\dot{\alpha}_1}\cdots
\bar{z}_{\dot{\alpha}_r}\label{SF}\\
(\alpha_i,\dot{\alpha}_i=0,1),\nonumber
\end{gather}
%\begin{sloppypar}\noindent
where the numbers 
$a^{\alpha_1\cdots\alpha_k\dot{\alpha}_1\cdots\dot{\alpha}_r}$
are unaffected at the permutations of indices. 

The infinitesimal operators $\sA_i$, defined on the symmetric
representation spaces, have the form
\begin{eqnarray}
\sA_1&=&-\frac{i}{2}\boldsymbol{\alpha}^l_m\xi_{m-1}-
\frac{i}{2}\boldsymbol{\alpha}^l_{m+1}\xi_{m+1},\nonumber\\
\sA_2&=&\frac{1}{2}\boldsymbol{\alpha}^l_m\xi_{m-1}-
\frac{1}{2}\boldsymbol{\alpha}^l_{m+1}\xi_{m+1},\label{OpA}\\
\sA_3&=&-im\xi_m,\nonumber
\end{eqnarray}
where
\[
\boldsymbol{\alpha}^l_m=\sqrt{(l+m)(l-m+1)},
\]
and $\xi_i$ are vectors of a canonical basis in the representation space.
Or, in the matrix notation:
\begin{equation}\label{A1}
\sA^j_{1}=-\frac{i}{2}
\ar\begin{bmatrix}
0 & \boldsymbol{\alpha}_{-l_j+1} & 0 & \dots & 0 & 0\\
\boldsymbol{\alpha}_{-l_j+1} &0&\boldsymbol{\alpha}_{-l_j+2} & \dots & 0 & 0\\
0 & \boldsymbol{\alpha}_{-l_j+2} & 0 & \dots & 0 & 0\\
\hdotsfor[2]{6}\\
\hdotsfor[2]{6}\\
0 & 0 & 0 & \dots & 0 & \boldsymbol{\alpha}_{l_j}\\
0 & 0 & 0 & \dots & \boldsymbol{\alpha}_{l_j} & 0
\end{bmatrix}
\end{equation}
\begin{equation}\label{A2}
\sA^j_{2}=\frac{1}{2}
\ar\begin{bmatrix}
0 & \boldsymbol{\alpha}_{-l_j+1} & 0 & \dots & 0 & 0\\
-\boldsymbol{\alpha}_{-l_j+1}&0&\boldsymbol{\alpha}_{-l_j+2} & \dots & 0 & 0\\
0 & -\boldsymbol{\alpha}_{-l_j+2} & 0 & \dots & 0 & 0\\
\hdotsfor[2]{6}\\
\hdotsfor[2]{6}\\
0 & 0 & 0 & \dots & 0 & \boldsymbol{\alpha}_{l_j}\\
0 & 0 & 0 & \dots & -\boldsymbol{\alpha}_{l_j} & 0
\end{bmatrix}
\end{equation}
\begin{equation}\label{A3}
\sA^j_{3}=
\ar\begin{bmatrix}
il_j & 0 & 0 & \dots & 0 & 0\\
0 & i(l_j-1) & 0 & \dots & 0 & 0\\
0 & 0 & i(l_j-2) & \dots & 0 & 0\\
\hdotsfor[2]{6}\\
\hdotsfor[2]{6}\\
0 & 0 & 0 & \dots & -i(l_j-1) & 0\\
0 & 0 & 0 & \dots & 0 & -il_j
\end{bmatrix}
\end{equation}

Further, for the operators $\sB_i=i\sA_i$ we have
\begin{eqnarray}
\sB_1&=&\frac{1}{2}\boldsymbol{\alpha}^l_m\xi_{m-1}+
\frac{1}{2}\boldsymbol{\alpha}^l_{m+1}\xi_{m+1},\nonumber\\
\sB_2&=&\frac{i}{2}\boldsymbol{\alpha}^l_m\xi_{m-1}-
\frac{i}{2}\boldsymbol{\alpha}^l_{m+1}\xi_{m+1},\label{OpB}\\
\sB_3&=&m\xi_m.\nonumber
\end{eqnarray}
Analogously, infinitesimal operators for conjugate representations,
in dual representation spaces, are defined by following formulae
\begin{eqnarray}
\widetilde{A}_1&=&\frac{i}{2}\boldsymbol{\alpha}^l_m\xi_{m-1}+
\frac{i}{2}\boldsymbol{\alpha}^l_{m+1}\xi_{m+1},\nonumber\\
\widetilde{A}_2&=&-\frac{1}{2}\boldsymbol{\alpha}^l_m\xi_{m-1}+
\frac{1}{2}\boldsymbol{\alpha}^l_{m+1}\xi_{m+1},\label{DopA}\\
\widetilde{A}_3&=&im\xi_m,\nonumber\\
\widetilde{B}_1&=&-\frac{1}{2}\boldsymbol{\alpha}^l_m\xi_{m-1}-
\frac{1}{2}\boldsymbol{\alpha}^l_{m+1}\xi_{m+1},\nonumber\\
\widetilde{B}_2&=&-\frac{i}{2}\boldsymbol{\alpha}^l_m\xi_{m-1}+
\frac{i}{2}\boldsymbol{\alpha}^l_{m+1}\xi_{m+1},\label{DopB}\\
\widetilde{B}_3&=&-m\xi_m.\nonumber
\end{eqnarray}
\subsection{Matrix elements of $SU(2)$ and $SL(2,\C)$}
One--parameter subgroups of $SU(2)$ are defined by the matrices
%\end{sloppypar}
\begin{equation}
a_1(t)={\renewcommand{\arraystretch}{1.7}
\begin{pmatrix}
\cos\dfrac{t}{2} & i\sin\dfrac{t}{2}\\
i\sin\dfrac{t}{2} & \cos\dfrac{t}{2}
\end{pmatrix},\quad
a_2(t)=\begin{pmatrix}
\cos\dfrac{t}{2} & -\sin\dfrac{t}{2}\\
\sin\dfrac{t}{2} & \cos\dfrac{t}{2}
\end{pmatrix}},\quad
a_3(t)=\begin{pmatrix}
e^{\frac{it}{2}} & 0\\
0 & e^{-\frac{it}{2}}
\end{pmatrix}.\label{PS1}
\end{equation}
An arbitrary matrix $u\in SU(2)$ written via Euler angles has a form
\begin{equation}\label{SU2}\ar
u=\ar\begin{pmatrix}
\alpha & \beta\\
-\overset{\ast}{\beta} & \overset{\ast}{\alpha}
\end{pmatrix}=
{\renewcommand{\arraystretch}{1.7}\begin{pmatrix}
\cos\dfrac{\theta}{2}e^{\frac{i(\varphi+\psi)}{2}} &
i\sin\dfrac{\theta}{2}e^{\frac{i(\varphi-\psi)}{2}}\\
i\sin\dfrac{\theta}{2}e^{\frac{i(\psi-\varphi)}{2}} &
\cos\dfrac{\theta}{2}e^{-\frac{i(\varphi+\psi)}{2}}
\end{pmatrix}},
\end{equation}
where $0\leq\varphi<2\pi$, $0<\theta<\pi$, $-2\pi\leq\psi<2\pi$, $\det u=1$.
Hence it follows that $|\alpha|=\cos\frac{\theta}{2}$,
$|\beta|=\sin\frac{\theta}{2}$ and
\begin{eqnarray}
\cos\theta&=&2|\alpha|^2-1,\label{SU3}\\
e^{i\varphi}&=&-\frac{\alpha\beta i}{|\alpha|\,|\beta|},\label{SU4}\\
e^{\frac{i\psi}{2}}&=&\frac{\alpha e^{-\frac{i\varphi}{2}}}{|\alpha|}.
\label{SU5}
\end{eqnarray}
\begin{sloppypar}\noindent
Diagonal matrices $\ar\begin{pmatrix} e^{\frac{i\varphi}{2}} & 0\\ 0 &
e^{-\frac{i\varphi}{2}}\end{pmatrix}$ form one--parameter subgroup in the
group $SU(2)$. Therefore, each matrix $u\in SU(2)$ belongs to a bilateral
adjacency class contained the matrix\end{sloppypar}
\[
{\renewcommand{\arraystretch}{1.7}
\begin{pmatrix}
\cos\dfrac{\theta}{2} & i\sin\dfrac{\theta}{2}\\
i\sin\dfrac{\theta}{2} & \cos\dfrac{\theta}{2}
\end{pmatrix}}.
\]
The matrix element $t^l_{mn}=e^{-i(m\varphi+n\psi)}\langle T_l(\theta)
\psi_n,\psi_m\rangle$ of the group $SU(2)$ in the polynomial basis
\[
\psi_n(\zeta)=\frac{\zeta^{l-n}}{\sqrt{\Gamma(l-n+1)\Gamma(l+n+1)}}
\quad -l\leq n\leq l,
\]
where
\[
T_l(\theta)\psi(\zeta)=\left(i\sin\frac{\theta}{2}\zeta+
\cos\frac{\theta}{2}\right)^{2l}\psi
\left(\frac{\cos\dfrac{\theta}{2}\zeta+i\sin\dfrac{\theta}{2}}
{i\sin\dfrac{\theta}{2}\zeta+\cos\dfrac{\theta}{2}}\right),
\]
has a form
\begin{multline}
t^l_{mn}(g)=e^{-i(m\varphi+n\psi)}\langle T_l(\theta)\psi_n,\psi_m\rangle
=\\[0.2cm]
\frac{e^{-i(m\varphi+n\psi)}\langle T_l(\theta)\zeta^{l-n}\zeta^{l-m}\rangle}
{\sqrt{\Gamma(l-m+1)\Gamma(l+m+1)\Gamma(l-n+1)\Gamma(l+n+1)}}=\\[0.2cm]
e^{-i(m\varphi+n\psi)}i^{m-n}\sqrt{\Gamma(l-m+1)\Gamma(l+m+1)\Gamma(l-n+1)
\Gamma(l+n+1)}\times\\
\cos^{2l}\frac{\theta}{2}\tg^{m-n}\frac{\theta}{2}\times\\
\sum^{\min(l-n,l+n)}_{j=\max(0,n-m)}
\frac{i^{2j}\tg^{2j}\dfrac{\theta}{2}}
{\Gamma(j+1)\Gamma(l-m-j+1)\Gamma(l+n-j+1)\Gamma(m-n+j+1)}.\label{Mat1}
\end{multline}
Further, using the formula
\begin{equation}\label{HG}
%{}_2F_1(\alpha,\beta;\gamma;z)
\hypergeom{2}{1}{\alpha,\beta}{\gamma}{z}
=\frac{\Gamma(\gamma)}{\Gamma(\alpha)
\Gamma(\beta)}\sum_{k\ge 0}\frac{\Gamma(\alpha+k)\Gamma(\beta+k)}
{\Gamma(\gamma+k)}\frac{z^k}{k!}
\end{equation}
we express the matrix element (\ref{Mat1}) via the hypergeometric
function:
\begin{multline}
t^l_{mn}(g)=\frac{i^{m-n}e^{-i(m\varphi+n\psi)}}{\Gamma(m-n+1)}
\sqrt{\frac{\Gamma(l+m+1)\Gamma(l-n+1)}{\Gamma(l-m+1)\Gamma(l+n+1)}}\times\\
\cos^{2l}\frac{\theta}{2}\tg^{m-n}\frac{\theta}{2}
\hypergeom{2}{1}{m-l+1,1-l-n}{m-n+1}{i^2\tg^2\dfrac{\theta}{2}},\label{Mat2}
\end{multline}
where $m\ge n$. At $m<n$ in the right part of (\ref{Mat2}) it needs to
replace $m$ and $n$ by $-m$ and $-n$, respectively. Since $l,m$ and $n$
are finite numbers, then the hypergeometric series is interrupted.

Further, replacing in the one--parameter subgroups (\ref{PS1}) the
parameter $t$ by $-it$ we obtain
\begin{equation}{\renewcommand{\arraystretch}{1.7}
b_1(t)=
\begin{pmatrix}
\ch\dfrac{t}{2} & \sh\dfrac{t}{2}\\
\sh\dfrac{t}{2} & \ch\dfrac{t}{2}
\end{pmatrix},\quad
b_2(t)=\begin{pmatrix}
\ch\dfrac{t}{2} & i\sh\dfrac{t}{2}\\
-i\sh\dfrac{t}{2} & \ch\dfrac{t}{2}
\end{pmatrix}},\quad
{\renewcommand{\arraystretch}{1.05}
b_3(t)=\begin{pmatrix}
e^{\frac{t}{2}} & 0\\
0 & e^{-\frac{t}{2}}
\end{pmatrix}}.\label{PS2}
\end{equation}
These subgroups correspond to hyperbolic rotations. 

The group $SL(2,\C)$ of all complex matrices
\[\ar
\begin{pmatrix}
\alpha & \beta\\
\gamma & \delta
\end{pmatrix}
\]
of 2-nd order with the determinant $\alpha\delta-\gamma\beta=1$, is
a {\it complexification} of the group $SU(2)$. The group $SU(2)$ is one of
the real forms of $SL(2,\C)$. The transition from $SU(2)$ to $SL(2,\C)$
is realized via the complexification of three real parameters
$\varphi,\,\theta,\,\psi$ (Euler angles). Let $\theta^c=\theta-i\tau$,
$\varphi^c=\varphi-i\epsilon$, $\psi^c=\psi-i\varepsilon$ be complex
Euler angles, where
\[
{\renewcommand{\arraystretch}{1.05}
\begin{array}{ccccc}
0 &\leq&\re\theta^c=\theta& \leq& \pi,\\
0 &\leq&\re\varphi^c=\varphi& <&2\pi,\\
-2\pi&\leq&\re\psi^c=\psi&<&2\pi,
\end{array}\quad\quad
\begin{array}{ccccc}
-\infty &<&\im\theta^c=\tau&<&+\infty,\\
-\infty&<&\im\varphi^c=\epsilon&<&+\infty,\\
-\infty&<&\im\psi^c=\varepsilon&<&+\infty.
\end{array}}
\]
Replacing in (\ref{SU2}) the angles $\varphi,\,\theta,\,\psi$ by the
complex angles $\varphi^c,\theta^c,\psi^c$ we come to the following matrix
%\[
\begin{gather}
{\renewcommand{\arraystretch}{1.7}
\mathfrak{g}=
\begin{pmatrix}
\cos\dfrac{\theta^c}{2}e^{\frac{i(\varphi^c+\psi^c)}{2}} &
i\sin\dfrac{\theta^c}{2}e^{\frac{i(\varphi^c-\psi^c)}{2}}\\
i\sin\dfrac{\theta^c}{2}e^{\frac{i(\psi^c-\varphi^c)}{2}} &
\cos\dfrac{\theta^c}{2}e^{-\frac{i(\varphi^c+\psi^c)}{2}}
\end{pmatrix}}=
\nonumber\\
%\]
%\begin{multline}
{\renewcommand{\arraystretch}{1.9}
\begin{pmatrix}
\left[\cos\dfrac{\theta}{2}\ch\dfrac{\tau}{2}+
i\sin\dfrac{\theta}{2}\sh\dfrac{\tau}{2}\right]
e^{\frac{\epsilon+\varepsilon+i(\varphi+\psi)}{2}} &
\left[\cos\dfrac{\theta}{2}\sh\dfrac{\tau}{2}+
i\sin\dfrac{\theta}{2}\ch\dfrac{\tau}{2}\right]
e^{\frac{\epsilon-\varepsilon+i(\varphi-\psi)}{2}} \\
%\end{array}
%\right.}\\
%{\renewcommand{\arraystretch}{1.1}
%\left.\begin{array}{c}
\left[\cos\dfrac{\theta}{2}\sh\dfrac{\tau}{2}+
i\sin\dfrac{\theta}{2}\ch\dfrac{\tau}{2}\right]
e^{\frac{\varepsilon-\epsilon+i(\psi-\varphi)}{2}} &
\left[\cos\dfrac{\theta}{2}\ch\dfrac{\tau}{2}+
i\sin\dfrac{\theta}{2}\sh\dfrac{\tau}{2}\right]
e^{\frac{-\epsilon-\varepsilon-i(\varphi+\psi)}{2}}
\end{pmatrix},}\label{SL1}
\end{gather}
since $\cos\dfrac{1}{2}(\theta-i\tau)=\cos\dfrac{\theta}{2}\ch\dfrac{\tau}{2}+
i\sin\dfrac{\theta}{2}\sh\dfrac{\tau}{2}$, and 
$\sin\dfrac{1}{2}(\theta-i\tau)=\sin\dfrac{\theta}{2}\ch\dfrac{\tau}{2}-
i\cos\dfrac{\theta}{2}\sh\dfrac{\tau}{2}$. It is easy to verify that the
matrix (\ref{SL1}) coincides with a matrix of the fundamental reprsentation
of the group $SL(2,\C)$ (in Euler parametrization):
\begin{multline}
\mathfrak{g}(\varphi,\,\epsilon,\,\theta,\,\tau,\,\psi,\,\varepsilon)=\\[0.2cm]
{\renewcommand{\arraystretch}{1.05}
\begin{pmatrix}
e^{i\frac{\varphi}{2}} & 0\\
0 & e^{-i\frac{\varphi}{2}}
\end{pmatrix}}{\renewcommand{\arraystretch}{1.4}\!\!\begin{pmatrix}
e^{\frac{\epsilon}{2}} & 0\\
0 & e^{-\frac{\epsilon}{2}}
\end{pmatrix}}\!\!{\renewcommand{\arraystretch}{1.7}\begin{pmatrix}
\cos\dfrac{\theta}{2} & i\sin\dfrac{\theta}{2}\\
i\sin\dfrac{\theta}{2} & \cos\dfrac{\theta}{2}
\end{pmatrix}
%{\renewcommand{\arraystretch}{1.05}
\begin{pmatrix}
\ch\dfrac{\tau}{2} & \sh\dfrac{\tau}{2}\\
\sh\dfrac{\tau}{2} & \ch\dfrac{\tau}{2}
\end{pmatrix}}\!\!{\renewcommand{\arraystretch}{1.4}\begin{pmatrix}
e^{i\frac{\psi}{2}} & 0\\
0 & e^{-i\frac{\psi}{2}}
\end{pmatrix}}\!\!{\renewcommand{\arraystretch}{1.05}\begin{pmatrix}
e^{\frac{\varepsilon}{2}} & 0\\
0 & e^{-\frac{\varepsilon}{2}}
\end{pmatrix}.}\label{FUN}
\end{multline}

The matrix element $t^l_{mn}=e^{-m(\epsilon+i\varphi)-n(\varepsilon+
i\psi)}\langle T_l(\theta,\tau)\psi_\lambda,\psi_{\dot{\lambda}}\rangle$
of the finite--dimensional repsesentation of $SL(2,\C)$ at $l=\dot{l}$
in the polynomial basis
\[
\psi_\lambda(\zeta,\bar{\zeta})=\frac{\zeta^{l-n}\bar{\zeta}^{l-m}}
{\sqrt{\Gamma(l-n+1)\Gamma(l+n+1)\Gamma(l-m+1)\Gamma(l+m+1)}},
\]
%where
%\begin{multline}
%T_l(\theta,\tau)\psi(\zeta,\bar{zeta})=\left(i\sin\dfrac{\theta}{2}\zeta+
%\cos\dfrac{\theta}{2}\right)^{2l}\left(\sh\dfrac{\tau}{2}\bar{\zeta}+
%\ch\dfrac{\tau}{2}\right)^{2l}\times\\
%\psi\left(\frac{\cos\dfrac{\theta}{2}\zeta+
%i\sin\dfrac{\theta}{2}}{i\sin\dfrac{\theta}{2}\zeta+\cos\dfrac{\theta}{2}},\,
%\frac{\ch\dfrac{\tau}{2}\bar{\zeta}+\sh\dfrac{\tau}{2}}
%{\sh\dfrac{\tau}{2}\bar{\zeta}+\ch\dfrac{\tau}{2}}\right)\nonumber
%\end{multline}
has a form
\begin{multline}
t^l_{mn}(\mathfrak{g})=e^{-m(\epsilon+i\varphi)-n(\varepsilon+i\psi)}
Z^l_{mn}=e^{-m(\epsilon+i\varphi)-n(\varepsilon+i\psi)}\times\\[0.2cm]
\sum^l_{k=-l}i^{m-k}
\sqrt{\Gamma(l-m+1)\Gamma(l+m+1)\Gamma(l-k+1)\Gamma(l+k+1)}\times\\
\cos^{2l}\frac{\theta}{2}\tg^{m-k}\frac{\theta}{2}\times\\[0.2cm]
\sum^{\min(l-m,l+k)}_{j=\max(0,k-m)}
\frac{i^{2j}\tg^{2j}\dfrac{\theta}{2}}
{\Gamma(j+1)\Gamma(l-m-j+1)\Gamma(l+k-j+1)\Gamma(m-k+j+1)}\times\\[0.2cm]
\sqrt{\Gamma(l-n+1)\Gamma(l+n+1)\Gamma(l-k+1)\Gamma(l+k+1)}
\ch^{2l}\frac{\tau}{2}\tnh^{n-k}\frac{\tau}{2}\times\\[0.2cm]
\sum^{\min(l-n,l+k)}_{s=\max(0,k-n)}
\frac{\tnh^{2s}\dfrac{\tau}{2}}
{\Gamma(s+1)\Gamma(l-n-s+1)\Gamma(l+k-s+1)\Gamma(n-k+s+1)}.\label{HS}
\end{multline}
We will call the functions $Z^l_{mn}$ in (\ref{HS}) as
{\it hyperspherical functions}. Using (\ref{HG}) we can write the
hyperspherical functions $Z^l_{mn}$ via the hypergeometric series:
\begin{multline}
Z^l_{mn}=\cos^{2l}\frac{\theta}{2}\ch^{2l}\frac{\tau}{2}
\sum^l_{k=-l}i^{m-k}\tg^{m-k}\frac{\theta}{2}
\tnh^{n-k}\frac{\tau}{2}\times\\[0.2cm]
\hypergeom{2}{1}{m-l+1,1-l-k}{m-k+1}{i^2\tg^2\dfrac{\theta}{2}}
\hypergeom{2}{1}{n-l+1,1-l-k}{n-k+1}{\tnh^2\dfrac{\tau}{2}}.\label{HS1}
\end{multline}
Therefore, matrix elements can be expressed by means of the function
({\it a generalized hyperspherical function})
\begin{equation}\label{HS2}
\fT^l_{mn}(\mathfrak{g})=e^{-m(\epsilon+i\varphi)}Z^l_{mn}
e^{-n(\varepsilon+i\psi)},
\end{equation}
where
\begin{equation}\label{HS3}
Z^l_{mn}=\sum^l_{k=-l}P^l_{mk}(\cos\theta)\mathfrak{P}^l_{kn}(\ch\tau),
\end{equation}
here $P^l_{mn}(\cos\theta)$ is a generalized spherical function on the
group $SU(2)$ (see \cite{GMS}), and $\mathfrak{P}^l_{mn}$ is an analog of
the generalized spherical function for the group $QU(2)$ (so--called
Jacobi function \cite{Vil68}). $QU(2)$ is a group of quasiunitary
unimodular matrices of second order. As well as the group $SU(2)$, the
group $QU(2)$ is one of the real forms of $SL(2,\C)$
($QU(2)$ is noncompact).

Infinitesimal operators of the group $\fG_+$ can be expressed via the
Euler angles as follows (for detailed deriving of the formulae see
\cite{Var022})
\begin{eqnarray}
\sA_1&=&\cos\psi^c\frac{\partial}{\partial\theta}+
\frac{\sin\psi^c}{\sin\theta^c}\frac{\partial}{\partial\varphi}-
\ctg\theta^c\sin\psi^c\frac{\partial}{\partial\psi},\label{SL12}\\
\sA_2&=&-\sin\psi^c\frac{\partial}{\partial\theta}+
\frac{\cos\psi^c}{\sin\theta^c}\frac{\partial}{\partial\varphi}-
\ctg\theta^c\cos\psi^c\frac{\partial}{\partial\psi},\label{SL24'}\\
\sA_3&=&\frac{\partial}{\partial\psi},\label{SL8}\\
\sB_1&=&\cos\psi^c\frac{\partial}{\partial\tau}+
\frac{\sin\psi^c}{\sin\theta^c}\frac{\partial}{\partial\epsilon}-
\ctg\theta^c\sin\psi^c\frac{\partial}{\partial\varepsilon},\label{SL13}\\
\sB_2&=&-\sin\psi^c\frac{\partial}{\partial\tau}+
\frac{\cos\psi^c}{\sin\theta^c}\frac{\partial}{\partial\epsilon}-
\ctg\theta^c\cos\psi^c\frac{\partial}{\partial\varepsilon},\label{SL24''}\\
\sB_3&=&\frac{\partial}{\partial\varepsilon}.\label{SL8'}
\end{eqnarray}
It is easy to verify that operators $\sA_i$, $\sB_i$,
defined by the formulae (\ref{SL8}),
(\ref{SL8'}), (\ref{SL12}), (\ref{SL13}) and (\ref{SL24'}), (\ref{SL24''}),
are satisfy the commutation relations (\ref{Com1}).
\subsection{Casimir operators and differential equations for hyperspherical
functions}
Taking into account the expressions (\ref{SL12})-(\ref{SL8'}) we can write
the operators (\ref{SL25}) in the form
\begin{eqnarray}
\sX_1&=&\cos\psi^c\frac{\partial}{\partial\theta^c}+
\frac{\sin\psi^c}{\sin\theta^c}\frac{\partial}{\partial\varphi^c}-
\ctg\theta^c\sin\psi^c\frac{\partial}{\partial\psi^c},\label{X1}\\
\sX_2&=&-\sin\psi^c\frac{\partial}{\partial\theta^c}+
\frac{\cos\psi^c}{\sin\theta^c}\frac{\partial}{\partial\varpi^c}-
\ctg\theta^c\cos\psi^c\frac{\partial}{\partial\psi^c},\label{X2}\\
\sX_3&=&\frac{\partial}{\partial\psi^c},\label{X3}\\
\sY_1&=&\cos\dot{\psi}^c\frac{\partial}{\partial\dot{\theta}^c}+
\frac{\sin\dot{\psi}^c}{\sin\dot{\theta}^c}
\frac{\partial}{\partial\dot{\varphi}^c}-
\ctg\dot{\theta}^c\sin\dot{\psi}^c\frac{\partial}{\partial\dot{\psi}^c},
\label{Y1}\\
\sY_2&=&-\sin\dot{\psi}^c\frac{\partial}{\partial\dot{\theta}^c}+
\frac{\cos\dot{\psi}^c}{\sin\dot{\theta}^c}
\frac{\partial}{\partial\dot{\varphi}^c}-
\ctg\dot{\theta}^c\cos\dot{\psi}^c\frac{\partial}{\partial\dot{\psi}^c},
\label{Y2}\\
\sY_3&=&\frac{\partial}{\partial\dot{\psi}^c},\label{Y3}
\end{eqnarray}
where
\begin{gather}
\frac{\partial}{\partial\theta^c}=\frac{1}{2}\left(
\frac{\partial}{\partial\theta}+i\frac{\partial}{\partial\tau}\right),\;\;
\frac{\partial}{\partial\varphi^c}=\frac{1}{2}\left(
\frac{\partial}{\partial\varphi}+i\frac{\partial}{\partial\epsilon}\right),\;\;
\frac{\partial}{\partial\psi^c}=\frac{1}{2}\left(
\frac{\partial}{\partial\psi}+i\frac{\partial}{\partial\varepsilon}\right),
\nonumber\\
\frac{\partial}{\partial\dot{\theta}^c}=\frac{1}{2}\left(
\frac{\partial}{\partial\theta}-i\frac{\partial}{\partial\tau}\right),\;\;
\frac{\partial}{\partial\dot{\varphi}^c}=\frac{1}{2}\left(
\frac{\partial}{\partial\varphi}-i\frac{\partial}{\partial\epsilon}\right),\;\;
\frac{\partial}{\partial\dot{\psi}^c}=\frac{1}{2}\left(
\frac{\partial}{\partial\psi}-i\frac{\partial}{\partial\varepsilon}\right),
\nonumber
\end{gather}
As known, for the Lorentz group there are two independent Casimir operators
\begin{eqnarray}
\sX^2&=&\sX^2_1+\sX^2_2+\sX^2_3=\frac{1}{4}(\sA^2-\sB^2+2i\sA\sB),\nonumber\\
\sY^2&=&\sY^2_1+\sY^2_2+\sY^2_3=\frac{1}{4}(\sA^2-\sB^2-2i\sA\sB).\label{KO}
\end{eqnarray}
Substituting (\ref{X1})-(\ref{Y3}) into (\ref{KO}) we obtain in Euler
parametrization for the Casimir operators the following expressions
\begin{eqnarray}
\sX^2&=&\frac{\partial^2}{\partial\theta^c{}^2}+
\ctg\theta^c\frac{\partial}{\partial\theta^c}+\frac{1}{\sin^2\theta^c}\left[
\frac{\partial^2}{\partial\varphi^c{}^2}-
2\cos\theta^c\frac{\partial}{\partial\varphi^c}
\frac{\partial}{\partial\psi^c}+
\frac{\partial^2}{\partial\psi^c{}^2}\right],\nonumber\\
\sY^2&=&\frac{\partial^2}{\partial\dot{\theta}^c{}^2}+
\ctg\dot{\theta}^c\frac{\partial}{\partial\dot{\theta}^c}+
\frac{1}{\sin^2\dot{\theta}^c}\left[
\frac{\partial^2}{\partial\dot{\varphi}^c{}^2}-
2\cos\dot{\theta}^c\frac{\partial}{\partial\dot{\varphi}^c}
\frac{\partial}{\partial\dot{\psi}^c}+
\frac{\partial^2}{\partial\dot{\psi}^c{}^2}\right].\label{KO2}
\end{eqnarray}
Matrix elements of unitary irreducible representations of the Lorentz
group are eigenfunctions of the operators (\ref{KO2}):
\begin{eqnarray}
\left[\sX^2+l(l+1)\right]\fM^l_{mn}(\varphi^c,\theta^c,\psi^c)&=&0,\nonumber\\
\left[\sY^2+\dot{l}(\dot{l}+1)\right]\fM^{\dot{l}}_{\dot{m}\dot{n}}
(\dot{\varphi}^c,\dot{\theta}^c,\dot{\psi}^c)&=&0,\label{EQ}
\end{eqnarray}
where
\begin{eqnarray}
\fM^l_{mn}(\varphi^c,\theta^c,\psi^c)&=&e^{-i(m\varphi^c+n\psi^c)}
Z^l_{mn}(\theta^c),\nonumber\\
\fM^{\dot{l}}_{\dot{m}\dot{n}}(\dot{\varphi}^c,\dot{\theta}^c,\dot{\psi}^c)&=&
e^{-i(\dot{m}\dot{\varphi}^c+\dot{n}\dot{\varphi}^c)}
Z^{\dot{l}}_{\dot{m}\dot{n}}(\dot{\theta}^c).\label{HF3}
\end{eqnarray}
Substituting the hyperspherical functions (\ref{HF3}) into (\ref{EQ}) and
taking into account the operators (\ref{KO2}) we obtain 
\begin{eqnarray}
\left[\frac{d^2}{d\theta^c{}^2}+\ctg\theta^c\frac{d}{d\theta^c}-
\frac{m^2+n^2-2mn\cos\theta^c}{\sin^2\theta^c}+l(l+1)\right]
Z^l_{mn}(\theta^c)&=&0,\nonumber\\
\left[\frac{d^2}{d\dot{\theta}^c{}^2}+\ctg\dot{\theta}^c
\frac{d}{d\dot{\theta}^c}-\frac{\dot{m}+\dot{n}-
2\dot{m}\dot{n}\cos\dot{\theta}^c}{\sin^2\dot{\theta}^c}+
\dot{l}(\dot{l}+1)\right]Z^{\dot{l}}_{\dot{m}\dot{n}}(\dot{\theta}^c)
&=&0.\nonumber
\end{eqnarray}
Finally, after substitutions $z=\cos\theta^c$ and 
$\overset{\ast}{z}=\cos\dot{\theta}^c$ we come to the following
differential equations
\begin{eqnarray}
\left[(1-z^2)\frac{d^2}{dz^2}-2z\frac{d}{dz}-
\frac{m^2+n^2-3mnz}{1-z^2}+l(l+1)\right]Z^l_{mn}(\arc z)&=&0,\nonumber\\
\left[(1-\overset{\ast}{z}{}^2)\frac{d^2}{d\overset{\ast}{z}{}^2}-
2\overset{\ast}{z}\frac{d}{d\overset{\ast}{z}}-
\frac{\dot{m}^2+\dot{n}^2-2\dot{m}\dot{n}\overset{\ast}{z}}
{1-\overset{\ast}{z}{}^2}+\dot{l}(\dot{l}+1)\right]
Z^{\dot{l}}_{\dot{m}\dot{n}}(\arc\overset{\ast}{z})&=&0.\nonumber
\end{eqnarray}
\subsection{Recurrence relations between hyperspherical functions}
Between generalized hyperspherical functions $\fT^l_{mn}$ (and also the
hyperspherical functions $Z^l_{mn}$) there exists a wide variety of
recurrence relations. 
Part of them relates the hyperspherical functions
of one and the same order (with identical $l$), other part relates the
functions of different orders (for more details see \cite{Var022}).

In virtue of the Van der Waerden representation (\ref{Waerden}) the
recurrence formulae for the hyperspherical functions of one and the same
order follow from the equalities
\begin{eqnarray}
&&\sX_-\dot{\fT}^l_{mn}=\dot{\boldsymbol{\alpha}}_n\dot{\fT}^l_{m,n-1},\quad
\sX_+\dot{\fT}^l_{mn}=\dot{\boldsymbol{\alpha}}_{n+1}
\dot{\fT}^l_{m,n+1},\label{SL26'}\\
&&\sY_-\fT^l_{mn}=\boldsymbol{\alpha}_n\fT^l_{m,n-1},\quad
\sY_+\fT^l_{mn}=\boldsymbol{\alpha}_{n+1}
\fT^l_{m,n+1},\label{SL26''}
\end{eqnarray}
where
\[
\dot{\boldsymbol{\alpha}}_n=\sqrt{(\dot{l}+\dot{n})(\dot{l}-\dot{n}+1)},
\quad
\boldsymbol{\alpha}_n=\sqrt{(l+n)(l-n+1)}.
\]
%From (\ref{SL25}) and (\ref{SL26}) it follows that
%\begin{eqnarray}
%\sX_+&=&\frac{1}{2}\left(\sA_1+i\sA_2+i\sB_1-\sB_2\right),\nonumber\\
%\sX_-&=&\frac{1}{2}\left(\sA_1-i\sA_2+i\sB_1+\sB_2\right),\nonumber\\
%\sY_+&=&\frac{1}{2}\left(\sA_1+i\sA_2-i\sB_1+\sB_2\right),\nonumber\\
%\sY_-&=&\frac{1}{2}\left(\sA_1-i\sA_2-i\sB_1-\sB_2\right).\nonumber
%\end{eqnarray}
Using the formulae (\ref{SL12}), (\ref{SL13}) and (\ref{SL24'}), (\ref{SL24''})
we obtain
\begin{eqnarray}
\sX_+&=&\frac{e^{-i\psi^c}}{2}\left[\frac{\partial}{\partial\theta}+
\frac{i}{\sin\theta^c}\frac{\partial}{\partial\varphi}
-i\ctg\theta^c\frac{\partial}{\partial\psi}+i\frac{\partial}{\partial\tau}-
\frac{1}{\sin\theta^c}\frac{\partial}{\partial\epsilon}+
\ctg\theta^c\frac{\partial}{\partial\varepsilon}\right],\label{SL27}\\
\sX_-&=&\frac{e^{i\psi^c}}{2}\left[\frac{\partial}{\partial\theta}-
\frac{i}{\sin\theta^c}\frac{\partial}{\partial\varphi}
+i\ctg\theta^c\frac{\partial}{\partial\psi}+i\frac{\partial}{\partial\tau}+
\frac{1}{\sin\theta^c}\frac{\partial}{\partial\epsilon}-
\ctg\theta^c\frac{\partial}{\partial\varepsilon}\right],\label{SL28}\\
\sY_+&=&\frac{e^{-i\psi^c}}{2}\left[\frac{\partial}{\partial\theta}+
\frac{i}{\sin\theta^c}\frac{\partial}{\partial\varphi}-
i\ctg\theta^c\frac{\partial}{\partial\psi}-i\frac{\partial}{\partial\tau}+
\frac{1}{\sin\theta^c}\frac{\partial}{\partial\epsilon}-
\ctg\theta^c\frac{\partial}{\partial\varepsilon}\right],\label{SL29}\\
\sY_-&=&\frac{e^{i\psi^c}}{2}\left[\frac{\partial}{\partial\theta}-
\frac{i}{\sin\theta^c}\frac{\partial}{\partial\varphi}
+i\ctg\theta^c\frac{\partial}{\partial\psi}-i\frac{\partial}{\partial\tau}-
\frac{1}{\sin\theta^c}\frac{\partial}{\partial\epsilon}+
\ctg\theta^c\frac{\partial}{\partial\varepsilon}\right].\label{SL30}
\end{eqnarray}
Further, substituting the function $\fT^l_{mn}=e^{-m(\epsilon-i\varphi)}
\do{Z}^l_{mn}(\theta,\tau)e^{-n(\varepsilon-i\psi)}$ 
into the relations (\ref{SL26'})
and taking into account the operators (\ref{SL27}) and (\ref{SL28})
we find that
\begin{eqnarray}
\frac{\partial\dot{Z}^l_{mn}}{\partial\theta}+
i\frac{\partial\dot{Z}^l_{mn}}{\partial\tau}-
\frac{2(m-n\cos\theta^c)}{\sin\theta^c}\dot{Z}^l_{mn}&=&
2\boldsymbol{\alpha}^\prime_n\dot{Z}^l_{m,n-1},\label{SL31}\\
\frac{\partial\dot{Z}^l_{mn}}{\partial\theta}+
i\frac{\partial\dot{Z}^l_{mn}}{\partial\tau}+
\frac{2(m-n\cos\theta^c)}{\sin\theta^c}\dot{Z}^l_{mn}&=&
2\boldsymbol{\alpha}^\prime_{n+1}\dot{Z}^l_{m,n+1}.\label{SL32}
\end{eqnarray}
Since the functions $\dot{Z}^l_{mn}(\theta,\tau)$ are symmetric, that is
$\dot{Z}^l_{mn}(\theta,\tau)=\dot{Z}^l_{nm}(\theta,\tau)$, then substituting
$\dot{Z}^l_{nm}(\theta,\tau)$ in lieu of $\dot{Z}^l_{mn}$ into the formulae
(\ref{SL31})--(\ref{SL32}) and replacing $m$ by $n$, and $n$ by $m$,
we obtain
\begin{eqnarray}
\frac{\partial Z^l_{mn}}{\partial\theta}+
i\frac{\partial Z^l_{mn}}{\partial\tau}-
\frac{2(n-m\cos\theta^c)}{\sin\theta^c}Z^l_{mn}&=&
2\boldsymbol{\alpha}^\prime_mZ^l_{m-1,n},\label{SL33}\\
\frac{\partial Z^l_{mn}}{\partial\theta}+
i\frac{\partial Z^l_{mn}}{\partial\tau}+
\frac{2(n-m\cos\theta^c)}{\sin\theta^c}Z^l_{mn}&=&
2\boldsymbol{\alpha}^\prime_{m+1}Z^l_{m+1,n}.\label{SL34}
\end{eqnarray}
Analogously, for the relations (\ref{SL26''}) we have
\begin{eqnarray}
\frac{\partial\dot{Z}^l_{mn}}{\partial\theta}
-i\frac{\partial\dot{Z}^l_{mn}}{\partial\tau}+
\frac{2(n-m\cos\theta^c)}{\sin\theta^c}\dot{Z}^l_{mn}&=&
2\boldsymbol{\alpha}_m\dot{Z}^l_{m-1,n},\label{SL39}\\
\frac{\partial\dot{Z}^l_{mn}}{\partial\theta}
-i\frac{\partial\dot{Z}^l_{mn}}{\partial\tau}-
\frac{2(n-m\cos\theta^c)}{\sin\theta^c}\dot{Z}^l_{mn}&=&
2\boldsymbol{\alpha}_{m+1}\dot{Z}^l_{m+1,n}.\label{SL40}
\end{eqnarray}

\section{Two-dimensional complex sphere and separation of variables}
\subsection{Generalized Gel'fand-Yaglom equations}
In the three-dimensional Euclidean space $\R^3$ the functions
$\psi_1(x_1,x_2,x_3)$, $\psi_2(x_1,x_2,x_3)$, $\ldots$, $\psi_N(x_1,x_2,x_3)$
satisfy the following system of invariant (under action of the group $SU(2)$)
equations \cite{GMS}:
\begin{equation}\label{3DE}
\sum^3_{i=1}L_i\frac{\partial
\boldsymbol{\psi}}{\partial x_i}+\kappa\boldsymbol{\psi}=0,
\end{equation}
where $L_i$ are $n$-dimensional matrices and $\kappa$ is a number.
Follows to the general group complexification 
$SL(2,\C)\sim\mbox{\sf complex}(SU(2))$ let us introduce a complex analog
of the equations (\ref{3DE}) in the three-dimensional complex space
$\C^3$:
\begin{eqnarray}
\sum^3_{i=1}\sL_i\frac{\partial\boldsymbol{\psi}}{\partial x_i}-
i\sum^3_{i=1}\sL_i\frac{\partial\boldsymbol{\psi}}{\partial x^\ast_i}
+\kappa^c\boldsymbol{\psi}&=&0,\nonumber\\
\sum^3_{i=1}\sL^\ast_i\frac{\partial\dot{\boldsymbol{\psi}}}
{\partial\widetilde{x}_i}+
i\sum^3_{i=1}\sL^\ast_i\frac{\partial\dot{\boldsymbol{\psi}}}
{\partial\widetilde{x}^\ast_i}+
\dot{\kappa}^c\dot{\boldsymbol{\psi}}&=&0,\label{Complex}
\end{eqnarray}
where $\sL_i$, $\sL^\ast_i$ are $n$-dimensional matrices and
$\kappa^c$ is a complex number. We will call these equations as
{\it generalized Gel'fand-Yaglom equations}.

The elements of the matrices $\sL_i$, $\sL^\ast_i$ are

\begin{equation}\label{L1}
{\renewcommand{\arraystretch}{1.6}
\sL_1:\quad\left\{\begin{array}{ccc}
a^{k^\prime k}_{l-1,l,m-1,m}&=&
-\dfrac{c_{l-1,l}}{2}\sqrt{(l+m)(l+m-1)},\\
a^{k^\prime k}_{l,l,m-1,m}&=&
\dfrac{c_{ll}}{2}\sqrt{(l+m)(l-m+1)},\\
a^{k^\prime k}_{l+1,l,m-1,m}&=&
\dfrac{c_{l+1,l}}{2}\sqrt{(l-m+1)(l-m+2)},\\
a^{k^\prime k}_{l-1,l,m+1,m}&=&
\dfrac{c_{l-1,l}}{2}\sqrt{(l-m)(l-m-1)},\\
a^{k^\prime k}_{l,l,m+1,m}&=&
\dfrac{c_{ll}}{2}\sqrt{(l+m+1)(l-m)},\\
a^{k^\prime k}_{l+1,l,m+1,m}&=&
-\dfrac{c_{l+1,l}}{2}\sqrt{(l+m+1)(l+m+2)}.
\end{array}\right.}
\end{equation}
\begin{equation}\label{L2}
{\renewcommand{\arraystretch}{1.6}
\sL_2:\quad\left\{\begin{array}{ccc}
b^{k^\prime k}_{l-1,l,m-1,m}&=&
-\dfrac{ic_{l-1,l}}{2}\sqrt{(l+m)(l+m-1)},\\
b^{k^\prime k}_{l,l,m-1,m}&=&
\dfrac{ic_{ll}}{2}\sqrt{(l+m)(l-m+1)},\\
b^{k^\prime k}_{l+1,l,m-1,m}&=&
\dfrac{ic_{l+1,l}}{2}\sqrt{(l-m+1)(l-m+2)},\\
b^{k^\prime k}_{l-1,l,m+1,m}&=&
-\dfrac{ic_{l-1,l}}{2}\sqrt{(l-m)(l-m-1)},\\
b^{k^\prime k}_{l,l,m+1,m}&=&
-\dfrac{ic_{ll}}{2}\sqrt{(l+m+1)(l-m)},\\
b^{k^\prime k}_{l+1,l,m+1,m}&=&
\dfrac{ic_{l+1,l}}{2}\sqrt{(l+m+1)(l+m+2)}.
\end{array}\right.}
\end{equation}
\begin{equation}\label{L3}
{\renewcommand{\arraystretch}{1.6}
\sL_3:\quad\left\{\begin{array}{ccc}
c^{k^\prime k}_{l-1,l,m}&=&
c^{k^\prime k}_{l-1,l}\sqrt{l^2-m^2},\\
c^{k^\prime k}_{l,l,m}&=&c^{k^\prime k}_{ll}m,\\
c^{k^\prime k}_{l+1,l,m}&=&
c^{k^\prime k}_{l+1,l}\sqrt{(l+1)^2-m^2}.
\end{array}\right.}
\end{equation}
\begin{equation}\label{L1'}
{\renewcommand{\arraystretch}{1.6}
\sL^\ast_1:\quad\left\{\begin{array}{ccc}
d^{\dot{k}^\prime\dot{k}}_{\dot{l}-1,\dot{l},\dot{m}-1,\dot{m}}&=&
-\dfrac{c_{\dot{l}-1,\dot{l}}}{2}
\sqrt{(\dot{l}+\dot{m})(\dot{l}-\dot{m}-1)},\\
d^{\dot{k}^\prime\dot{k}}_{\dot{l},\dot{l},\dot{m}-1,\dot{m}}&=&
\dfrac{c_{\dot{l}\dot{l}}}{2}
\sqrt{(\dot{l}+\dot{m})(\dot{l}-\dot{m}+1)},\\
d^{\dot{k}^\prime\dot{k}}_{\dot{l}+1,\dot{l},\dot{m}-1,\dot{m}}&=&
\dfrac{c_{\dot{l}+1,\dot{l}}}{2}
\sqrt{(\dot{l}-\dot{m}+1)(\dot{l}-\dot{m}+2)},\\
d^{\dot{k}^\prime\dot{k}}_{\dot{l}-1,\dot{l},\dot{m}+1,\dot{m}}&=&
\dfrac{c_{\dot{l}-1,\dot{l}}}{2}
\sqrt{(\dot{l}-\dot{m})(\dot{l}-\dot{m}-1)},\\
d^{\dot{k}^\prime\dot{k}}_{\dot{l},\dot{l},\dot{m}+1,\dot{m}}&=&
\dfrac{c_{\dot{l}\dot{l}}}{2}
\sqrt{(\dot{l}+\dot{m}+1)(\dot{l}-\dot{m})},\\
d^{\dot{k}^\prime\dot{k}}_{\dot{l}+1,\dot{l},\dot{m}+1,\dot{m}}&=&
-\dfrac{c_{\dot{l}+1,\dot{l}}}{2}
\sqrt{(\dot{l}+\dot{m}+1)(\dot{l}+\dot{m}+2)}.
\end{array}\right.}
\end{equation}
\begin{equation}\label{L2'}
{\renewcommand{\arraystretch}{1.6}
\sL^\ast_2:\quad\left\{\begin{array}{ccc}
e^{\dot{k}^\prime\dot{k}}_{\dot{l}-1,\dot{l},\dot{m}-1,\dot{m}}&=&
-\dfrac{ic_{\dot{l}-1,\dot{l}}}{2}
\sqrt{(\dot{l}+\dot{m})(\dot{l}-\dot{m}-1)},\\
e^{\dot{k}^\prime\dot{k}}_{\dot{l},\dot{l},\dot{m}-1,\dot{m}}&=&
\dfrac{ic_{\dot{l}\dot{l}}}{2}
\sqrt{(\dot{l}+\dot{m})(\dot{l}-\dot{m}+1)},\\
e^{\dot{k}^\prime\dot{k}}_{\dot{l}+1,\dot{l},\dot{m}-1,\dot{m}}&=&
\dfrac{ic_{\dot{l}+1,\dot{l}}}{2}
\sqrt{(\dot{l}-\dot{m}+1)(\dot{l}-\dot{m}+2)},\\
e^{\dot{k}^\prime\dot{k}}_{\dot{l}-1,\dot{l},\dot{m}+1,\dot{m}}&=&
\dfrac{-ic_{\dot{l}-1,\dot{l}}}{2}
\sqrt{(\dot{l}-\dot{m})(\dot{l}-\dot{m}-1)},\\
e^{\dot{k}^\prime\dot{k}}_{\dot{l},\dot{l},\dot{m}+1,\dot{m}}&=&
\dfrac{-ic_{\dot{l}\dot{l}}}{2}
\sqrt{(\dot{l}+\dot{m}+1)(\dot{l}-\dot{m})},\\
e^{\dot{k}^\prime\dot{k}}_{\dot{l}+1,\dot{l},\dot{m}+1,\dot{m}}&=&
-\dfrac{ic_{\dot{l}+1,\dot{l}}}{2}
\sqrt{(\dot{l}+\dot{m}+1)(\dot{l}+\dot{m}+2)}.
\end{array}\right.}
\end{equation}
\begin{equation}\label{L3'}
{\renewcommand{\arraystretch}{1.6}
\sL^\ast_3:\quad\left\{\begin{array}{ccc}
f^{\dot{k}^\prime\dot{k}}_{\dot{l}-1,\dot{l},\dot{m}}&=&
c^{\dot{k}^\prime\dot{k}}_{\dot{l}-1,\dot{l}}
\sqrt{\dot{l}^2-\dot{m}^2},\\
f^{\dot{k}^\prime\dot{k}}_{\dot{l}\dot{l},\dot{m}}&=&
c^{\dot{k}^\prime\dot{k}}_{\dot{l}\dot{l}}\dot{m},\\
f^{\dot{k}^\prime\dot{k}}_{\dot{l}+1,\dot{l},\dot{m}}&=&
c^{\dot{k}^\prime\dot{k}}_{\dot{l}+1,\dot{l}}
\sqrt{(\dot{l}+1)^2-\dot{m}^2}.
\end{array}\right.}
\end{equation}
The commutation relations between the matrices $\sL_i$, $\sL^\ast_i$
and infinitesimal operators (\ref{OpA}), (\ref{OpB}), (\ref{DopA}),
(\ref{DopB}) are
\begin{equation}\label{AL}
\begin{array}{rcl}
\ld\sA_1,\sL_1\rd &=& 0,\\
\ld\sA_2,\sL_1\rd &=&-\sL_3,\\
\ld\sA_3,\sL_1\rd &=& \sL_2,
\end{array}\;\;\;
\begin{array}{rcl}
\ld\sA_1,\sL_2\rd &=& \sL_3,\\
\ld\sA_2,\sL_2\rd &=& 0,\\
\ld\sA_3,\sL_2\rd &=&-\sL_1,
\end{array}\;\;\;
\begin{array}{rcl}
\ld\sA_1,\sL_3\rd &=&-\sL_2,\\
\ld\sA_2,\sL_3\rd &=& \sL_1,\\
\ld\sA_3,\sL_3\rd &=& 0.
\end{array}
\end{equation}
\begin{equation}\label{BL}
\begin{array}{rcl}
\ld\sB_1,\sL_1\rd &=& 0,\\
\ld\sB_2,\sL_1\rd &=&-i\sL_3,\\
\ld\sB_3,\sL_1\rd &=& i\sL_2,
\end{array}\;\;\;
\begin{array}{rcl}
\ld\sB_1,\sL_2\rd &=& i\sL_3,\\
\ld\sB_2,\sL_2\rd &=& 0,\\
\ld\sB_3,\sL_2\rd &=&-i\sL_1,
\end{array}\;\;\;
\begin{array}{rcl}
\ld\sB_1,\sL_3\rd &=&-i\sL_2,\\
\ld\sB_2,\sL_3\rd &=& i\sL_1,\\
\ld\sB_3,\sL_3\rd &=& 0.
\end{array}
\end{equation}
\begin{equation}\label{DAL}
{\renewcommand{\arraystretch}{1.6}
\begin{array}{rcl}
\ld\widetilde{\sA}_1,\sL^\ast_1\rd &=& 0,\\
\ld\widetilde{\sA}_2,\sL^\ast_1\rd &=& \sL^\ast_3,\\
\ld\widetilde{\sA}_3,\sL^\ast_1\rd &=&-\sL^\ast_2,
\end{array}\;\;\;
\begin{array}{rcl}
\ld\widetilde{\sA}_1,\sL^\ast_2\rd &=&-\sL^\ast_3,\\
\ld\widetilde{\sA}_2,\sL^\ast_2\rd &=& 0,\\
\ld\widetilde{\sA}_3,\sL^\ast_2\rd &=& \sL^\ast_1,
\end{array}\;\;\;
\begin{array}{rcl}
\ld\widetilde{\sA}_1,\sL^\ast_3\rd &=& \sL^\ast_2,\\
\ld\widetilde{\sA}_2,\sL^\ast_3\rd &=&-\sL^\ast_1,\\
\ld\widetilde{\sA}_3,\sL^\ast_3\rd &=& 0.
\end{array}}
\end{equation}
\begin{equation}\label{DBL}
{\renewcommand{\arraystretch}{1.6}
\begin{array}{rcl}
\ld\widetilde{\sB}_1,\sL^\ast_1\rd &=& 0,\\
\ld\widetilde{\sB}_2,\sL^\ast_1\rd &=& i\sL^\ast_3,\\
\ld\widetilde{\sB}_3,\sL^\ast_1\rd &=&-i\sL^\ast_2,
\end{array}\;\;\;
\begin{array}{rcl}
\ld\widetilde{\sB}_1,\sL^\ast_2\rd &=&-i\sL^\ast_3,\\
\ld\widetilde{\sB}_2,\sL^\ast_2\rd &=& 0,\\
\ld\widetilde{\sB}_3,\sL^\ast_2\rd &=& i\sL^\ast_1,
\end{array}\;\;\;
\begin{array}{rcl}
\ld\widetilde{\sB}_1,\sL^\ast_3\rd &=& i\sL^\ast_2,\\
\ld\widetilde{\sB}_2,\sL^\ast_3\rd &=&-i\sL^\ast_1,\\
\ld\widetilde{\sB}_3,\sL^\ast_3\rd &=& 0.
\end{array}}
\end{equation} 

Let us establish now an important relationship between the complexification
(\ref{Complex}) and the four--dimensional Gel'fand--Yaglom formalism.
As known,
one of the most powerful higher spin formalisms is a Gel'fand--Yaglom
approach \cite{GY48} based primarily on the representation theory of
the Lorentz group. In contrast to the Bargmann--Wigner and
Joos--Weinberg formalisms, the main advantage of the Gel'fand--Yaglom
formalism lies in the fact that it admits naturally a Lagrangian
formulation. Indeed, an initial point of this theory is the following
lagrangian \cite{GY48,GMS,AB}:
\begin{equation}\label{Lag}
\cL=-\frac{1}{2}\left(\bar{\psi}\Gamma_\mu\frac{\partial\psi}{\partial x_\mu}-
\frac{\partial\bar{\psi}}{\partial x_\mu}\Gamma_\mu\psi\right)-
\kappa\bar{\psi}\psi,
\end{equation}
where $\Gamma_\mu$ are $n$--dimensional matrices, $n$ equals to the
number of components of the wave function $\psi$. Varying independently
the functions $\psi$ and $\bar{\psi}$ one gets general Dirac--like
(Gel'fand--Yaglom \cite{GY48}) equations
\begin{eqnarray}
\Gamma_\mu\frac{\partial\psi}{\partial x_\mu}+\kappa\psi&=&0,\nonumber\\
\Gamma^{\sT}_\mu\frac{\partial\bar{\psi}}{\partial x_\mu}-
\kappa\bar{\psi}&=&0.\label{GY}
\end{eqnarray}
As it shown in \cite{GMS} (see also \cite{AD72,PS83}) the matrix
$\Gamma_0$ in 4D Gel'fand--Yaglom equations (\ref{GY}) can be
written in the form
\begin{equation}\label{G1}
\Gamma_0=\text{diag}\left(C^0\otimes I_1,C^1\otimes I_3,\ldots,
C^s\otimes I_{2s+1},\ldots\right)
\end{equation}
for integer spin and
\begin{equation}\label{G2}
\Gamma_0=\text{diag}\left(C^{\frac{1}{2}}\otimes I_2,
C^{\frac{3}{2}}\otimes I_4,\ldots,C^s\otimes I_{2s+1},\ldots\right)
\end{equation}
for half--integer spin,
where $C^s$ is a spin block. If the spin block $C^s$ has non--null
roots, then the particle possesses the spin $s$. The spin block
$C^s$ in (\ref{G1})--(\ref{G2}) consists of the elements
$c^s_{\boldsymbol{\tau}\boldsymbol{\tau}^\prime}$, where
$\boldsymbol{\tau}_{l_1,l_2}$ and $\boldsymbol{\tau}_{l^\prime_1,l^\prime_2}$
are interlocking irreducible representations of the Lorentz group,
that is, such representations, for which
$l^\prime_1=l_1\pm\frac{1}{2}$, $l^\prime_2=l_2\pm\frac{1}{2}$.
At this point the block $C^s$ contains only the elements
$c^s_{\boldsymbol{\tau}\boldsymbol{\tau}^\prime}$ corresponding to such
interlocking representations $\boldsymbol{\tau}_{l_1,l_2}$,
$\boldsymbol{\tau}_{l^\prime_1,l^\prime_2}$ which satisfy the conditions
\[
|l_1-l_2|\leq s\leq l_1+l_2,\quad
|l^\prime_1-l^\prime_2|\leq s \leq l^\prime_1+l^\prime_2.
\]
The two most full schemes of the interlocking irreducible representations
of the Lorentz group (Gel'fand--Yaglom chains) for integer
(scheme (\ref{Bose})) and half--integer (scheme (\ref{Fermi})) spins are
\begin{equation}\label{Bose}
\dgARROWLENGTH=0.5em
\dgHORIZPAD=1.7em
\dgVERTPAD=2.2ex
\begin{diagram}
\node[5]{(s,0)}\arrow{e,-}\arrow{s,-}\node{\cdots}\\
\node[5]{\vdots}\arrow{s,-}\\
\node[3]{(2,0)}\arrow{e,-}\arrow{s,-}\node{\cdots}\arrow{e,-}
\node{\left(\frac{s+2}{2},\frac{s-2}{2}\right)}\arrow{s,-}\arrow{e,-}
\node{\cdots}\\
\node[2]{(1,0)}\arrow{s,-}\arrow{e,-}
\node{\left(\frac{3}{2},\frac{1}{2}\right)}\arrow{s,-}\arrow{e,-}
\node{\cdots}\arrow{e,-}
\node{\left(\frac{s+1}{2},\frac{s-1}{2}\right)}\arrow{s,-}\arrow{e,-}
\node{\cdots}\\
\node{(0,0)}\arrow{e,-}
\node{\left(\frac{1}{2},\frac{1}{2}\right)}\arrow{s,-}\arrow{e,-}
\node{(1,1)}\arrow{s,-}\arrow{e,-}\node{\cdots}\arrow{e,-}
\node{\left(\frac{s}{2},\frac{s}{2}\right)}\arrow{s,-}\arrow{e,-}
\node{\cdots}\\
\node[2]{(0,1)}\arrow{e,-}
\node{\left(\frac{1}{2},\frac{3}{2}\right)}\arrow{s,-}\arrow{e,-}
\node{\cdots}\arrow{e,-}
\node{\left(\frac{s-1}{2},\frac{s+1}{2}\right)}\arrow{s,-}\arrow{e,-}
\node{\cdots}\\
\node[3]{(0,2)}\arrow{e,-}\node{\cdots}\arrow{e,-}
\node{\left(\frac{s-2}{2},\frac{s+2}{2}\right)}\arrow{s,-}\arrow{e,-}
\node{\cdots}\\
\node[5]{\vdots}\arrow{s,-}\\
\node[5]{(0,s)}\arrow{e,-}\node{\cdots}
\end{diagram}
\end{equation}
\begin{equation}\label{Fermi}
\dgARROWLENGTH=0.5em
\dgHORIZPAD=1.7em %1.5em
\dgVERTPAD=2.2ex %2ex
\begin{diagram}
\node[4]{(s,0)}\arrow{s,-}\arrow{e,-}\node{\cdots}\\
\node[4]{\vdots}\arrow{s,-}\\
\node[2]{\left(\frac{3}{2},0\right)}\arrow{s,-}\arrow{e,-}
\node{\cdots}\arrow{e,-}
\node{\left(\frac{2s+3}{4},\frac{2s-3}{4}\right)}\arrow{s,-}\arrow{e,-}
\node{\cdots}\\
\node{\left(\frac{1}{2},0\right)}\arrow{s,-}\arrow{e,-}
\node{\left(1,\frac{1}{2}\right)}\arrow{s,-}\arrow{e,-}
\node{\cdots}\arrow{e,-}
\node{\left(\frac{2s+1}{4},\frac{2s-1}{4}\right)}\arrow{s,-}\arrow{e,-}
\node{\cdots}\\
\node{\left(0,\frac{1}{2}\right)}\arrow{e,-}
\node{\left(\frac{1}{2},1\right)}\arrow{s,-}\arrow{e,-}
\node{\cdots}\arrow{e,-}
\node{\left(\frac{2s-1}{4},\frac{2s+1}{4}\right)}\arrow{s,-}\arrow{e,-}
\node{\cdots}\\
\node[2]{\left(0,\frac{3}{2}\right)}\arrow{e,-}
\node{\cdots}\arrow{e,-}
\node{\left(\frac{2s-3}{4},\frac{2s+3}{4}\right)}\arrow{s,-}\arrow{e,-}
\node{\cdots}\\
\node[4]{\vdots}\arrow{s,-}\\
\node[4]{(0,s)}\arrow{e,-}\node{\cdots}
\end{diagram}
\end{equation}
\subsection{Bivector space $\R^6$}
We will establish a relationship between complex equations (\ref{Complex})
and 4D Gel'fand--Yaglom formalism by means of the mapping of the
equations (\ref{GY}) onto a bivector space $\R^6$.

Let $\R^{p,q}$ be the $n$--dimensional pseudo--Euclidean space,
$p+q=n$. Let us evolve in $\R^{p,q}$ all the tensors satisfying the
following two conditions: 1) a rank of the tensors is even;
2) covariant and contravariant indexes are divided into separate
skewsymmetric pairs. Such tensors can be exemplified by bivectors
(skewsymmetric tensors of the second rank). The set of all bivector
tensor fields in $\R^{p,q}$ is called {\it a bivector set}, and its
representation in a given point of $\R^{p,q}$ is called
{\it a local bivector set}. In any tensor from the bivector set
we take the each skewsymmetric pair $\alpha\beta$ as one collective
index. At this point from the two possible pairs $\alpha\beta$ and
$\beta\alpha$ we fix only one, for example, $\alpha\beta$.
The number of all collective indexes is equal to
$N=\frac{n(n-1)}{2}$. In a given point, the bivector set of the space
$\R^{p,q}$ with contravariant components defines in the collective
indexes a vector set, the each vector of this set has $N$ components.
Identifying these vectors with the points of $N$--dimensional
manifold, we see that it be an affine manifold $E^N$ if and only if
this manifold admits a Klein geometry with the group
\begin{gather}
\eta^{a^\prime}=A^{a^\prime}_a\eta^a,\quad
\eta^a=A^a_{a^\prime}\eta^{a^\prime},\nonumber\\
\det A^{a^\prime}_a\neq 0,\quad
A^a_{b^\prime}A^{b^\prime}_c=\delta^a_c,\nonumber
\end{gather}
where
\[
A^{a^\prime}_a\longrightarrow 
A^{\alpha^\prime}_{[\alpha}A^{\beta^\prime}_{\beta]}.
\]
Thus, any local bivector set of the space $\R^{p,q}$ ($p+q=n$) can be
mapped onto the affine space $E^N$ . Therefore, $E^N$ related with
the each point of the space $\R^{p,q}$. The space $E^N$ is called
{\it a bivector space}. It should be noted that the bivector space is
a particular case of the most general mathematical construction called
a Grassmannian manifold (manifold of $m$--dimensional planes of the
affine space). In the case $m=2$ the manifold of two--dimensional planes
is isometric to the bivector space, and the Grassmann coordinates in
this case are called Pluecker coordinates.

The metrization of the bivector space $E^N$ is given by the formula
(see \cite{Pet69})
\begin{equation}\label{Metric}
g_{ab}\longrightarrow g_{\alpha\beta\gamma\delta}\equiv
g_{\alpha\gamma}g_{\beta\delta}-g_{\alpha\delta}g_{\beta\gamma},
\end{equation}
where $g_{\alpha\beta}$ is a metric tensor of the space $\R^{p,q}$,
and the collective indexes are skewsymmetric pairs
$\alpha\beta\rightarrow a$, $\gamma\delta\rightarrow b$. After
introduction of $g_{ab}$, the bivector affine space $E^N$ is transformed
to a metric space $\R^N$.

In the case of Minkowski spacetime $\R^{1,3}$ with the metric tensor
\[
g_{\alpha\beta}=\begin{pmatrix}
-1 & 0 & 0 & 0\\
0  & -1& 0 & 0\\
0  & 0 & -1& 0\\
0  & 0 & 0 & 1
\end{pmatrix}
\]
in virtue of (\ref{Metric}) we obtain for the bivector space $\R^6$:
\begin{equation}\label{MetB}
g_{ab}=\begin{pmatrix}
-1& 0 & 0 & 0 & 0 & 0\\
0 & -1& 0 & 0 & 0 & 0\\
0 & 0 & -1& 0 & 0 & 0\\
0 & 0 & 0 & 1 & 0 & 0\\
0 & 0 & 0 & 0 & 1 & 0\\
0 & 0 & 0 & 0 & 0 & 1
\end{pmatrix},
\end{equation}
where the order of collective indexes in $\R^6$ is
$23\rightarrow 0$, $10\rightarrow 1$, $20\rightarrow 2$,
$30\rightarrow 3$, $31\rightarrow 4$, $12\rightarrow 5$.
As it shown in \cite{Kag26}, the Lorentz transformations can be represented
by linear transformations of the space $\R^6$. Let us write an
invariance condition of the system (\ref{Complex}). Let 
$\fg: x^\prime=\fg^{-1} x$ be a transformation of the bivector space $\R^6$,
that is, $x^\prime=\sum^6_{b=1}g_{ba} x_b$, where
$x=(x_1,x_2,x_3,\widetilde{x}_1,\widetilde{x}_2,\widetilde{x}_3)$ and
$g_{ba}$ is the metric tensor (\ref{MetB}). We can write the tensor
(\ref{MetB}) in the form
$g_{ab}=\begin{pmatrix}
g^-_{ik} & \\
& g^+_{ik}
\end{pmatrix}$,
then $x^\prime=\sum^3_{k=1}g^-_{ki}x_k$,
$\widetilde{x}^\prime=\sum^3_{k=1}g^+_{ki}\widetilde{x}_k$.
Replacing $\psi$ via $\sT^{-1}_{\fg}\psi^\prime$ and differentiation on
$x_k$ ($\widetilde{x}_k$) by differentiation on $x^\prime_k$
($\widetilde{x}^\prime_k$) via the formulae
\[
\frac{\partial}{\partial x_k}=\sum g^-_{ik}
\frac{\partial}{\partial x^\prime_i},\quad
\frac{\partial}{\partial\widetilde{x}_k}=\sum g^+_{ik}
\frac{\partial}{\partial\widetilde{x}^\prime_i}
\]
we obtain
\begin{multline}
\sum^3_{i=1}\left[
g^-_{i1}\sL_1\frac{\partial(\sT^{-1}_{\fg}\psi^\prime)}{\partial x^\prime_i}+
g^-_{i2}\sL_2\frac{\partial(\sT^{-1}_{\fg}\psi^\prime)}{\partial x^\prime_i}+
g^-_{i3}\sL_3\frac{\partial(\sT^{-1}_{\fg}\psi^\prime)}{\partial x^\prime_i}-
\right.\\
\shoveright{\left.
-ig^-_{i1}\sL_1\frac{\partial(\sT^{-1}_{\fg}\psi^\prime)}
{\partial\overset{\ast}{x}_i{}^\prime}-
ig^-_{i2}\sL_2\frac{\partial(\sT^{-1}_{\fg}\psi^\prime)}
{\partial\overset{\ast}{x}_i{}^\prime}-
ig^-_{i3}\sL_3\frac{\partial(\sT^{-1}_{\fg}\psi^\prime)}
{\partial\overset{\ast}{x}_i{}^\prime}\right]+
\kappa^c\sT^{-1}_{\fg}\psi^\prime=0,}\\
\shoveleft{
\sum^3_{i=1}\left[
g^+_{i1}\sL^\ast_1\frac{\partial(\overset{\ast}{\sT}_{\fg}\!\!{}^{-1}
\dot{\psi}^\prime)}{\partial\widetilde{x}^\prime_i}+
g^+_{i2}\sL^\ast_2\frac{\partial(\overset{\ast}{\sT}_{\fg}\!\!{}^{-1}
\dot{\psi}^\prime)}{\partial\widetilde{x}^\prime_i}+
g^+_{i3}\sL^\ast_3\frac{\partial(\overset{\ast}{\sT}_{\fg}\!\!{}^{-1}
\dot{\psi}^\prime)}{\partial\widetilde{x}^\prime_i}+\right.}\nonumber\\
\left.
+ig^+_{i1}\sL^\ast_1\frac{\partial(\overset{\ast}{\sT}_{\fg}\!\!{}^{-1}
\dot{\psi}^\prime)}{\partial\widetilde{x^\ast}_i{}^\prime}+
ig^+_{i2}\sL^\ast_2\frac{\partial(\overset{\ast}{\sT}_{\fg}\!\!{}^{-1}
\dot{\psi}^\prime)}{\partial\widetilde{x^\ast}_i{}^\prime}+
ig^+_{i3}\sL^\ast_3\frac{\partial(\overset{\ast}{\sT}_{\fg}\!\!{}^{-1}
\dot{\psi}^\prime)}{\partial\widetilde{x^\ast}_i{}^\prime}\right]+
\dot{\kappa}^c\overset{\ast}{\sT}\!\!{}^{-1}_{\fg}\dot{\psi}^\prime=0.\nonumber
\end{multline}
Or, since $\sT_{\fg}$ is a constant matrix, we have
\begin{multline}
\sum_i\left[
g^-_{i1}\sL_1\sT^{-1}_{\fg}\frac{\partial\psi^\prime}{\partial x^\prime_i}+
g^-_{i2}\sL_2\sT^{-1}_{\fg}\frac{\partial\psi^\prime}{\partial x^\prime_i}+
g^-_{i3}\sL_3\sT^{-1}_{\fg}\frac{\partial\psi^\prime}{\partial x^\prime_i}-
\right.\\
\shoveright{\left.
-ig^-_{i1}\sL_1\sT^{-1}_{\fg}\frac{\partial\psi^\prime}
{\partial x^\ast_i{}^\prime}-
ig^-_{i2}\sL_2\sT^{-1}_{\fg}\frac{\partial\psi^\prime}
{\partial x^\ast_i{}^\prime}-
ig^-_{i3}\sL_3\sT^{-1}_{\fg}\frac{\partial\psi^\prime}
{\partial x^\ast_i{}^\prime}\right]+
\kappa^c\sT^{-1}_{\fg}\psi^\prime=0,}\\
\shoveleft{
\sum_i\left[
g^+_{i1}\sL^\ast_1\overset{\ast}{\sT}_{\fg}\!\!{}^{-1}
\frac{\partial\dot{\psi}^\prime}{\partial\widetilde{x}^\prime_i}+
g^+_{i2}\sL^\ast_2\overset{\ast}{\sT}_{\fg}\!\!{}^{-1}
\frac{\partial\dot{\psi}^\prime}{\partial\widetilde{x}^\prime_i}+
g^+_{i3}\sL^\ast_3\overset{\ast}{\sT}_{\fg}\!\!{}^{-1}
\frac{\partial\dot{\psi}^\prime}{\partial\widetilde{x}^\prime_i}+\right.}\\
\left.
+ig^+_{i1}\sL^\ast_1\overset{\ast}{\sT}_{\fg}\!\!{}^{-1}
\frac{\partial\dot{\psi}^\prime}{\partial\widetilde{x^\ast}_i{}^\prime}+
ig^+_{i2}\sL^\ast_2\overset{\ast}{\sT}_{\fg}\!\!{}^{-1}
\frac{\partial\dot{\psi}^\prime}{\partial\widetilde{x^\ast}_i{}^\prime}+
ig^+_{i3}\sL^\ast_3\overset{\ast}{\sT}_{\fg}\!\!{}^{-1}
\frac{\partial\dot{\psi}^\prime}{\partial\widetilde{x^\ast}_i{}^\prime}\right]+
\dot{\kappa}^c\overset{\ast}{\sT}_{\fg}\!\!{}^{-1}\dot{\psi}^\prime=0.\nonumber
\end{multline}
For coincidence of the latter system with (\ref{Complex}) we must multiply
this system by $\sT_{\fg}$($\overset{\ast}{\sT}_{\fg}$) from the left:
\begin{eqnarray}
\sum_i\sum_k g^-_{ik}\sT_{\fg}\sL_k\sT^{-1}_{\fg}
\frac{\partial\psi^\prime}{\partial x^\prime_i}
-i\sum_i\sum_k g^-_{ik}\sT_{\fg}\sL_k\sT^{-1}_{\fg}
\frac{\partial\psi^\prime}{\partial x^\ast_i{}^\prime}+
\kappa^c\psi^\prime&=&0,\nonumber\\
\sum_i\sum_k g^+_{ik}\overset{\ast}{\sT}_{\fg}\sL^\ast_k
\overset{\ast}{\sT}_{\fg}\!\!{}^{-1}\frac{\partial\dot{\psi}^\prime}
{\partial\widetilde{x}^\prime_i}+
i\sum_i\sum_k g^+_{ik}\overset{\ast}{\sT}_{\fg}\sL^\ast_k
\overset{\ast}{\sT}_{\fg}\!\!{}^{-1}\frac{\partial\dot{\psi}^\prime}
{\partial\widetilde{x^\ast}_i{}^\prime}+
\dot{\kappa}^c\dot{\psi}^\prime&=&0.\nonumber
\end{eqnarray}
The requirement of invariance means that for any transformation $\fg$
between the matrices $\sL_k$($\sL^\ast_k$) we must have the relations
\begin{eqnarray}
\sum_k g^-_{ik}\sT_{\fg}\sL_k\sT^{-1}_{\fg}&=&\sL_i,\nonumber\\
\sum_k g^+_{ik}\overset{\ast}{\sT}_{\fg}\sL^\ast_k
\overset{\ast}{\sT}_{\fg}\!\!\!{}^{-1}&=&\sL^\ast_i.\label{IC}
\end{eqnarray}

In such a way, we see that the 6--dimensional bivector space $\R^6$ is
associated with the each point of the Minkowski spacetime $\R^{1,3}$.
Using this fact we can establish now a relationship between the
$\Lambda$--matrices of the equations (\ref{Complex}) and
$\Gamma$--matrices of the Gel'fand--Yaglom equations (\ref{GY}).
We have here two essentially different cases:\\
1) The dimension of the $\Gamma$--matrices is even, $n\equiv 0\pmod{2}$.
In this case the $\Gamma$--matrices have the form
\begin{equation}\label{Type1}
\overset{i}{\Gamma}_{2^m}=\begin{pmatrix}
\underset{i}{\Delta}{}_{2^{m-1}} & 0\\
0 & \overset{\ast}{\underset{i}{\Delta}}{}_{2^{m-1}}
\end{pmatrix},\quad\text{or}\quad
\overset{i}{\Gamma}_{2^m}=\begin{pmatrix}
0 & \underset{i}{\Delta}{}_{2^{m-1}}\\
\overset{\ast}{\underset{i}{\Delta}}{}_{2^{m-1}} & 0
\end{pmatrix}
\end{equation}
The $\Lambda$--matrices are expressed via (\ref{Type1}) as follows
\begin{eqnarray}
&&\underset{1}{\sL}{}_{2^{m-1}}=\underset{1}{\Delta}{}_{2^{m-1}}
\underset{0}{\Delta}{}_{2^{m-1}},\;\;
\underset{2}{\sL}{}_{2^{m-1}}=\underset{2}{\Delta}{}_{2^{m-1}}
\underset{0}{\Delta}{}_{2^{m-1}},\;\;
\underset{3}{\sL}{}_{2^{m-1}}=\underset{3}{\Delta}{}_{2^{m-1}}
\underset{0}{\Delta}{}_{2^{m-1}},\nonumber\\
&&\underset{4}{\sL}{}_{2^{m-1}}=\overset{\ast}{\underset{2}{\Delta}}{}_{2^{m-1}}
\overset{\ast}{\underset{3}{\Delta}}{}_{2^{m-1}},\;\;
\underset{5}{\sL}{}_{2^{m-1}}=\overset{\ast}{\underset{3}{\Delta}}{}_{2^{m-1}}
\overset{\ast}{\underset{1}{\Delta}}{}_{2^{m-1}},\;\;
\underset{6}{\sL}{}_{2^{m-1}}=\overset{\ast}{\underset{1}{\Delta}}{}_{2^{m-1}}
\overset{\ast}{\underset{2}{\Delta}}{}_{2^{m-1}},\label{Lambda1}
\end{eqnarray}
2) $n\equiv 1\pmod{2}$. In this case a relationship between the
$\Lambda$-- and $\Gamma$--matrices is
\begin{eqnarray}
&&\underset{1}{\sL}{}_n=\overset{1}{\Gamma}_n\overset{0}{\Gamma}_n,\;\;\;
\underset{2}{\sL}{}_n=\overset{2}{\Gamma}_n\overset{0}{\Gamma}_n,\;\;\;
\underset{3}{\sL}{}_n=\overset{3}{\Gamma}_n\overset{0}{\Gamma}_n,\nonumber\\
&&\underset{4}{\sL}{}_n=\overset{2}{\Gamma}_n\overset{3}{\Gamma}_n,\;\;\;
\underset{5}{\sL}{}_n=\overset{3}{\Gamma}_n\overset{1}{\Gamma}_n,\;\;\;
\underset{6}{\sL}{}_n=\overset{1}{\Gamma}_n\overset{2}{\Gamma}_n.
\label{Lambda2}
\end{eqnarray}
\subsection{Separation of variables in generalized Gel'fand-Yaglom equations}
Let us construct in $\C^3$ a two--dimensional complex sphere from the
quantities $z_k=x_k+iy_k$, $\overset{\ast}{z}_k=x_k-iy_k$
as follows (see {\bf Fig.1})
\begin{equation}\label{CS}
\bz^2=z^2_1+z^2_2+z^2_3=\bx^2-\by^2+2i\bx\by=r^2
\end{equation}
and its complex conjugate (dual) sphere
\begin{equation}\label{DS}
\overset{\ast}{\bz}{}^2=\overset{\ast}{z}_1{}^2+\overset{\ast}{z}_2{}^2+
\overset{\ast}{z}_3{}^2=\bx^2-\by^2-2i\bx\by=\overset{\ast}{r}{}^2.
\end{equation}
It is well-known that both quantities $\bx^2-\by^2$, $\bx\by$ are
invariant with respect to the Lorentz transformations, since a surface of
the complex sphere is invariant (Casimir operators of the Lorentz group are
constructed from such quantities, see also (\ref{KO})).
It is easy to see that three--dimensional complex space $\C^3$ is
isometric to a real space $\R^{3,3}$ with a basis
$\left\{i\e_1,i\e_2,i\e_3,\e_4,\e_5,\e_6\right\}$. At this point
a metric tensor of $\R^{3,3}$ has the form (\ref{MetB}). Hence it
immediately follows that $\C^3$ is isometric to the bivector space $\R^6$.

%\documentclass[12pt]{article}
%\usepackage{amsmath}
%\usepackage{eufrak}
%\textwidth=15cm
%\textheight=15cm
%\thispagestyle{empty}
%\catcode`\@=11

%\font\twelvemsa=msam10 scaled 1200
%\newcommand{\square}{\mbox{\twelvemsa \char'003}}

%\font\twelvemsb=msbm10 scaled 1200
%\font\tenmsb=msbm10
%\font\sevenmsb=msbm7
%\font\fivemsb=msbm5
%\newfam\msbfam
%\textfont\msbfam=\twelvemsb
%\scriptfont\msbfam=\sevenmsb
%\scriptscriptfont\msbfam=\fivemsb

%\def\Bbb{\ifmmode\let\next\Bbb@\else
%\def\next{\errmessage{Use \string\Bbb\space only in math mode}}\fi\next}
%\def\Bbb@#1{{\Bbb@@{#1}}}
%\def\Bbb@@#1{\fam\msbfam#1}

%\catcode`\@=12
%\gdef\C{{\Bbb C}}
%\gdef\dZ{{\Bbb Z}}
%\gdef\A{{\Bbb A}}
%\gdef\K{{\Bbb K}}
%\gdef\R{{\Bbb R}}
%\gdef\BH{{\Bbb H}}
%\gdef\F{{\Bbb F}}
%
%\newcommand{\fC}{\mathfrak{C}}
%\newcommand{\fR}{\mathfrak{R}}
%\newcommand{\fH}{\mathfrak{H}}

%\begin{document}
%\begin[b]{float}
\[
\unitlength=0.35mm
%\linethickness{0.4pt}
\begin{picture}(100.00,175.00)(0,25)
\put(50,50){\vector(0,1){51}}
\put(50,150){\vector(0,-1){48}}
\put(48,50){$\bullet$}
\put(48,150){$\bullet$}
\put(54,75){$r^\ast$}
\put(54,125){$r$}
\put(58,13){$\bullet$}
\put(58,183){$\bullet$}
\put(60,15){\vector(1,1){23}}
\put(60,15){\vector(1,-1){23}}
\put(60,185){\vector(1,1){23}}
\put(60,185){\vector(1,-1){23}}
\put(50,10){$P^\prime$}
\put(50,180){$P$}
\put(-85,50){$\mathfrak{M}^{\dot{l}_0}_{\dot{m}\dot{n}}
(\dot{\varphi}^c,\dot{\theta}^c,0)$}
\put(-85,150){$\mathfrak{M}^{l_0}_{mn}(\varphi^c,\theta^c,0)$}
\put(85,-12){$\boldsymbol{e}_{\dot{\varphi}^c}$}
\put(80,210){$\boldsymbol{e}_{\varphi^c}$}
\put(80,30){$\boldsymbol{e}_{\dot{\theta}^c}$}
\put(80,170){$\boldsymbol{e}_{\theta^c}$}
\put(105,18){$\dot{\varphi}^c=\varphi+i\epsilon$}
\put(105,2){$\dot{\theta}^c=\theta+i\tau$}
\put(105,198){$\varphi^c=\varphi-i\epsilon$}
\put(105,182){$\theta^c=\theta-i\tau$}
%ellips
\put(50,34){$\cdot$}
\put(49.5,34){$\cdot$}
\put(49,34){$\cdot$}
\put(48.5,34){$\cdot$}
\put(48,34.01){$\cdot$}
\put(47.5,34.02){$\cdot$}
\put(47,34.03){$\cdot$}
\put(46.5,34.04){$\cdot$}
\put(46,34.05){$\cdot$}
\put(45.5,34.06){$\cdot$}
\put(45,34.08){$\cdot$}
\put(44.5,34.10){$\cdot$}
\put(44,34.11){$\cdot$}
\put(43.5,34.13){$\cdot$}
\put(43,34.16){$\cdot$}
\put(42.5,34.18){$\cdot$}
\put(42,34.20){$\cdot$}
\put(41.5,34.23){$\cdot$}
\put(41,34.26){$\cdot$}
\put(40.5,34.29){$\cdot$}
\put(40,34.32){$\cdot$}
\put(39.5,34.36){$\cdot$}
\put(39,34.39){$\cdot$}
\put(38.5,34.43){$\cdot$}
\put(38,34.47){$\cdot$}
\put(37.5,34.51){$\cdot$}
\put(37,34.55){$\cdot$}
\put(36.5,34.59){$\cdot$}
\put(36,34.64){$\cdot$}
\put(35.5,34.69){$\cdot$}
\put(35,34.74){$\cdot$}
\put(34.5,34.79){$\cdot$}
\put(34,34.84){$\cdot$}
\put(33.5,34.90){$\cdot$}
\put(33,34.95){$\cdot$}
\put(32.5,35.01){$\cdot$}
\put(32,35.07){$\cdot$}
\put(31.5,35.13){$\cdot$}
\put(31,35.20){$\cdot$}
\put(30.5,35.27){$\cdot$}
\put(30,35.33){$\cdot$}
\put(29.5,35.40){$\cdot$}
\put(29,35.48){$\cdot$}
\put(28.5,35.55){$\cdot$}
\put(28,35.63){$\cdot$}
\put(27.5,35.71){$\cdot$}
\put(27,35.79){$\cdot$}
\put(26.5,35.88){$\cdot$}
\put(26,35.96){$\cdot$}
\put(25.5,36.05){$\cdot$}
\put(25,36.14){$\cdot$}
\put(24.5,36.24){$\cdot$}
\put(24,36.33){$\cdot$}
\put(23.5,36.43){$\cdot$}
\put(23,36.53){$\cdot$}
\put(22.5,36.64){$\cdot$}
\put(22,36.74){$\cdot$}
\put(21.5,36.85){$\cdot$}
\put(21,36.97){$\cdot$}
\put(20.5,37.08){$\cdot$}
\put(20,37.20){$\cdot$}
\put(19.5,37.32){$\cdot$}
\put(19,37.45){$\cdot$}
\put(18.5,37.57){$\cdot$}
\put(18,37.70){$\cdot$}
\put(17.5,37.84){$\cdot$}
\put(17,37.98){$\cdot$}
\put(16.5,38.12){$\cdot$}
\put(16,38.27){$\cdot$}
\put(15.5,38.42){$\cdot$}
\put(15,38.57){$\cdot$}
\put(14.5,38.73){$\cdot$}
\put(14,38.90){$\cdot$}
\put(13.5,39.06){$\cdot$}
\put(13,39.24){$\cdot$}
\put(12.5,39.42){$\cdot$}
\put(12,39.60){$\cdot$}
\put(11.5,39.79){$\cdot$}
\put(11,39.99){$\cdot$}
\put(10.5,40.19){$\cdot$}
\put(10,40.40){$\cdot$}
\put(9.5,40.62){$\cdot$}
\put(9,40.84){$\cdot$}
\put(8.5,41.07){$\cdot$}
\put(8,41.32){$\cdot$}
\put(7.5,41.57){$\cdot$}
\put(7,41.83){$\cdot$}
\put(6.5,42.11){$\cdot$}
\put(6,42.40){$\cdot$}
\put(5.5,42.70){$\cdot$}
\put(5,43.02){$\cdot$}
\put(4.5,43.36){$\cdot$}
\put(4,43.73){$\cdot$}
\put(3.5,44.12){$\cdot$}
\put(3,44.54){$\cdot$}
\put(2.5,45.00){$\cdot$}
\put(2,45.52){$\cdot$}
\put(1.5,46.11){$\cdot$}
\put(1,46.82){$\cdot$}
\put(0.5,47.74){$\cdot$}
\put(0,50){$\cdot$}
\put(50,34){$\cdot$}
\put(50.5,34){$\cdot$}
\put(51,34){$\cdot$}
\put(51.5,34){$\cdot$}
\put(52,34.01){$\cdot$}
\put(52.5,34.02){$\cdot$}
\put(53,34.03){$\cdot$}
\put(53.5,34.04){$\cdot$}
\put(54,34.05){$\cdot$}
\put(54.5,34.06){$\cdot$}
\put(55,34.08){$\cdot$}
\put(55.5,34.10){$\cdot$}
\put(56,34.11){$\cdot$}
\put(56.5,34.13){$\cdot$}
\put(57,34.16){$\cdot$}
\put(57.5,34.18){$\cdot$}
\put(58,34.20){$\cdot$}
\put(58.5,34.23){$\cdot$}
\put(59,34.26){$\cdot$}
\put(59.5,34.29){$\cdot$}
\put(60,34.32){$\cdot$}
\put(60.5,34.36){$\cdot$}
\put(61,34.39){$\cdot$}
\put(61.5,34.43){$\cdot$}
\put(62,34.47){$\cdot$}
\put(62.5,34.51){$\cdot$}
\put(63,34.55){$\cdot$}
\put(63.5,34.59){$\cdot$}
\put(64,34.64){$\cdot$}
\put(64.5,34.69){$\cdot$}
\put(65,34.74){$\cdot$}
\put(65.5,34.79){$\cdot$}
\put(66,34.84){$\cdot$}
\put(66.5,34.90){$\cdot$}
\put(67,34.95){$\cdot$}
\put(67.5,35.01){$\cdot$}
\put(68,35.07){$\cdot$}
\put(68.5,35.13){$\cdot$}
\put(69,35.20){$\cdot$}
\put(69.5,35.27){$\cdot$}
\put(70,35.33){$\cdot$}
\put(70.5,35.40){$\cdot$}
\put(71,35.48){$\cdot$}
\put(71.5,35.55){$\cdot$}
\put(72,35.63){$\cdot$}
\put(72.5,35.71){$\cdot$}
\put(73,35.79){$\cdot$}
\put(73.5,35.88){$\cdot$}
\put(74,35.96){$\cdot$}
\put(74.5,36.05){$\cdot$}
\put(75,36.14){$\cdot$}
\put(75.5,36.24){$\cdot$}
\put(76,36.33){$\cdot$}
\put(76.5,36.43){$\cdot$}
\put(77,36.53){$\cdot$}
\put(77.5,36.64){$\cdot$}
\put(78,36.74){$\cdot$}
\put(78.5,36.85){$\cdot$}
\put(79,36.97){$\cdot$}
\put(79.5,37.08){$\cdot$}
\put(80,37.20){$\cdot$}
\put(80.5,37.32){$\cdot$}
\put(81,37.45){$\cdot$}
\put(81.5,37.57){$\cdot$}
\put(82,37.70){$\cdot$}
\put(82.5,37.84){$\cdot$}
\put(83,37.98){$\cdot$}
\put(83.5,38.12){$\cdot$}
\put(84,38.27){$\cdot$}
\put(84.5,38.42){$\cdot$}
\put(85,38.57){$\cdot$}
\put(85.5,38.73){$\cdot$}
\put(86,38.90){$\cdot$}
\put(86.5,39.06){$\cdot$}
\put(87,39.24){$\cdot$}
\put(87.5,39.42){$\cdot$}
\put(88,39.60){$\cdot$}
\put(88.5,39.79){$\cdot$}
\put(89,39.99){$\cdot$}
\put(89.5,40.19){$\cdot$}
\put(90,40.40){$\cdot$}
\put(90.5,40.62){$\cdot$}
\put(91,40.84){$\cdot$}
\put(91.5,41.07){$\cdot$}
\put(92,41.32){$\cdot$}
\put(92.5,41.57){$\cdot$}
\put(93,41.83){$\cdot$}
\put(93.5,42.11){$\cdot$}
\put(94,42.40){$\cdot$}
\put(94.5,42.70){$\cdot$}
\put(95,43.02){$\cdot$}
\put(95.5,43.36){$\cdot$}
\put(96,43.73){$\cdot$}
\put(96.5,44.12){$\cdot$}
\put(97,44.54){$\cdot$}
\put(97.5,45.00){$\cdot$}
\put(98,45.52){$\cdot$}
\put(98.5,46.11){$\cdot$}
\put(99,46.82){$\cdot$}
\put(99.5,47.74){$\cdot$}
\put(100,50){$\cdot$}

\put(50,134){$\cdot$}
\put(49.5,134){$\cdot$}
\put(49,134){$\cdot$}
\put(48.5,134){$\cdot$}
\put(48,134.01){$\cdot$}
\put(47.5,134.02){$\cdot$}
\put(47,134.03){$\cdot$}
\put(46.5,134.04){$\cdot$}
\put(46,134.05){$\cdot$}
\put(45.5,134.06){$\cdot$}
\put(45,134.08){$\cdot$}
\put(44.5,134.10){$\cdot$}
\put(44,134.11){$\cdot$}
\put(43.5,134.13){$\cdot$}
\put(43,134.16){$\cdot$}
\put(42.5,134.18){$\cdot$}
\put(42,134.20){$\cdot$}
\put(41.5,134.23){$\cdot$}
\put(41,134.26){$\cdot$}
\put(40.5,134.29){$\cdot$}
\put(40,134.32){$\cdot$}
\put(39.5,134.36){$\cdot$}
\put(39,134.39){$\cdot$}
\put(38.5,134.43){$\cdot$}
\put(38,134.47){$\cdot$}
\put(37.5,134.51){$\cdot$}
\put(37,134.55){$\cdot$}
\put(36.5,134.59){$\cdot$}
\put(36,134.64){$\cdot$}
\put(35.5,134.69){$\cdot$}
\put(35,134.74){$\cdot$}
\put(34.5,134.79){$\cdot$}
\put(34,134.84){$\cdot$}
\put(33.5,134.90){$\cdot$}
\put(33,134.95){$\cdot$}
\put(32.5,135.01){$\cdot$}
\put(32,135.07){$\cdot$}
\put(31.5,135.13){$\cdot$}
\put(31,135.20){$\cdot$}
\put(30.5,135.27){$\cdot$}
\put(30,135.33){$\cdot$}
\put(29.5,135.40){$\cdot$}
\put(29,135.48){$\cdot$}
\put(28.5,135.55){$\cdot$}
\put(28,135.63){$\cdot$}
\put(27.5,135.71){$\cdot$}
\put(27,135.79){$\cdot$}
\put(26.5,135.88){$\cdot$}
\put(26,135.96){$\cdot$}
\put(25.5,136.05){$\cdot$}
\put(25,136.14){$\cdot$}
\put(24.5,136.24){$\cdot$}
\put(24,136.33){$\cdot$}
\put(23.5,136.43){$\cdot$}
\put(23,136.53){$\cdot$}
\put(22.5,136.64){$\cdot$}
\put(22,136.74){$\cdot$}
\put(21.5,136.85){$\cdot$}
\put(21,136.97){$\cdot$}
\put(20.5,137.08){$\cdot$}
\put(20,137.20){$\cdot$}
\put(19.5,137.32){$\cdot$}
\put(19,137.45){$\cdot$}
\put(18.5,137.57){$\cdot$}
\put(18,137.70){$\cdot$}
\put(17.5,137.84){$\cdot$}
\put(17,137.98){$\cdot$}
\put(16.5,138.12){$\cdot$}
\put(16,138.27){$\cdot$}
\put(15.5,138.42){$\cdot$}
\put(15,138.57){$\cdot$}
\put(14.5,138.73){$\cdot$}
\put(14,138.90){$\cdot$}
\put(13.5,139.06){$\cdot$}
\put(13,139.24){$\cdot$}
\put(12.5,139.42){$\cdot$}
\put(12,139.60){$\cdot$}
\put(11.5,139.79){$\cdot$}
\put(11,139.99){$\cdot$}
\put(10.5,140.19){$\cdot$}
\put(10,140.40){$\cdot$}
\put(9.5,140.62){$\cdot$}
\put(9,140.84){$\cdot$}
\put(8.5,141.07){$\cdot$}
\put(8,141.32){$\cdot$}
\put(7.5,141.57){$\cdot$}
\put(7,141.83){$\cdot$}
\put(6.5,142.11){$\cdot$}
\put(6,142.40){$\cdot$}
\put(5.5,142.70){$\cdot$}
\put(5,143.02){$\cdot$}
\put(4.5,143.36){$\cdot$}
\put(4,143.73){$\cdot$}
\put(3.5,144.12){$\cdot$}
\put(3,144.54){$\cdot$}
\put(2.5,145.00){$\cdot$}
\put(2,145.52){$\cdot$}
\put(1.5,146.11){$\cdot$}
\put(1,146.82){$\cdot$}
\put(0.5,147.74){$\cdot$}
\put(0,150){$\cdot$}
\put(50,134){$\cdot$}
\put(50.5,134){$\cdot$}
\put(51,134){$\cdot$}
\put(51.5,134){$\cdot$}
\put(52,134.01){$\cdot$}
\put(52.5,134.02){$\cdot$}
\put(53,134.03){$\cdot$}
\put(53.5,134.04){$\cdot$}
\put(54,134.05){$\cdot$}
\put(54.5,134.06){$\cdot$}
\put(55,134.08){$\cdot$}
\put(55.5,134.10){$\cdot$}
\put(56,134.11){$\cdot$}
\put(56.5,134.13){$\cdot$}
\put(57,134.16){$\cdot$}
\put(57.5,134.18){$\cdot$}
\put(58,134.20){$\cdot$}
\put(58.5,134.23){$\cdot$}
\put(59,134.26){$\cdot$}
\put(59.5,134.29){$\cdot$}
\put(60,134.32){$\cdot$}
\put(60.5,134.36){$\cdot$}
\put(61,134.39){$\cdot$}
\put(61.5,134.43){$\cdot$}
\put(62,134.47){$\cdot$}
\put(62.5,134.51){$\cdot$}
\put(63,134.55){$\cdot$}
\put(63.5,134.59){$\cdot$}
\put(64,134.64){$\cdot$}
\put(64.5,134.69){$\cdot$}
\put(65,134.74){$\cdot$}
\put(65.5,134.79){$\cdot$}
\put(66,134.84){$\cdot$}
\put(66.5,134.90){$\cdot$}
\put(67,134.95){$\cdot$}
\put(67.5,135.01){$\cdot$}
\put(68,135.07){$\cdot$}
\put(68.5,135.13){$\cdot$}
\put(69,135.20){$\cdot$}
\put(69.5,135.27){$\cdot$}
\put(70,135.33){$\cdot$}
\put(70.5,135.40){$\cdot$}
\put(71,135.48){$\cdot$}
\put(71.5,135.55){$\cdot$}
\put(72,135.63){$\cdot$}
\put(72.5,135.71){$\cdot$}
\put(73,135.79){$\cdot$}
\put(73.5,135.88){$\cdot$}
\put(74,135.96){$\cdot$}
\put(74.5,136.05){$\cdot$}
\put(75,136.14){$\cdot$}
\put(75.5,136.24){$\cdot$}
\put(76,136.33){$\cdot$}
\put(76.5,136.43){$\cdot$}
\put(77,136.53){$\cdot$}
\put(77.5,136.64){$\cdot$}
\put(78,136.74){$\cdot$}
\put(78.5,136.85){$\cdot$}
\put(79,136.97){$\cdot$}
\put(79.5,137.08){$\cdot$}
\put(80,137.20){$\cdot$}
\put(80.5,137.32){$\cdot$}
\put(81,137.45){$\cdot$}
\put(81.5,137.57){$\cdot$}
\put(82,137.70){$\cdot$}
\put(82.5,137.84){$\cdot$}
\put(83,137.98){$\cdot$}
\put(83.5,138.12){$\cdot$}
\put(84,138.27){$\cdot$}
\put(84.5,138.42){$\cdot$}
\put(85,138.57){$\cdot$}
\put(85.5,138.73){$\cdot$}
\put(86,138.90){$\cdot$}
\put(86.5,139.06){$\cdot$}
\put(87,139.24){$\cdot$}
\put(87.5,139.42){$\cdot$}
\put(88,139.60){$\cdot$}
\put(88.5,139.79){$\cdot$}
\put(89,139.99){$\cdot$}
\put(89.5,140.19){$\cdot$}
\put(90,140.40){$\cdot$}
\put(90.5,140.62){$\cdot$}
\put(91,140.84){$\cdot$}
\put(91.5,141.07){$\cdot$}
\put(92,141.32){$\cdot$}
\put(92.5,141.57){$\cdot$}
\put(93,141.83){$\cdot$}
\put(93.5,142.11){$\cdot$}
\put(94,142.40){$\cdot$}
\put(94.5,142.70){$\cdot$}
\put(95,143.02){$\cdot$}
\put(95.5,143.36){$\cdot$}
\put(96,143.73){$\cdot$}
\put(96.5,144.12){$\cdot$}
\put(97,144.54){$\cdot$}
\put(97.5,145.00){$\cdot$}
\put(98,145.52){$\cdot$}
\put(98.5,146.11){$\cdot$}
\put(99,146.82){$\cdot$}
\put(99.5,147.74){$\cdot$}
\put(100,150){$\cdot$}
\put(50,66){$\cdot$}
\put(49.5,65.99){$\cdot$}
\put(49,65.99){$\cdot$}
\put(48.5,65.99){$\cdot$}
\put(48,65.98){$\cdot$}
\put(45,65.91){$\cdot$}
\put(44.5,65.90){$\cdot$}
\put(44,65.88){$\cdot$}
\put(43.5,65.86){$\cdot$}
\put(43,65.84){$\cdot$}
\put(40,65.68){$\cdot$}
\put(39.5,65.64){$\cdot$}
\put(39,65.60){$\cdot$}
\put(38.5,65.57){$\cdot$}
\put(38,65.53){$\cdot$}
\put(35,65.26){$\cdot$}
\put(34.5,65.21){$\cdot$}
\put(34,65.16){$\cdot$}
\put(33.5,65.10){$\cdot$}
\put(33,65.04){$\cdot$}
\put(30,64.66){$\cdot$}
\put(29.5,64.59){$\cdot$}
\put(29,64.52){$\cdot$}
\put(28.5,64.44){$\cdot$}
\put(28,64.37){$\cdot$}
\put(25,63.86){$\cdot$}
\put(24.5,63.76){$\cdot$}
\put(24,63.66){$\cdot$}
\put(23.5,63.57){$\cdot$}
\put(23,63.47){$\cdot$}
\put(20,62.80){$\cdot$}
\put(19.5,62.68){$\cdot$}
\put(19,62.55){$\cdot$}
\put(18.5,62.42){$\cdot$}
\put(18,62.29){$\cdot$}
\put(15,61.43){$\cdot$}
\put(14.5,61.27){$\cdot$}
\put(14,61.10){$\cdot$}
\put(13.5,60.93){$\cdot$}
\put(13,60.76){$\cdot$}
\put(10,59.60){$\cdot$}
\put(9.5,59.38){$\cdot$}
\put(9,59.16){$\cdot$}
\put(8.5,58.92){$\cdot$}
\put(8,58.68){$\cdot$}
\put(5,56.97){$\cdot$}
\put(4.5,56.63){$\cdot$}
\put(4,56.27){$\cdot$}
\put(3.5,55.88){$\cdot$}
\put(3,55.46){$\cdot$}
\put(53,65.98){$\cdot$}
\put(53.5,65.97){$\cdot$}
\put(54,65.96){$\cdot$}
\put(54.5,65.95){$\cdot$}
\put(55,65.93){$\cdot$}
\put(58,65.82){$\cdot$}
\put(58.5,65.79){$\cdot$}
\put(59,65.77){$\cdot$}
\put(59.5,65.74){$\cdot$}
\put(60,65.70){$\cdot$}
\put(63,65.49){$\cdot$}
\put(63.5,65.45){$\cdot$}
\put(64,65.40){$\cdot$}
\put(64.5,65.36){$\cdot$}
\put(65,65.31){$\cdot$}
\put(68,64.99){$\cdot$}
\put(68.5,64.93){$\cdot$}
\put(69,64.86){$\cdot$}
\put(69.5,64.79){$\cdot$}
\put(70,64.73){$\cdot$}
\put(73,64.29){$\cdot$}
\put(73.5,64.20){$\cdot$}
\put(74,64.12){$\cdot$}
\put(74.5,64.04){$\cdot$}
\put(75,63.95){$\cdot$}
\put(78,63.36){$\cdot$}
\put(78.5,63.25){$\cdot$}
\put(79,63.15){$\cdot$}
\put(79.5,63.03){$\cdot$}
\put(80,62.92){$\cdot$}
\put(83,62.16){$\cdot$}
\put(83.5,62.02){$\cdot$}
\put(84,61.88){$\cdot$}
\put(84.5,61.73){$\cdot$}
\put(85,61.58){$\cdot$}
\put(88,60.58){$\cdot$}
\put(88.5,60.40){$\cdot$}
\put(89,60.20){$\cdot$}
\put(89.5,60.01){$\cdot$}
\put(90,59.81){$\cdot$}
\put(93,58.43){$\cdot$}
\put(93.5,58.16){$\cdot$}
\put(94,57.88){$\cdot$}
\put(94.5,57.59){$\cdot$}
\put(95,57.29){$\cdot$}
\put(98,54.99){$\cdot$}
\put(98.5,54.48){$\cdot$}
\put(99,53.89){$\cdot$}
\put(99.5,53.18){$\cdot$}
\put(100,52.26){$\cdot$}
\put(50,166){$\cdot$}
\put(49.5,165.99){$\cdot$}
\put(49,165.99){$\cdot$}
\put(48.5,165.99){$\cdot$}
\put(48,165.98){$\cdot$}
\put(45,165.91){$\cdot$}
\put(44.5,165.90){$\cdot$}
\put(44,165.88){$\cdot$}
\put(43.5,165.86){$\cdot$}
\put(43,165.84){$\cdot$}
\put(40,165.68){$\cdot$}
\put(39.5,165.64){$\cdot$}
\put(39,165.60){$\cdot$}
\put(38.5,165.57){$\cdot$}
\put(38,165.53){$\cdot$}
\put(35,165.26){$\cdot$}
\put(34.5,165.21){$\cdot$}
\put(34,165.16){$\cdot$}
\put(33.5,165.10){$\cdot$}
\put(33,165.04){$\cdot$}
\put(30,164.66){$\cdot$}
\put(29.5,164.59){$\cdot$}
\put(29,164.52){$\cdot$}
\put(28.5,164.44){$\cdot$}
\put(28,164.37){$\cdot$}
\put(25,163.86){$\cdot$}
\put(24.5,163.76){$\cdot$}
\put(24,163.66){$\cdot$}
\put(23.5,163.57){$\cdot$}
\put(23,163.47){$\cdot$}
\put(20,162.80){$\cdot$}
\put(19.5,162.68){$\cdot$}
\put(19,162.55){$\cdot$}
\put(18.5,162.42){$\cdot$}
\put(18,162.29){$\cdot$}
\put(15,161.43){$\cdot$}
\put(14.5,161.27){$\cdot$}
\put(14,161.10){$\cdot$}
\put(13.5,160.93){$\cdot$}
\put(13,160.76){$\cdot$}
\put(10,159.60){$\cdot$}
\put(9.5,159.38){$\cdot$}
\put(9,159.16){$\cdot$}
\put(8.5,158.92){$\cdot$}
\put(8,158.68){$\cdot$}
\put(5,156.97){$\cdot$}
\put(4.5,156.63){$\cdot$}
\put(4,156.27){$\cdot$}
\put(3.5,155.88){$\cdot$}
\put(3,155.46){$\cdot$}
\put(53,165.98){$\cdot$}
\put(53.5,165.97){$\cdot$}
\put(54,165.96){$\cdot$}
\put(54.5,165.95){$\cdot$}
\put(55,165.93){$\cdot$}
\put(58,165.82){$\cdot$}
\put(58.5,165.79){$\cdot$}
\put(59,165.77){$\cdot$}
\put(59.5,165.74){$\cdot$}
\put(60,165.70){$\cdot$}
\put(63,165.49){$\cdot$}
\put(63.5,165.45){$\cdot$}
\put(64,165.40){$\cdot$}
\put(64.5,165.36){$\cdot$}
\put(65,165.31){$\cdot$}
\put(68,164.99){$\cdot$}
\put(68.5,164.93){$\cdot$}
\put(69,164.86){$\cdot$}
\put(69.5,164.79){$\cdot$}
\put(70,164.73){$\cdot$}
\put(73,164.29){$\cdot$}
\put(73.5,164.20){$\cdot$}
\put(74,164.12){$\cdot$}
\put(74.5,164.04){$\cdot$}
\put(75,163.95){$\cdot$}
\put(78,163.36){$\cdot$}
\put(78.5,163.25){$\cdot$}
\put(79,163.15){$\cdot$}
\put(79.5,163.03){$\cdot$}
\put(80,162.92){$\cdot$}
\put(83,162.16){$\cdot$}
\put(83.5,162.02){$\cdot$}
\put(84,161.88){$\cdot$}
\put(84.5,161.73){$\cdot$}
\put(85,161.58){$\cdot$}
\put(88,160.58){$\cdot$}
\put(88.5,160.40){$\cdot$}
\put(89,160.20){$\cdot$}
\put(89.5,160.01){$\cdot$}
\put(90,159.81){$\cdot$}
\put(93,158.43){$\cdot$}
\put(93.5,158.16){$\cdot$}
\put(94,157.88){$\cdot$}
\put(94.5,157.59){$\cdot$}
\put(95,157.29){$\cdot$}
\put(98,154.99){$\cdot$}
\put(98.5,154.48){$\cdot$}
\put(99,153.89){$\cdot$}
\put(99.5,153.18){$\cdot$}
\put(100,152.26){$\cdot$}
%shere I quadrant
\put(50,100){$\cdot$}
\put(49.5,99.99){$\cdot$}
\put(49,99.98){$\cdot$}
\put(48.5,99.97){$\cdot$}
\put(48,99.96){$\cdot$}
\put(47.5,99.94){$\cdot$}
\put(47,99.90){$\cdot$}
\put(46.5,99.88){$\cdot$}
\put(46,99.84){$\cdot$}
\put(45.5,99.80){$\cdot$}
\put(45,99.75){$\cdot$}
\put(44.5,99.70){$\cdot$}
\put(44,99.64){$\cdot$}
\put(43.5,99.58){$\cdot$}
\put(43,99.51){$\cdot$}
\put(42.5,99.43){$\cdot$}
\put(42,99.35){$\cdot$}
\put(41.5,99.27){$\cdot$}
\put(41,99.18){$\cdot$}
\put(40.5,99.09){$\cdot$}
\put(40,98.99){$\cdot$}
\put(39.5,98.88){$\cdot$}
\put(39,98.77){$\cdot$}
\put(38.5,98.66){$\cdot$}
\put(38,98.54){$\cdot$}
\put(37.5,98.41){$\cdot$}
\put(37,98.28){$\cdot$}
\put(36.5,98.14){$\cdot$}
\put(36,98.00){$\cdot$}
\put(35.5,97.85){$\cdot$}
\put(35,97.70){$\cdot$}
\put(34.5,97.54){$\cdot$}
\put(34,97.37){$\cdot$}
\put(33.5,97.20){$\cdot$}
\put(33,97.02){$\cdot$}
\put(32.5,96.84){$\cdot$}
\put(32,96.65){$\cdot$}
\put(31.5,96.45){$\cdot$}
\put(31,96.25){$\cdot$}
\put(30.5,96.04){$\cdot$}
\put(30,95.82){$\cdot$}
\put(29.5,95.60){$\cdot$}
\put(29,95.38){$\cdot$}
\put(28.5,95.14){$\cdot$}
\put(28,94.90){$\cdot$}
\put(27.5,94.65){$\cdot$}
\put(27,94.39){$\cdot$}
\put(26.5,94.13){$\cdot$}
\put(26,93.86){$\cdot$}
\put(25.5,93.59){$\cdot$}
\put(25,93.30){$\cdot$}
\put(24.5,93.01){$\cdot$}
\put(24,92.71){$\cdot$}
\put(23.5,92.40){$\cdot$}
\put(23,92.08){$\cdot$}
\put(22.5,91.76){$\cdot$}
\put(22,91.42){$\cdot$}
\put(21.5,91.08){$\cdot$}
\put(21,90.73){$\cdot$}
\put(20.5,90.37){$\cdot$}
\put(20,90.00){$\cdot$}
\put(19.5,89.62){$\cdot$}
\put(19,89.23){$\cdot$}
\put(18.5,88.83){$\cdot$}
\put(18,88.42){$\cdot$}
\put(17.5,87.99){$\cdot$}
\put(17,87.56){$\cdot$}
\put(16.5,87.12){$\cdot$}
\put(16,86.66){$\cdot$}
\put(15.5,86.19){$\cdot$}
\put(15,85.70){$\cdot$}
\put(14.5,85.21){$\cdot$}
\put(14,84.70){$\cdot$}
\put(13.5,84.17){$\cdot$}
\put(13,83.63){$\cdot$}
\put(12.5,83.07){$\cdot$}
\put(12,82.49){$\cdot$}
\put(11.5,81.90){$\cdot$}
\put(11,81.29){$\cdot$}
\put(10.5,80.65){$\cdot$}
\put(10,80.00){$\cdot$}
\put(9.5,79.32){$\cdot$}
\put(9,78.62){$\cdot$}
\put(8.5,77.88){$\cdot$}
\put(8,77.13){$\cdot$}
\put(7.5,76.34){$\cdot$}
\put(7,75.51){$\cdot$}
\put(6.5,74.65){$\cdot$}
\put(6,73.75){$\cdot$}
\put(5.5,72.80){$\cdot$}
\put(5,71.79){$\cdot$}
\put(4.5,70.73){$\cdot$}
\put(4,69.59){$\cdot$}
\put(3.5,68.38){$\cdot$}
\put(3,67.06){$\cdot$}
\put(2.5,65.61){$\cdot$}
\put(2,64.00){$\cdot$}
\put(1.5,62.15){$\cdot$}
\put(1,59.95){$\cdot$}
\put(0.5,57.05){$\cdot$}
\put(0,50){$\cdot$}
%  II quadrant
\put(50,100){$\cdot$}
\put(50.5,99.99){$\cdot$}
\put(51,99.98){$\cdot$}
\put(51.5,99.97){$\cdot$}
\put(52,99.96){$\cdot$}
\put(52.5,99.94){$\cdot$}
\put(53,99.90){$\cdot$}
\put(53.5,99.88){$\cdot$}
\put(54,99.84){$\cdot$}
\put(54.5,99.80){$\cdot$}
\put(55,99.75){$\cdot$}
\put(55.5,99.70){$\cdot$}
\put(56,99.64){$\cdot$}
\put(56.5,99.58){$\cdot$}
\put(57,99.51){$\cdot$}
\put(57.5,99.43){$\cdot$}
\put(58,99.35){$\cdot$}
\put(58.5,99.27){$\cdot$}
\put(59,99.18){$\cdot$}
\put(59.5,99.09){$\cdot$}
\put(60,98.99){$\cdot$}
\put(60.5,98.88){$\cdot$}
\put(61,98.77){$\cdot$}
\put(61.5,98.66){$\cdot$}
\put(62,98.54){$\cdot$}
\put(62.5,98.41){$\cdot$}
\put(63,98.28){$\cdot$}
\put(63.5,98.14){$\cdot$}
\put(64,98.00){$\cdot$}
\put(64.5,97.85){$\cdot$}
\put(65,97.70){$\cdot$}
\put(65.5,97.54){$\cdot$}
\put(66,97.37){$\cdot$}
\put(66.5,97.20){$\cdot$}
\put(67,97.02){$\cdot$}
\put(67.5,96.84){$\cdot$}
\put(68,96.65){$\cdot$}
\put(68.5,96.45){$\cdot$}
\put(69,96.25){$\cdot$}
\put(69.5,96.04){$\cdot$}
\put(70,95.82){$\cdot$}
\put(70.5,95.60){$\cdot$}
\put(71,95.38){$\cdot$}
\put(71.5,95.14){$\cdot$}
\put(72,94.90){$\cdot$}
\put(72.5,94.65){$\cdot$}
\put(73,94.39){$\cdot$}
\put(73.5,94.13){$\cdot$}
\put(74,93.86){$\cdot$}
\put(74.5,93.59){$\cdot$}
\put(75,93.30){$\cdot$}
\put(75.5,93.01){$\cdot$}
\put(76,92.71){$\cdot$}
\put(76.5,92.40){$\cdot$}
\put(77,92.08){$\cdot$}
\put(77.5,91.76){$\cdot$}
\put(78,91.42){$\cdot$}
\put(78.5,91.08){$\cdot$}
\put(79,90.73){$\cdot$}
\put(79.5,90.37){$\cdot$}
\put(80,90.00){$\cdot$}
\put(80.5,89.62){$\cdot$}
\put(81,89.23){$\cdot$}
\put(81.5,88.83){$\cdot$}
\put(82,88.42){$\cdot$}
\put(82.5,87.99){$\cdot$}
\put(83,87.56){$\cdot$}
\put(83.5,87.12){$\cdot$}
\put(84,86.66){$\cdot$}
\put(84.5,86.19){$\cdot$}
\put(85,85.70){$\cdot$}
\put(85.5,85.21){$\cdot$}
\put(86,84.70){$\cdot$}
\put(86.5,84.17){$\cdot$}
\put(87,83.63){$\cdot$}
\put(87.5,83.07){$\cdot$}
\put(88,82.49){$\cdot$}
\put(88.5,81.90){$\cdot$}
\put(89,81.29){$\cdot$}
\put(89.5,80.65){$\cdot$}
\put(90,80.00){$\cdot$}
\put(90.5,79.32){$\cdot$}
\put(91,78.62){$\cdot$}
\put(91.5,77.88){$\cdot$}
\put(92,77.13){$\cdot$}
\put(92.5,76.34){$\cdot$}
\put(93,75.51){$\cdot$}
\put(93.5,74.65){$\cdot$}
\put(94,73.75){$\cdot$}
\put(94.5,72.80){$\cdot$}
\put(95,71.79){$\cdot$}
\put(95.5,70.73){$\cdot$}
\put(96,69.59){$\cdot$}
\put(96.5,68.38){$\cdot$}
\put(97,67.06){$\cdot$}
\put(97.5,65.61){$\cdot$}
\put(98,64.00){$\cdot$}
\put(98.5,62.15){$\cdot$}
\put(99,59.95){$\cdot$}
\put(99.5,57.05){$\cdot$}
\put(100,50){$\cdot$}
%shere I quadrant 2
\put(50,200){$\cdot$}
\put(49.5,199.99){$\cdot$}
\put(49,199.98){$\cdot$}
\put(48.5,199.97){$\cdot$}
\put(48,199.96){$\cdot$}
\put(47.5,199.94){$\cdot$}
\put(47,199.90){$\cdot$}
\put(46.5,199.88){$\cdot$}
\put(46,199.84){$\cdot$}
\put(45.5,199.80){$\cdot$}
\put(45,199.75){$\cdot$}
\put(44.5,199.70){$\cdot$}
\put(44,199.64){$\cdot$}
\put(43.5,199.58){$\cdot$}
\put(43,199.51){$\cdot$}
\put(42.5,199.43){$\cdot$}
\put(42,199.35){$\cdot$}
\put(41.5,199.27){$\cdot$}
\put(41,199.18){$\cdot$}
\put(40.5,199.09){$\cdot$}
\put(40,198.99){$\cdot$}
\put(39.5,198.88){$\cdot$}
\put(39,198.77){$\cdot$}
\put(38.5,198.66){$\cdot$}
\put(38,198.54){$\cdot$}
\put(37.5,198.41){$\cdot$}
\put(37,198.28){$\cdot$}
\put(36.5,198.14){$\cdot$}
\put(36,198.00){$\cdot$}
\put(35.5,197.85){$\cdot$}
\put(35,197.70){$\cdot$}
\put(34.5,197.54){$\cdot$}
\put(34,197.37){$\cdot$}
\put(33.5,197.20){$\cdot$}
\put(33,197.02){$\cdot$}
\put(32.5,196.84){$\cdot$}
\put(32,196.65){$\cdot$}
\put(31.5,196.45){$\cdot$}
\put(31,196.25){$\cdot$}
\put(30.5,196.04){$\cdot$}
\put(30,195.82){$\cdot$}
\put(29.5,195.60){$\cdot$}
\put(29,195.38){$\cdot$}
\put(28.5,195.14){$\cdot$}
\put(28,194.90){$\cdot$}
\put(27.5,194.65){$\cdot$}
\put(27,194.39){$\cdot$}
\put(26.5,194.13){$\cdot$}
\put(26,193.86){$\cdot$}
\put(25.5,193.59){$\cdot$}
\put(25,193.30){$\cdot$}
\put(24.5,193.01){$\cdot$}
\put(24,192.71){$\cdot$}
\put(23.5,192.40){$\cdot$}
\put(23,192.08){$\cdot$}
\put(22.5,191.76){$\cdot$}
\put(22,191.42){$\cdot$}
\put(21.5,191.08){$\cdot$}
\put(21,190.73){$\cdot$}
\put(20.5,190.37){$\cdot$}
\put(20,190.00){$\cdot$}
\put(19.5,189.62){$\cdot$}
\put(19,189.23){$\cdot$}
\put(18.5,188.83){$\cdot$}
\put(18,188.42){$\cdot$}
\put(17.5,187.99){$\cdot$}
\put(17,187.56){$\cdot$}
\put(16.5,187.12){$\cdot$}
\put(16,186.66){$\cdot$}
\put(15.5,186.19){$\cdot$}
\put(15,185.70){$\cdot$}
\put(14.5,185.21){$\cdot$}
\put(14,184.70){$\cdot$}
\put(13.5,184.17){$\cdot$}
\put(13,183.63){$\cdot$}
\put(12.5,183.07){$\cdot$}
\put(12,182.49){$\cdot$}
\put(11.5,181.90){$\cdot$}
\put(11,181.29){$\cdot$}
\put(10.5,180.65){$\cdot$}
\put(10,180.00){$\cdot$}
\put(9.5,179.32){$\cdot$}
\put(9,178.62){$\cdot$}
\put(8.5,177.88){$\cdot$}
\put(8,177.13){$\cdot$}
\put(7.5,176.34){$\cdot$}
\put(7,175.51){$\cdot$}
\put(6.5,174.65){$\cdot$}
\put(6,173.75){$\cdot$}
\put(5.5,172.80){$\cdot$}
\put(5,171.79){$\cdot$}
\put(4.5,170.73){$\cdot$}
\put(4,169.59){$\cdot$}
\put(3.5,168.38){$\cdot$}
\put(3,167.06){$\cdot$}
\put(2.5,165.61){$\cdot$}
\put(2,164.00){$\cdot$}
\put(1.5,162.15){$\cdot$}
\put(1,159.95){$\cdot$}
\put(0.5,157.05){$\cdot$}
\put(0,150){$\cdot$}
%  II quadrant 2
\put(50,200){$\cdot$}
\put(50.5,199.99){$\cdot$}
\put(51,199.98){$\cdot$}
\put(51.5,199.97){$\cdot$}
\put(52,199.96){$\cdot$}
\put(52.5,199.94){$\cdot$}
\put(53,199.90){$\cdot$}
\put(53.5,199.88){$\cdot$}
\put(54,199.84){$\cdot$}
\put(54.5,199.80){$\cdot$}
\put(55,199.75){$\cdot$}
\put(55.5,199.70){$\cdot$}
\put(56,199.64){$\cdot$}
\put(56.5,199.58){$\cdot$}
\put(57,199.51){$\cdot$}
\put(57.5,199.43){$\cdot$}
\put(58,199.35){$\cdot$}
\put(58.5,199.27){$\cdot$}
\put(59,199.18){$\cdot$}
\put(59.5,199.09){$\cdot$}
\put(60,198.99){$\cdot$}
\put(60.5,198.88){$\cdot$}
\put(61,198.77){$\cdot$}
\put(61.5,198.66){$\cdot$}
\put(62,198.54){$\cdot$}
\put(62.5,198.41){$\cdot$}
\put(63,198.28){$\cdot$}
\put(63.5,198.14){$\cdot$}
\put(64,198.00){$\cdot$}
\put(64.5,197.85){$\cdot$}
\put(65,197.70){$\cdot$}
\put(65.5,197.54){$\cdot$}
\put(66,197.37){$\cdot$}
\put(66.5,197.20){$\cdot$}
\put(67,197.02){$\cdot$}
\put(67.5,196.84){$\cdot$}
\put(68,196.65){$\cdot$}
\put(68.5,196.45){$\cdot$}
\put(69,196.25){$\cdot$}
\put(69.5,196.04){$\cdot$}
\put(70,195.82){$\cdot$}
\put(70.5,195.60){$\cdot$}
\put(71,195.38){$\cdot$}
\put(71.5,195.14){$\cdot$}
\put(72,194.90){$\cdot$}
\put(72.5,194.65){$\cdot$}
\put(73,194.39){$\cdot$}
\put(73.5,194.13){$\cdot$}
\put(74,193.86){$\cdot$}
\put(74.5,193.59){$\cdot$}
\put(75,193.30){$\cdot$}
\put(75.5,193.01){$\cdot$}
\put(76,192.71){$\cdot$}
\put(76.5,192.40){$\cdot$}
\put(77,192.08){$\cdot$}
\put(77.5,191.76){$\cdot$}
\put(78,191.42){$\cdot$}
\put(78.5,191.08){$\cdot$}
\put(79,190.73){$\cdot$}
\put(79.5,190.37){$\cdot$}
\put(80,190.00){$\cdot$}
\put(80.5,189.62){$\cdot$}
\put(81,189.23){$\cdot$}
\put(81.5,188.83){$\cdot$}
\put(82,188.42){$\cdot$}
\put(82.5,187.99){$\cdot$}
\put(83,187.56){$\cdot$}
\put(83.5,187.12){$\cdot$}
\put(84,186.66){$\cdot$}
\put(84.5,186.19){$\cdot$}
\put(85,185.70){$\cdot$}
\put(85.5,185.21){$\cdot$}
\put(86,184.70){$\cdot$}
\put(86.5,184.17){$\cdot$}
\put(87,183.63){$\cdot$}
\put(87.5,183.07){$\cdot$}
\put(88,182.49){$\cdot$}
\put(88.5,181.90){$\cdot$}
\put(89,181.29){$\cdot$}
\put(89.5,180.65){$\cdot$}
\put(90,180.00){$\cdot$}
\put(90.5,179.32){$\cdot$}
\put(91,178.62){$\cdot$}
\put(91.5,177.88){$\cdot$}
\put(92,177.13){$\cdot$}
\put(92.5,176.34){$\cdot$}
\put(93,175.51){$\cdot$}
\put(93.5,174.65){$\cdot$}
\put(94,173.75){$\cdot$}
\put(94.5,172.80){$\cdot$}
\put(95,171.79){$\cdot$}
\put(95.5,170.73){$\cdot$}
\put(96,169.59){$\cdot$}
\put(96.5,168.38){$\cdot$}
\put(97,167.06){$\cdot$}
\put(97.5,165.61){$\cdot$}
\put(98,164.00){$\cdot$}
\put(98.5,162.15){$\cdot$}
\put(99,159.95){$\cdot$}
\put(99.5,157.05){$\cdot$}
\put(100,150){$\cdot$}
% shere III quadrant
\put(50,0){$\cdot$}
\put(49.5,0){$\cdot$}
\put(49,0.01){$\cdot$}
\put(48.5,0.02){$\cdot$}
\put(48,0.04){$\cdot$}
\put(47.5,0.06){$\cdot$}
\put(47,0.09){$\cdot$}
\put(46.5,0.12){$\cdot$}
\put(46,0.16){$\cdot$}
\put(45.5,0.2){$\cdot$}
\put(45,0.25){$\cdot$}
\put(44.5,0.3){$\cdot$}
\put(44,0.36){$\cdot$}
\put(43.5,0.42){$\cdot$}
\put(43,0.49){$\cdot$}
\put(42.5,0.56){$\cdot$}
\put(42,0.64){$\cdot$}
\put(41.5,0.73){$\cdot$}
\put(41,0.82){$\cdot$}
\put(40.5,0.91){$\cdot$}
\put(40,1.01){$\cdot$}
\put(39.5,1.11){$\cdot$}
\put(39,1.22){$\cdot$}
\put(38.5,1.34){$\cdot$}
\put(38,1.46){$\cdot$}
\put(37.5,1.59){$\cdot$}
\put(37,1.72){$\cdot$}
\put(36.5,1.86){$\cdot$}
\put(36,2.0){$\cdot$}
\put(35.5,2.15){$\cdot$}
\put(35,2.3){$\cdot$}
\put(34.5,2.46){$\cdot$}
\put(34,2.63){$\cdot$}
\put(33.5,2.8){$\cdot$}
\put(33,2.98){$\cdot$}
\put(32.5,3.16){$\cdot$}
\put(32,3.35){$\cdot$}
\put(31.5,3.55){$\cdot$}
\put(31,3.75){$\cdot$}
\put(30.5,3.96){$\cdot$}
\put(30,4.17){$\cdot$}
\put(29.5,4.39){$\cdot$}
\put(29,4.62){$\cdot$}
\put(28.5,4.86){$\cdot$}
\put(28,5.1){$\cdot$}
\put(27.5,5.35){$\cdot$}
\put(27,5.6){$\cdot$}
\put(26.5,5.87){$\cdot$}
\put(26,6.14){$\cdot$}
\put(25.5,6.41){$\cdot$}
\put(25,6.7){$\cdot$}
\put(24.5,6.99){$\cdot$}
\put(24,7.29){$\cdot$}
\put(23.5,7.6){$\cdot$}
\put(23,7.9){$\cdot$}
\put(22.5,8.24){$\cdot$}
\put(22,8.57){$\cdot$}
\put(21.5,8.92){$\cdot$}
\put(21,9.27){$\cdot$}
\put(20.5,9.63){$\cdot$}
\put(20,10.0){$\cdot$}
\put(19.5,10.38){$\cdot$}
\put(19,10.77){$\cdot$}
\put(18.5,11.17){$\cdot$}
\put(18,11.58){$\cdot$}
\put(17.5,12.0){$\cdot$}
\put(17,12.44){$\cdot$}
\put(16.5,12.88){$\cdot$}
\put(16,13.34){$\cdot$}
\put(15.5,13.81){$\cdot$}
\put(15,14.29){$\cdot$}
\put(14.5,14.79){$\cdot$}
\put(14,15.3){$\cdot$}
\put(13.5,15.83){$\cdot$}
\put(13,16.37){$\cdot$}
\put(12.5,16.93){$\cdot$}
\put(12,17.5){$\cdot$}
\put(11.5,18.1){$\cdot$}
\put(11,18.71){$\cdot$}
\put(10.5,19.34){$\cdot$}
\put(10,20.00){$\cdot$}
\put(9.5,20.68){$\cdot$}
\put(9,21.38){$\cdot$}
\put(8.5,22.11){$\cdot$}
\put(8,22.87){$\cdot$}
\put(7.5,23.66){$\cdot$}
\put(7,24.48){$\cdot$}
\put(6.5,25.35){$\cdot$}
\put(6,26.25){$\cdot$}
\put(5.5,27.2){$\cdot$}
\put(5,28.2){$\cdot$}
\put(4.5,29.27){$\cdot$}
\put(4,30.4){$\cdot$}
\put(3.5,31.62){$\cdot$}
\put(3,32.94){$\cdot$}
\put(2.5,34.39){$\cdot$}
\put(2,36.00){$\cdot$}
\put(1.5,37.84){$\cdot$}
\put(1,40.05){$\cdot$}
\put(0.5,42.95){$\cdot$}
\put(0,50){$\cdot$}
% shere IV quadrant
\put(50,0){$\cdot$}
\put(50.5,0){$\cdot$}
\put(51,0.01){$\cdot$}
\put(51.5,0.02){$\cdot$}
\put(52,0.04){$\cdot$}
\put(52.5,0.06){$\cdot$}
\put(53,0.09){$\cdot$}
\put(53.5,0.12){$\cdot$}
\put(54,0.16){$\cdot$}
\put(54.5,0.2){$\cdot$}
\put(55,0.25){$\cdot$}
\put(55.5,0.3){$\cdot$}
\put(56,0.36){$\cdot$}
\put(56.5,0.42){$\cdot$}
\put(57,0.49){$\cdot$}
\put(57.5,0.56){$\cdot$}
\put(58,0.64){$\cdot$}
\put(58.5,0.73){$\cdot$}
\put(59,0.82){$\cdot$}
\put(59.5,0.91){$\cdot$}
\put(60,1.01){$\cdot$}
\put(60.5,1.11){$\cdot$}
\put(61,1.22){$\cdot$}
\put(61.5,1.34){$\cdot$}
\put(62,1.46){$\cdot$}
\put(62.5,1.59){$\cdot$}
\put(63,1.72){$\cdot$}
\put(63.5,1.86){$\cdot$}
\put(64,2.0){$\cdot$}
\put(64.5,2.15){$\cdot$}
\put(65,2.3){$\cdot$}
\put(65.5,2.46){$\cdot$}
\put(66,2.63){$\cdot$}
\put(66.5,2.8){$\cdot$}
\put(67,2.98){$\cdot$}
\put(67.5,3.16){$\cdot$}
\put(68,3.35){$\cdot$}
\put(68.5,3.55){$\cdot$}
\put(69,3.75){$\cdot$}
\put(69.5,3.96){$\cdot$}
\put(70,4.17){$\cdot$}
\put(70.5,4.39){$\cdot$}
\put(71,4.62){$\cdot$}
\put(71.5,4.86){$\cdot$}
\put(72,5.1){$\cdot$}
\put(72.5,5.35){$\cdot$}
\put(73,5.6){$\cdot$}
\put(73.5,5.87){$\cdot$}
\put(74,6.14){$\cdot$}
\put(74.5,6.41){$\cdot$}
\put(75,6.7){$\cdot$}
\put(75.5,6.99){$\cdot$}
\put(76,7.29){$\cdot$}
\put(76.5,7.6){$\cdot$}
\put(77,7.9){$\cdot$}
\put(77.5,8.24){$\cdot$}
\put(78,8.57){$\cdot$}
\put(78.5,8.92){$\cdot$}
\put(79,9.27){$\cdot$}
\put(79.5,9.63){$\cdot$}
\put(80,10.0){$\cdot$}
\put(80.5,10.38){$\cdot$}
\put(81,10.77){$\cdot$}
\put(81.5,11.17){$\cdot$}
\put(82,11.58){$\cdot$}
\put(82.5,12.0){$\cdot$}
\put(83,12.44){$\cdot$}
\put(83.5,12.88){$\cdot$}
\put(84,13.34){$\cdot$}
\put(84.5,13.81){$\cdot$}
\put(85,14.29){$\cdot$}
\put(85.5,14.79){$\cdot$}
\put(86,15.3){$\cdot$}
\put(86.5,15.83){$\cdot$}
\put(87,16.37){$\cdot$}
\put(87.5,16.93){$\cdot$}
\put(88,17.5){$\cdot$}
\put(88.5,18.1){$\cdot$}
\put(89,18.71){$\cdot$}
\put(89.5,19.34){$\cdot$}
\put(90,20.00){$\cdot$}
\put(90.5,20.68){$\cdot$}
\put(91,21.38){$\cdot$}
\put(91.5,22.11){$\cdot$}
\put(92,22.87){$\cdot$}
\put(92.5,23.66){$\cdot$}
\put(93,24.48){$\cdot$}
\put(93.5,25.35){$\cdot$}
\put(94,26.25){$\cdot$}
\put(94.5,27.2){$\cdot$}
\put(95,28.2){$\cdot$}
\put(95.5,29.27){$\cdot$}
\put(96,30.4){$\cdot$}
\put(96.5,31.62){$\cdot$}
\put(97,32.94){$\cdot$}
\put(97.5,34.39){$\cdot$}
\put(98,36.00){$\cdot$}
\put(98.5,37.84){$\cdot$}
\put(99,40.05){$\cdot$}
\put(99.5,42.95){$\cdot$}
\put(100,50){$\cdot$}
% shere III quadrant 2
\put(50,100){$\cdot$}
\put(49.5,100){$\cdot$}
\put(49,100.01){$\cdot$}
\put(48.5,100.02){$\cdot$}
\put(48,100.04){$\cdot$}
\put(47.5,100.06){$\cdot$}
\put(47,100.09){$\cdot$}
\put(46.5,100.12){$\cdot$}
\put(46,100.16){$\cdot$}
\put(45.5,100.2){$\cdot$}
\put(45,100.25){$\cdot$}
\put(44.5,100.3){$\cdot$}
\put(44,100.36){$\cdot$}
\put(43.5,100.42){$\cdot$}
\put(43,100.49){$\cdot$}
\put(42.5,100.56){$\cdot$}
\put(42,100.64){$\cdot$}
\put(41.5,100.73){$\cdot$}
\put(41,100.82){$\cdot$}
\put(40.5,100.91){$\cdot$}
\put(40,101.01){$\cdot$}
\put(39.5,101.11){$\cdot$}
\put(39,101.22){$\cdot$}
\put(38.5,101.34){$\cdot$}
\put(38,101.46){$\cdot$}
\put(37.5,101.59){$\cdot$}
\put(37,101.72){$\cdot$}
\put(36.5,101.86){$\cdot$}
\put(36,102.0){$\cdot$}
\put(35.5,102.15){$\cdot$}
\put(35,102.3){$\cdot$}
\put(34.5,102.46){$\cdot$}
\put(34,102.63){$\cdot$}
\put(33.5,102.8){$\cdot$}
\put(33,102.98){$\cdot$}
\put(32.5,103.16){$\cdot$}
\put(32,103.35){$\cdot$}
\put(31.5,103.55){$\cdot$}
\put(31,103.75){$\cdot$}
\put(30.5,103.96){$\cdot$}
\put(30,104.17){$\cdot$}
\put(29.5,104.39){$\cdot$}
\put(29,104.62){$\cdot$}
\put(28.5,104.86){$\cdot$}
\put(28,105.1){$\cdot$}
\put(27.5,105.35){$\cdot$}
\put(27,105.6){$\cdot$}
\put(26.5,105.87){$\cdot$}
\put(26,106.14){$\cdot$}
\put(25.5,106.41){$\cdot$}
\put(25,106.7){$\cdot$}
\put(24.5,106.99){$\cdot$}
\put(24,107.29){$\cdot$}
\put(23.5,107.6){$\cdot$}
\put(23,107.9){$\cdot$}
\put(22.5,108.24){$\cdot$}
\put(22,108.57){$\cdot$}
\put(21.5,108.92){$\cdot$}
\put(21,109.27){$\cdot$}
\put(20.5,109.63){$\cdot$}
\put(20,110.0){$\cdot$}
\put(19.5,110.38){$\cdot$}
\put(19,110.77){$\cdot$}
\put(18.5,111.17){$\cdot$}
\put(18,111.58){$\cdot$}
\put(17.5,112.0){$\cdot$}
\put(17,112.44){$\cdot$}
\put(16.5,112.88){$\cdot$}
\put(16,113.34){$\cdot$}
\put(15.5,113.81){$\cdot$}
\put(15,114.29){$\cdot$}
\put(14.5,114.79){$\cdot$}
\put(14,115.3){$\cdot$}
\put(13.5,115.83){$\cdot$}
\put(13,116.37){$\cdot$}
\put(12.5,116.93){$\cdot$}
\put(12,117.5){$\cdot$}
\put(11.5,118.1){$\cdot$}
\put(11,118.71){$\cdot$}
\put(10.5,119.34){$\cdot$}
\put(10,120.00){$\cdot$}
\put(9.5,120.68){$\cdot$}
\put(9,121.38){$\cdot$}
\put(8.5,122.11){$\cdot$}
\put(8,122.87){$\cdot$}
\put(7.5,123.66){$\cdot$}
\put(7,124.48){$\cdot$}
\put(6.5,125.35){$\cdot$}
\put(6,126.25){$\cdot$}
\put(5.5,127.2){$\cdot$}
\put(5,128.2){$\cdot$}
\put(4.5,129.27){$\cdot$}
\put(4,130.4){$\cdot$}
\put(3.5,131.62){$\cdot$}
\put(3,132.94){$\cdot$}
\put(2.5,134.39){$\cdot$}
\put(2,136.00){$\cdot$}
\put(1.5,137.84){$\cdot$}
\put(1,140.05){$\cdot$}
\put(0.5,142.95){$\cdot$}
\put(0,150){$\cdot$}
% shere IV quadrant 2
\put(50,100){$\cdot$}
\put(50.5,100){$\cdot$}
\put(51,100.01){$\cdot$}
\put(51.5,100.02){$\cdot$}
\put(52,100.04){$\cdot$}
\put(52.5,100.06){$\cdot$}
\put(53,100.09){$\cdot$}
\put(53.5,100.12){$\cdot$}
\put(54,100.16){$\cdot$}
\put(54.5,100.2){$\cdot$}
\put(55,100.25){$\cdot$}
\put(55.5,100.3){$\cdot$}
\put(56,100.36){$\cdot$}
\put(56.5,100.42){$\cdot$}
\put(57,100.49){$\cdot$}
\put(57.5,100.56){$\cdot$}
\put(58,100.64){$\cdot$}
\put(58.5,100.73){$\cdot$}
\put(59,100.82){$\cdot$}
\put(59.5,100.91){$\cdot$}
\put(60,101.01){$\cdot$}
\put(60.5,101.11){$\cdot$}
\put(61,101.22){$\cdot$}
\put(61.5,101.34){$\cdot$}
\put(62,101.46){$\cdot$}
\put(62.5,101.59){$\cdot$}
\put(63,101.72){$\cdot$}
\put(63.5,101.86){$\cdot$}
\put(64,102.0){$\cdot$}
\put(64.5,102.15){$\cdot$}
\put(65,102.3){$\cdot$}
\put(65.5,102.46){$\cdot$}
\put(66,102.63){$\cdot$}
\put(66.5,102.8){$\cdot$}
\put(67,102.98){$\cdot$}
\put(67.5,103.16){$\cdot$}
\put(68,103.35){$\cdot$}
\put(68.5,103.55){$\cdot$}
\put(69,103.75){$\cdot$}
\put(69.5,103.96){$\cdot$}
\put(70,104.17){$\cdot$}
\put(70.5,104.39){$\cdot$}
\put(71,104.62){$\cdot$}
\put(71.5,104.86){$\cdot$}
\put(72,105.1){$\cdot$}
\put(72.5,105.35){$\cdot$}
\put(73,105.6){$\cdot$}
\put(73.5,105.87){$\cdot$}
\put(74,106.14){$\cdot$}
\put(74.5,106.41){$\cdot$}
\put(75,106.7){$\cdot$}
\put(75.5,106.99){$\cdot$}
\put(76,107.29){$\cdot$}
\put(76.5,107.6){$\cdot$}
\put(77,107.9){$\cdot$}
\put(77.5,108.24){$\cdot$}
\put(78,108.57){$\cdot$}
\put(78.5,108.92){$\cdot$}
\put(79,109.27){$\cdot$}
\put(79.5,109.63){$\cdot$}
\put(80,110.0){$\cdot$}
\put(80.5,110.38){$\cdot$}
\put(81,110.77){$\cdot$}
\put(81.5,111.17){$\cdot$}
\put(82,111.58){$\cdot$}
\put(82.5,112.0){$\cdot$}
\put(83,112.44){$\cdot$}
\put(83.5,112.88){$\cdot$}
\put(84,113.34){$\cdot$}
\put(84.5,113.81){$\cdot$}
\put(85,114.29){$\cdot$}
\put(85.5,114.79){$\cdot$}
\put(86,115.3){$\cdot$}
\put(86.5,115.83){$\cdot$}
\put(87,116.37){$\cdot$}
\put(87.5,116.93){$\cdot$}
\put(88,117.5){$\cdot$}
\put(88.5,118.1){$\cdot$}
\put(89,118.71){$\cdot$}
\put(89.5,119.34){$\cdot$}
\put(90,120.00){$\cdot$}
\put(90.5,120.68){$\cdot$}
\put(91,121.38){$\cdot$}
\put(91.5,122.11){$\cdot$}
\put(92,122.87){$\cdot$}
\put(92.5,123.66){$\cdot$}
\put(93,124.48){$\cdot$}
\put(93.5,125.35){$\cdot$}
\put(94,126.25){$\cdot$}
\put(94.5,127.2){$\cdot$}
\put(95,128.2){$\cdot$}
\put(95.5,129.27){$\cdot$}
\put(96,130.4){$\cdot$}
\put(96.5,131.62){$\cdot$}
\put(97,132.94){$\cdot$}
\put(97.5,134.39){$\cdot$}
\put(98,136.00){$\cdot$}
\put(98.5,137.84){$\cdot$}
\put(99,140.05){$\cdot$}
\put(99.5,142.95){$\cdot$}
\put(100,150){$\cdot$}
\end{picture}
\]
\vspace{2ex}
\begin{center}
\begin{minipage}{25pc}{\small
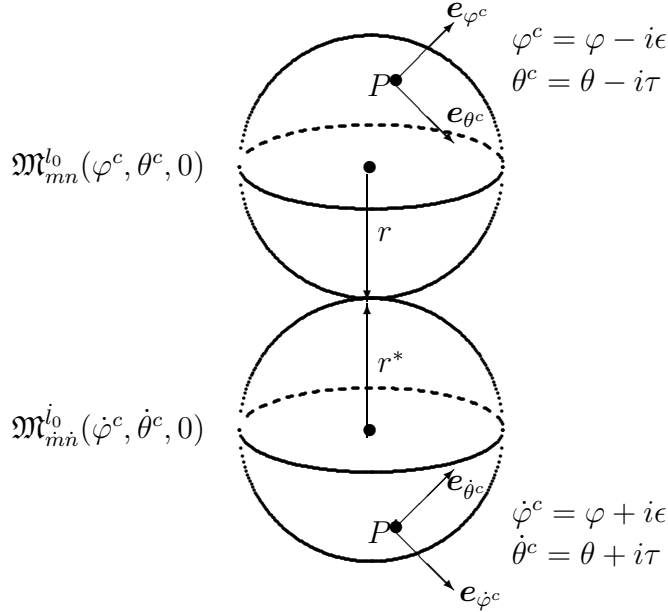
{\bf Fig.1} Two--dimensional complex sphere $z^2_1+z^2_2+z^2_3=r^2$ in
three--dimensional complex space $\C^3$. The space $\C^3$ is isometric
to the bivector space $\R^6$. The dual (complex conjugate) sphere
$\overset{\ast}{z}_1{}^2+\overset{\ast}{z}_2{}^2+\overset{\ast}{z}_3{}^2=
\overset{\ast}{r}{}^2$ is a mirror image of the complex sphere with
respect to the hyperplane. The hyperspherical functions
$\fM^l_{mn}(\varphi^c,\theta^c,0)$ 
($\fM^{\dot{l}}_{\dot{m}\dot{n}}(\dot{\varphi}^c,\dot{\theta}^c,0)$)
are defined on the surface of the complex (dual) sphere.}
\end{minipage}
\end{center}
\medskip
%\end{float}
%\end{document}

Therefore, with the each point of the Minkowski spacetime $\R^{1,3}$
we can associate the two--dimensional complex sphere and its conjugate.

Let us introduce now hyperspherical coordinates on the surfaces of the
complex and dual spheres
\begin{equation}\label{HC}
\begin{array}{ccl}
z_1&=&r\sin\theta^c\cos\varphi^c,\\
z_2&=&r\sin\theta^c\sin\varphi^c,\\
z_3&=&r\cos\theta^c,
\end{array}\quad
\begin{array}{ccl}
z^\ast_1&=&r^\ast\sin\dot{\theta}^c\cos\dot{\varphi}^c,\\
z^\ast_2&=&r^\ast\sin\dot{\theta}^c\sin\dot{\varphi}^c,\\
z^\ast_3&=&r^\ast\cos\dot{\theta}^c,
\end{array}
\end{equation}
where $\theta^c$, $\varphi^c$ are the complex Euler angles.
Casimir operators on the 2-dimensional complex sphere (correspondingly,
on the dual sphere) have the form
\begin{eqnarray}
\sX^2&=&\frac{\partial^2}{\partial\theta^c{}^2}+
\ctg\theta^c\frac{\partial}{\partial\theta^c}+\frac{1}{\sin^2\theta^c}
\frac{\partial^2}{\partial\varphi^c{}^2},\nonumber\\
\sY^2&=&\frac{\partial^2}{\partial\dot{\theta}^c{}^2}+
\ctg\dot{\theta}^c\frac{\partial}{\partial\dot{\theta}^c}+
\frac{1}{\sin^2\dot{\varphi}^c{}^2}
\frac{\partial^2}{\partial\dot{\varphi}^c{}^2}.\label{KO3}
\end{eqnarray}
Hyperspherical functions $\fM^l_{mn}(\varphi^c,\theta^c,0)$ and
$\fM^{\dot{l}}_{\dot{m}\dot{n}}(\dot{\varphi}^c,\dot{\theta}^c,0)$,
defined on the surface of the two-dimensional complex sphere, 
are eigenfunctions
of the operators $\sX^2$ and $\sY^2$:
\begin{eqnarray}
\left[\sX^2+l(l+1)\right]\fM^l_{mn}(\varphi^c,\theta^c,0)&=&0,\nonumber\\
\left[\sY^2+\dot{l}(\dot{l}+1)\right]
\fM^{\dot{l}}_{\dot{m}\dot{n}}(\dot{\varphi}^c,\dot{\theta}^c,0)&=&0
\label{EQ2}
\end{eqnarray}
\begin{sloppypar}\noindent
Substituting the functions $\fM^l_{mn}(\varphi^c,\theta^c,0)=
e^{-im\varphi^c}Z^l_{mn}(\theta^c)$, $\fM^{\dot{l}}_{\dot{m}\dot{n}}
(\dot{\varphi}^c,\dot{\theta}^c,0)=e^{-i\dot{m}\dot{\varphi}^c}
Z^{\dot{l}}_{\dot{m}\dot{n}}(\dot{\theta}^c)$ into (\ref{EQ2}) and
taking into account the operators (\ref{KO3}) we obtain the following
equations\end{sloppypar}
\begin{eqnarray}
\left[\frac{d^2}{d\theta^c{}^2}+\ctg\theta^c\frac{d}{d\theta^c}-
\frac{m^2}{\sin^2\theta^c}+l(l+1)\right]Z^l_{mn}(\theta^c)&=&0,\nonumber\\
\left[\frac{d^2}{d\dot{\theta}^c{}^2}+\ctg\dot{\theta}^c
\frac{d}{d\dot{\theta}^c}-\frac{m^2}{\sin^2\dot{\theta}^c}+
\dot{l}(\dot{l}+1)\right]Z^{\dot{l}}_{\dot{m}\dot{n}}(\dot{\theta}^c)&=&0.
\nonumber
\end{eqnarray}
Or, introducing the substitutions $z=\cos\theta^c$,
$\overset{\ast}{z}=\cos\dot{\theta}^c$ we find that
\begin{eqnarray}
\left[(1-z^2)\frac{d^2}{dz^2}-2z\frac{d}{dz}-\frac{m^2}{1-z^2}
+l(l+1)\right]Z^l_{mn}(\arc z)&=&0,\nonumber\\
\left[(1-\overset{\ast}{z}{}^2)\frac{d^2}{d\overset{\ast}{z}{}^2}-
2\overset{\ast}{z}\frac{d}{d\overset{\ast}{z}}-
\frac{\dot{m}^2}{1-\overset{\ast}{z}{}^2}+\dot{l}(\dot{l}+1)\right]
Z^{\dot{l}}_{\dot{m}\dot{n}}(\arc\overset{\ast}{z})&=&0.\nonumber
\end{eqnarray}
The scalar product of the functions $\fM^{\dot{l}}_{\dot{m}\dot{n}}
(\dot{\varphi}^c,\dot{\theta}^c,0)$ and $\fM^l_{mn}(\varphi^c,\theta^c,0)$,
defined on the complex shere, is given by an expression
\[
\int d\Omega\fM^{\dot{l}}_{\dot{m}\dot{n}}\fM^l_{mn}=
\int d\varphi d\epsilon d\theta d\tau\sin\theta^c\sin\dot{\theta}^c
\fM^{\dot{l}}_{\dot{m}\dot{n}}(\dot{\varphi}^c,\dot{\theta}^c,0)
\fM^l_{mn}(\varphi^c,\theta^c,0)
\]
with the integration limits
\[
\begin{array}{ccccc}
0 & \leq & \varphi & < & 2\pi,\\
0 & \leq & \theta & < & \pi,
\end{array}\quad
\begin{array}{ccccc}
-\infty & < & \epsilon & < & +\infty,\\
-\infty & < & \tau & < & +\infty.
\end{array}
\]
\begin{sloppypar}
Let us show that a solution of the generalized Gel'fand--Yaglom equations
(\ref{Complex}), or the equations (\ref{GY}), can be found in the form
of the series on generalized hyperspherical functions considered in the
previous section.

With this end in view let us transform the system (\ref{Complex}) as
follows. First of all, let us define the derivatives
$\dfrac{\partial}{\partial x_i}$, $\dfrac{\partial}{\partial x^\ast_i}$
on the surface of the complex sphere (\ref{CS}) and write them in the
hyperspherical coordinates (\ref{HC}) as
\end{sloppypar}
\begin{eqnarray}
\frac{\partial}{\partial x_1}&=&-\frac{\sin\varphi^c}{r\sin\theta^c}
\frac{\partial}{\partial\varphi}+\frac{\cos\varphi^c\cos\theta^c}{r}
\frac{\partial}{\partial\theta}+\cos\varphi^c\sin\theta^c\frac{\partial}
{\partial r},\label{CD1}\\
\frac{\partial}{\partial x_2}&=&\frac{\cos\varphi^c}{r\sin\theta^c}
\frac{\partial}{\partial\varphi}+\frac{\sin\varphi^c\cos\theta^c}{r}
\frac{\partial}{\partial\theta}+\sin\varphi^c\sin\theta^c\frac{\partial}
{\partial r},\label{CD2}\\
\frac{\partial}{\partial x_3}&=&-\frac{\sin\theta^c}{r}\frac{\partial}
{\partial\theta}+\cos\theta^c\frac{\partial}{\partial r}.\label{CD3}
\end{eqnarray}
\begin{eqnarray}
\frac{\partial}{\partial x^\ast_1}&=&i\frac{\partial}{\partial x_1}=
-\frac{\sin\varphi^c}{r\sin\theta^c}\frac{\partial}{\partial\epsilon}+
\frac{\cos\varphi^c\sin\theta^c}{r}\frac{\partial}{\partial\tau}+
i\cos\varphi^c\sin\theta^c\frac{\partial}{\partial r},\label{CD4}\\
\frac{\partial}{\partial x^\ast_2}&=&i\frac{\partial}{\partial x_2}=
\frac{\cos\varphi^c}{r\sin\theta^c}\frac{\partial}{\partial\epsilon}+
\frac{\sin\varphi^c\cos\theta^c}{r}\frac{\partial}{\partial\tau}+
i\sin\varphi^c\sin\theta^c\frac{\partial}{\partial r},\label{CD5}\\
\frac{\partial}{\partial x^\ast_3}&=&i\frac{\partial}{\partial x_3}=
-\frac{\sin\theta^c}{r}\frac{\partial}{\partial\tau}+
i\cos\theta^c\frac{\partial}{\partial r}.\label{CD6}
\end{eqnarray}
Analogously, on the surface of the dual sphere we have
\begin{eqnarray}
\frac{\partial}{\partial\widetilde{x}_1}&=&
-\frac{\sin\dot{\varphi}^c}{r^\ast\sin\dot{\theta}^c}\frac{\partial}
{\partial\varphi}+\frac{\cos\dot{\varphi}^c\cos\dot{\theta}^c}{r^\ast}
\frac{\partial}{\partial\theta}+\cos\dot{\varphi}^c\sin\dot{\theta}^c
\frac{\partial}{\partial r^\ast},\label{CDD1}\\
\frac{\partial}{\partial\widetilde{x}_2}&=&
\frac{\cos\dot{\varphi}^c}{r^\ast\sin\dot{\theta}^c}\frac{\partial}
{\partial\varphi}+\frac{\sin\dot{\varphi}^c\cos\dot{\theta}^c}{r^\ast}
\frac{\partial}{\partial\theta}+\sin\dot{\varphi}^c\sin\dot{\theta}^c
\frac{\partial}{\partial r^\ast},\label{CDD2}\\
\frac{\partial}{\partial\widetilde{x}_3}&=&
-\frac{\sin\dot{\theta}^c}{r^\ast}\frac{\partial}{\partial\theta}+
\cos\dot{\theta}^c\frac{\partial}{\partial r^\ast}.\label{CDD3}
\end{eqnarray}
\begin{eqnarray}
\frac{\partial}{\partial\widetilde{x}^\ast_1}&=&
-i\frac{\partial}{\partial\widetilde{x}_1}=
\frac{\sin\dot{\varphi}^c}{r^\ast\sin\dot{\theta}^c}\frac{\partial}
{\partial\epsilon}-\frac{\cos\dot{\varphi}^c\cos\dot{\theta}^c}{r^\ast}
\frac{\partial}{\partial\tau}-i\cos\dot{\varphi}^c\sin\dot{\theta}^c
\frac{\partial}{\partial r^\ast},\label{CDD4}\\
\frac{\partial}{\partial\widetilde{x}^\ast_2}&=&
-i\frac{\partial}{\partial\widetilde{x}_2}=
-\frac{\cos\dot{\varphi}^c}{r^\ast\sin\dot{\theta}^c}\frac{\partial}
{\partial\epsilon}-\frac{\sin\dot{\varphi}^c\cos\dot{\theta}^c}{r^\ast}
\frac{\partial}{\partial\tau}-i\sin\dot{\varphi}^c\sin\dot{\theta}^c
\frac{\partial}{\partial r^\ast},\label{CDD5}\\
\frac{\partial}{\partial\widetilde{x}^\ast_3}&=&
-i\frac{\partial}{\partial\widetilde{x}_3}=
\frac{\sin\dot{\theta}^c}{r^\ast}\frac{\partial}{\partial\tau}
-i\cos\dot{\theta}^c\frac{\partial}{\partial r^\ast}.\label{CDD6}
\end{eqnarray}
Substituting the functions 
$\boldsymbol{\psi}=\sT^{-1}_{\fg}\boldsymbol{\psi}^\prime$
($\dot{\boldsymbol{\psi}}=\overset{\ast}{\sT}_{\fg}\!\!{}^{-1}
\dot{\boldsymbol{\psi}}^\prime$) and the derivatives (\ref{CD1})--(\ref{CD6}),
(\ref{CDD1})--(\ref{CDD6}) into the system (\ref{Complex}), and
multiply by $\sT_{\fg}=\sT(\varphi^c,\theta^c,0)$
($\overset{\ast}{\sT}_{\fg}=
\overset{\ast}{\sT}(\dot{\varphi}^c,\dot{\theta}^c,0)$) from the left,
we obtain
\begin{multline}
\sT_{\fg}\sL_1\left[-\frac{\sin\varphi^c}{r\sin\theta^c}
\frac{\partial(\sT^{-1}_{\fg}\boldsymbol{\psi}^\prime)}
{\partial\varphi}+i\frac{\sin\varphi^c}{r\sin\theta^c}
\frac{\partial(\sT^{-1}_{\fg}\boldsymbol{\psi}^\prime)}
{\partial\epsilon}+\right.\\
\shoveright{\left.+\frac{\cos\varphi^c\cos\theta^c}{r}
\frac{\partial(\sT^{-1}_{\fg}\boldsymbol{\psi}^\prime)}
{\partial\theta}-
i\frac{\cos\varphi^c\cos\theta^c}{r}
\frac{\partial(\sT^{-1}_{\fg}\boldsymbol{\psi}^\prime)}
{\partial\tau}+2\cos\varphi^c\sin\theta^c
\frac{\partial(\sT^{-1}_{\fg}\boldsymbol{\psi}^\prime)}
{\partial r}\right]+}\nonumber\\
\sT_{\fg}\sL_2\left[\frac{\cos\varphi^c}{r\sin\theta^c}
\frac{\partial(\sT^{-1}_{\fg}\boldsymbol{\psi}^\prime)}
{\partial\varphi}-i\frac{\cos\varphi^c}{r\sin\theta^c}
\frac{\partial(\sT^{-1}_{\fg}\boldsymbol{\psi}^\prime)}
{\partial\epsilon}+\right.\\
\shoveleft{\left.+\frac{\sin\varphi^c\cos\theta^c}{r}
\frac{\partial(\sT^{-1}_{\fg}\boldsymbol{\psi}^\prime)}
{\partial\theta}-
i\frac{\sin\varphi^c\cos\theta^c}{r}
\frac{\partial(\sT^{-1}_{\fg}\boldsymbol{\psi}^\prime)}
{\partial\tau}+2\sin\varphi^c\sin\theta^c
\frac{\partial(\sT^{-1}_{\fg}\boldsymbol{\psi}^\prime)}
{\partial r}\right]+}\nonumber\\
+\sT_{\fg}\sL_3\left[-\frac{\sin\theta^c}{r}
\frac{\partial(\sT^{-1}_{\fg}\boldsymbol{\psi}^\prime)}
{\partial\theta}+i\frac{\sin\theta^c}{r}
\frac{\partial(\sT^{-1}_{\fg}\boldsymbol{\psi}^\prime)}
{\partial\tau}+2\cos\theta^c
\frac{\partial(\sT^{-1}_{\fg}\boldsymbol{\psi}^\prime)}
{\partial r}\right]+\kappa^c\boldsymbol{\psi}^\prime=0,\nonumber
\end{multline}
\begin{multline}
\overset{\ast}{\sT}_{\fg}\sL^\ast_1\left[
-\frac{\sin\dot{\varphi}^c}{r^\ast\sin\dot{\theta}^c}
\frac{\partial(\overset{\ast}{\sT}_{\fg}\!\!{}^{-1}
\dot{\boldsymbol{\psi}}^\prime)}{\partial\varphi}+
i\frac{\sin\dot{\varphi}^c}{r^\ast\sin\dot{\theta}^c}
\frac{\partial(\overset{\ast}{\sT}_{\fg}\!\!{}^{-1}
\dot{\boldsymbol{\psi}}^\prime)}{\partial\epsilon}+\right.\\
\shoveright{\left.+\frac{\cos\dot{\varphi}^c\cos\dot{\theta}^c}{r^\ast}
\frac{\partial(\overset{\ast}{\sT}_{\fg}\!\!{}^{-1}
\dot{\boldsymbol{\psi}}^\prime)}{\partial\theta}-
i\frac{\cos\dot{\varphi}^c\cos\dot{\theta}^c}{r^\ast}
\frac{\partial(\overset{\ast}{\sT}_{\fg}\!\!{}^{-1}
\dot{\boldsymbol{\psi}}^\prime)}{\partial\tau}+
2\cos\dot{\varphi}^c\sin\dot{\theta}^c
\frac{\partial(\overset{\ast}{\sT}_{\fg}\!\!{}^{-1}
\dot{\boldsymbol{\psi}}^\prime)}{\partial r^\ast}\right]+}\\
+\overset{\ast}{\sT}_{\fg}\sL^\ast_2\left[
\frac{\cos\dot{\varphi}^c}{r^\ast\sin\dot{\theta}^c}
\frac{\partial(\overset{\ast}{\sT}_{\fg}\!\!{}^{-1}
\dot{\boldsymbol{\psi}}^\prime)}{\partial\varphi}-
i\frac{\cos\dot{\varphi}^c}{r^\ast\sin\dot{\theta}^c}
\frac{\partial(\overset{\ast}{\sT}_{\fg}\!\!{}^{-1}
\dot{\boldsymbol{\psi}}^\prime)}{\partial\epsilon}+\right.\\
\shoveleft{\left.+\frac{\sin\dot{\varphi}^c\cos\dot{\theta}^c}{r^\ast}
\frac{\partial(\overset{\ast}{\sT}_{\fg}\!\!{}^{-1}
\dot{\boldsymbol{\psi}}^\prime)}{\partial\theta}-
i\frac{\sin\dot{\varphi}^c\cos\dot{\theta}^c}{r^\ast}
\frac{\partial(\overset{\ast}{\sT}_{\fg}\!\!{}^{-1}
\dot{\boldsymbol{\psi}}^\prime)}{\partial\tau}+
2\sin\dot{\varphi}^c\sin\dot{\theta}^c
\frac{\partial(\overset{\ast}{\sT}_{\fg}\!\!{}^{-1}
\dot{\boldsymbol{\psi}}^\prime)}{\partial r^\ast}\right]+}\\
+\overset{\ast}{\sT}_{\fg}\sL^\ast_3\left[
-\frac{\sin\dot{\theta}^c}{r^\ast}
\frac{\partial(\overset{\ast}{\sT}_{\fg}\!\!{}^{-1}
\dot{\boldsymbol{\psi}}^\prime)}{\partial\theta}+
i\frac{\sin\dot{\theta}^c}{r^\ast}
\frac{\partial(\overset{\ast}{\sT}_{\fg}\!\!{}^{-1}
\dot{\boldsymbol{\psi}}^\prime)}{\partial\tau}+2\cos\dot{\theta}^c
\frac{\partial(\overset{\ast}{\sT}_{\fg}\!\!{}^{-1}
\dot{\boldsymbol{\psi}}^\prime)}{\partial r^\ast}\right]+
\dot{\kappa}^c\dot{\boldsymbol{\psi}}^\prime=0.\label{Complex2}
\end{multline}
In virtue of the invariance conditions (\ref{IC}) we have
\begin{gather}
\sT_{\fg}\left[-\sL_1\sin\varphi^c+\sL_2\cos\varphi^c\right]
\sT^{-1}_{\fg}=\sL_1,\nonumber\\
\sT_{\fg}\left[\sL_1\cos\varphi^c\cos\theta^c+\sL_2\sin\varphi^c
\cos\theta^c-\sL_3\sin\theta^c\right]\sT^{-1}_{\fg}=\sL_2,\nonumber\\
\sT_{\fg}\left[2\sL_1\cos\varphi^c\sin\theta^c+
2\sL_2\sin\varphi^c\sin\theta^c+2\sL_3\cos\theta^c\right]\sT^{-1}_{\fg}=
2\sL_3,\nonumber\\
\sT_{\fg}\left[i\sL_1\sin\varphi^c-i\sL_2\cos\varphi^c\right]\sT^{-1}_{\fg}=
i\sL_1,\nonumber\\
\sT_{\fg}\left[-i\sL_1\cos\varphi^c\cos\theta^c-i\sL_2\sin\varphi^c
\cos\theta^c+i\sL_3\sin\theta^c\right]=i\sL_2,\nonumber\\
\overset{\ast}{\sT}_{\fg}\left[-\sL^\ast_1\sin\dot{\varphi}^c+
\sL^\ast_2\cos\dot{\varphi}^c\right]\overset{\ast}{\sT}_{\fg}\!\!{}^{-1}=
\sL^\ast_1,\nonumber\\
\overset{\ast}{\sT}_{\fg}\left[\sL^\ast_1\cos\dot{\varphi}^c
\cos\dot{\theta}^c+\sL^\ast_2\sin\dot{\varphi}^c\cos\dot{\theta}^c-
\sL^\ast_3\sin\dot{\theta}^c\right]\overset{\ast}{\sT}_{\fg}\!\!{}^{-1}=
\sL^\ast_2,\nonumber\\
\overset{\ast}{\sT}_{\fg}\left[2\sL^\ast_1\cos\dot{\varphi}^c
\sin\dot{\theta}^c+2\sL^\ast_2\sin\dot{\varphi}^c\sin\dot{\theta}^c+
2\sL^\ast_3\cos\dot{\theta}^c\right]\overset{\ast}{\sT}_{\fg}\!\!{}^{-1}=
\sL^\ast_3,\nonumber\\
\overset{\ast}{\sT}_{\fg}\left[i\sL^\ast_1\sin\dot{\varphi}^c-
i\sL^\ast_2\cos\dot{\varphi}^c\right]\overset{\ast}{\sT}_{\fg}\!\!{}^{-1}=
-i\sL^\ast_1,\nonumber\\
\overset{\ast}{\sT}_{\fg}\left[-i\sL^\ast_1\cos\dot{\varphi}^c
\cos\dot{\theta}^c-i\sL^\ast_2\sin\dot{\varphi}^c\cos\dot{\theta}^c+
i\sL^\ast_3\sin\dot{\theta}^c\right]\overset{\ast}{\sT}_{\fg}\!\!{}^{-1}=
-i\sL^\ast_2.\nonumber
\end{gather}
Taking into account the latter relations we can write the system
(\ref{Complex2}) as follows
\begin{multline}
\frac{1}{r\sin\theta^c}\sL_1\sT_{\fg}\frac{\partial(\sT^{-1}_{\fg}
\boldsymbol{\psi}^\prime)}{\partial\varphi}+
\frac{i}{r\sin\theta^c}\sL_1\sT_{\fg}\frac{\partial(\sT^{-1}_{\fg}
\boldsymbol{\psi}^\prime)}{\partial\epsilon}-\\
\shoveright{-\frac{1}{r}\sL_2\sT_{\fg}\frac{\partial(\sT^{-1}_{\fg}
\boldsymbol{\psi}^\prime)}{\partial\theta}-\frac{i}{r}\sL_2\sT_{\fg}
\frac{\partial(\sT^{-1}_{\fg}\boldsymbol{\psi}^\prime)}{\partial\tau}+
2\sL_3\sT_{\fg}
\frac{\partial(\sT^{-1}_{\fg}\boldsymbol{\psi}^\prime)}{\partial r}+
\kappa^c\boldsymbol{\psi}^\prime=0,}\\
\shoveleft{\frac{1}{r^\ast\sin\dot{\theta}^c}\sL^\ast_1
\overset{\ast}{\sT}_{\fg}
\frac{\partial(\overset{\ast}{\sT}_{\fg}\!\!{}^{-1}\dot{\boldsymbol{\psi}}^\prime)}
{\partial\varphi}-\frac{i}{r^\ast\sin\dot{\theta}^c}\sL^\ast_1
\overset{\ast}{\sT}_{\fg}
\frac{\partial(\overset{\ast}{\sT}_{\fg}\!\!{}^{-1}\dot{\boldsymbol{\psi}}^\prime)}
{\partial\epsilon}-}\\
%\shoveright{
-\frac{1}{r^\ast}\sL^\ast_2\overset{\ast}{\sT}_{\fg}
\frac{\partial(\overset{\ast}{\sT}_{\fg}\!\!{}^{-1}\dot{\boldsymbol{\psi}}^\prime)}
{\partial\theta}+\frac{i}{r^\ast}\sL^\ast_2\overset{\ast}{\sT}_{\fg}
\frac{\partial(\overset{\ast}{\sT}_{\fg}\!\!{}^{-1}\dot{\boldsymbol{\psi}}^\prime)}
{\partial\tau}+2\sL^\ast_3\overset{\ast}{\sT}_{\fg}
\frac{\partial(\overset{\ast}{\sT}_{\fg}\!\!{}^{-1}\dot{\boldsymbol{\psi}}^\prime)}
{\partial r^\ast}+\dot{\kappa}^c\dot{\boldsymbol{\psi}}^\prime=0.
%}
\label{Complex3}
\end{multline}
The matrices $\sT^{-1}_{\fg}$, $\overset{\ast}{\sT}_{\fg}\!\!{}^{-1}$ depend
on $\varphi$, $\epsilon$, $\theta$, $\tau$. By this reason we must
differentiate in $\sT^{-1}_{\fg}\boldsymbol{\psi}^\prime$
($\overset{\ast}{\sT}_{\fg}\!\!{}^{-1}\dot{\boldsymbol{\psi}}^\prime$) 
the both factors. After
differentiation, the system (\ref{Complex3}) takes a form
\begin{multline}
\frac{1}{r\sin\theta^c}\sL_1\frac{\partial\boldsymbol{\psi}^\prime}
{\partial\varphi}+\frac{i}{r\sin\theta^c}\sL_1
\frac{\partial\boldsymbol{\psi}^\prime}{\partial\epsilon}-
\frac{1}{r}\sL_2\frac{\partial\boldsymbol{\psi}^\prime}{\partial\theta}-\\
%\shoveright{
-\frac{i}{r}\sL_2\frac{\partial\boldsymbol{\psi}^\prime}
{\partial\tau}+2\sL_3\frac{\partial\boldsymbol{\psi}^\prime}{\partial r}
+\left[\frac{1}{r\sin\theta^c}\sL_1\sT_{\fg}
\frac{\partial\sT^{-1}_{\fg}}{\partial\varphi}+\right.
%}
\\
\shoveright{+\left.\frac{i}{r\sin\theta^c}\sL_1\sT_{\fg}
\frac{\partial\sT^{-1}_{\fg}}{\partial\epsilon}-\frac{1}{r}\sL_2\sT_{\fg}
\frac{\partial\sT^{-1}_{\fg}}{\partial\theta}-\frac{i}{r}\sL_2\sT_{\fg}
\frac{\partial\sT^{-1}_{\fg}}{\partial\tau}+\kappa^cI\right]
\boldsymbol{\psi}^\prime=0,}\\
\shoveleft{\frac{1}{r^\ast\sin\dot{\theta}^c}\sL^\ast_1
\frac{\partial\dot{\boldsymbol{\psi}}^\prime}{\partial\varphi}-
\frac{i}{r^\ast\sin\dot{\theta}^c}\sL^\ast_1
\frac{\partial\dot{\boldsymbol{\psi}}^\prime}{\partial\epsilon}-
\frac{1}{r^\ast}\sL^\ast_2
\frac{\partial\dot{\boldsymbol{\psi}}^\prime}{\partial\theta}+}\\
%\shoveright{
+\frac{i}{r^\ast}\sL^\ast_2
\frac{\partial\dot{\boldsymbol{\psi}}^\prime}{\partial\tau}+2\sL^\ast_3
\frac{\partial\dot{\boldsymbol{\psi}}^\prime}{\partial r^\ast}+
\left[\frac{1}{r^\ast\sin\dot{\theta}^c}\sL^\ast_1\overset{\ast}{\sT}_{\fg}
\frac{\partial\overset{\ast}{\sT}_{\fg}\!\!{}^{-1}}{\partial\varphi}-\right.
%}
\\
%\shoveright{
\left.-\frac{i}{r^\ast\sin\dot{\theta}^c}\sL^\ast_1
\overset{\ast}{\sT}_{\fg}
\frac{\partial\overset{\ast}{\sT}_{\fg}\!\!{}^{-1}}{\partial\epsilon}-
\frac{1}{r^\ast}\sL^\ast_2\overset{\ast}{\sT}_{\fg}
\frac{\partial\overset{\ast}{\sT}_{\fg}\!\!{}^{-1}}{\partial\theta}+
\frac{i}{r^\ast}\sL^\ast_2\overset{\ast}{\sT}_{\fg}
\frac{\partial\overset{\ast}{\sT}_{\fg}\!\!{}^{-1}}{\partial\tau}+
\dot{\kappa}^cI\right]\dot{\boldsymbol{\psi}}^\prime=0.
%}
\label{Complex4}
\end{multline}
Let us show that the products 
$\sT_{\fg}\frac{\partial\sT^{-1}_{\fg}}{\partial\varphi}$, $\ldots$,
$\overset{\ast}{\sT}_{\fg}\frac{\partial\overset{\ast}{\sT}_{\fg}\!\!{}^{-1}}
{\partial\tau}$ are expressed via linear combinations of the operators
$\sA_i$, $\sB_i$, $\widetilde{\sA}_i$, $\widetilde{\sB}_i$. Indeed, a matrix
of the fundamental representation 
$\fg(\varphi,\epsilon,\theta,\tau,0,0)=
\sT_{\fg}(\varphi^c,\theta^c,0)$ of the group $\fG_+$ has a form
(see (\ref{SL1}))
\begin{gather}
\fg^T(\varphi,\epsilon,\theta,\tau,0,0)=
{\renewcommand{\arraystretch}{1.7}
\begin{pmatrix}
\cos\dfrac{\theta^c}{2}e^{\frac{i\varphi^c}{2}} &
i\sin\dfrac{\theta^c}{2}e^{-\frac{i\varphi^c}{2}}\\
i\sin\dfrac{\theta^c}{2}e^{\frac{i\varphi^c}{2}} &
\cos\dfrac{\theta^c}{2}e^{-\frac{i\varphi^c}{2}}
\end{pmatrix}=}\nonumber\\
{\renewcommand{\arraystretch}{1.9}
\begin{pmatrix}
e^{\frac{\epsilon+i\varphi}{2}}\left[\cos\dfrac{\theta}{2}\ch\dfrac{\tau}{2}+
i\sin\dfrac{\theta}{2}\sh\dfrac{\tau}{2}\right] &
e^{\frac{-\epsilon-i\varphi}{2}}\left[\cos\dfrac{\theta}{2}\sh\dfrac{\tau}{2}+
i\sin\dfrac{\theta}{2}\ch\dfrac{\tau}{2}\right] \\
e^{\frac{\epsilon+i\varphi}{2}}\left[\cos\dfrac{\theta}{2}\sh\dfrac{\tau}{2}+
i\sin\dfrac{\theta}{2}\ch\dfrac{\tau}{2}\right] &
e^{\frac{-\epsilon-i\varphi}{2}}\left[\cos\dfrac{\theta}{2}\ch\dfrac{\tau}{2}+
i\sin\dfrac{\theta}{2}\sh\dfrac{\tau}{2}\right]
\end{pmatrix}}\nonumber
\end{gather}
and its inverse matrix is
\begin{gather}
\left(\fg^T\right)^{-1}=\nonumber\\
{\renewcommand{\arraystretch}{1.9}
\begin{pmatrix}
e^{\frac{-\epsilon-i\varphi}{2}}\left[\cos\dfrac{\theta}{2}\ch\dfrac{\tau}{2}+
i\sin\dfrac{\theta}{2}\sh\dfrac{\tau}{2}\right] &
-e^{\frac{-\epsilon-i\varphi}{2}}\left[\cos\dfrac{\theta}{2}\sh\dfrac{\tau}{2}+
i\sin\dfrac{\theta}{2}\ch\dfrac{\tau}{2}\right] \\
-e^{\frac{\epsilon+i\varphi}{2}}\left[\cos\dfrac{\theta}{2}\sh\dfrac{\tau}{2}+
i\sin\dfrac{\theta}{2}\ch\dfrac{\tau}{2}\right] &
e^{\frac{\epsilon+i\varphi}{2}}\left[\cos\dfrac{\theta}{2}\ch\dfrac{\tau}{2}+
i\sin\dfrac{\theta}{2}\sh\dfrac{\tau}{2}\right]
\end{pmatrix}.}\nonumber
\end{gather}
In accordance with (\ref{OpA}), (\ref{OpB}) and (\ref{A1})--(\ref{A3})
infinitesimal operators for the fundamental representation $\fg$ are
\begin{gather}
\sA_1=-\frac{i}{2}\begin{bmatrix}
0 & 1\\
1 & 0
\end{bmatrix},\quad
\sA_2=\frac{1}{2}\begin{bmatrix}
0 & 1\\
-1 & 0
\end{bmatrix},\quad
\sA_3=\frac{1}{2}\begin{bmatrix}
i & 0\\
0 & -i
\end{bmatrix},\nonumber\\
\sB_1=\frac{1}{2}\begin{bmatrix}
0 & 1\\
1 & 0
\end{bmatrix},\quad
\sB_2=\frac{1}{2}\begin{bmatrix}
0 & i\\
-i & 0
\end{bmatrix},\quad
\sB_3=\frac{1}{2}\begin{bmatrix}
-1 & 0\\
0 & 1
\end{bmatrix}.\label{IF}
\end{gather}
Then
\begin{eqnarray}
\fg^T\frac{\partial\left(\fg^T\right)^{-1}}{\partial\varphi}&=&
\frac{i}{2}\begin{pmatrix}
-\cos\theta^c & i\sin\theta^c\\
-i\sin\theta^c & \cos\theta^c
\end{pmatrix}=-\sA_3\cos\theta^c-\sA_2\sin\theta^c,\label{PR1}\\
\fg^T\frac{\partial\left(\fg^T\right)^{-1}}{\partial\epsilon}&=&
\frac{1}{2}\begin{pmatrix}
-\cos\theta^c & i\sin\theta^c\\
-i\sin\theta^c & \cos\theta^c
\end{pmatrix}=\sB_3\cos\theta^c+\sB_2\sin\theta^c,\label{PR2}\\
\fg^T\frac{\partial\left(\fg^T\right)^{-1}}{\partial\theta}&=&
\frac{1}{2}\begin{pmatrix}
0 & -i\\
-i & 0
\end{pmatrix}=\sA_1,\label{PR3}\\
\fg^T\frac{\partial\left(\fg^T\right)^{-1}}{\partial\tau}&=&
\frac{1}{2}\begin{pmatrix}
0 & 1\\
1 & 0
\end{pmatrix}=\sB_1.\label{PR4}
\end{eqnarray}
\begin{sloppypar}\noindent
Further, a matrix of the conjugate fundamental representation
$\dot{\fg}(\varphi,\epsilon,\theta,\tau,0,0)=
\overset{\ast}{\sT}_{\fg}(\dot{\varphi}^c,\dot{\theta}^c,0)$ of the
group $\fG_+$ has a form\end{sloppypar}
\begin{gather}
{\renewcommand{\arraystretch}{1.7}
\dot{\fg}^T(\varphi,\epsilon,\theta,\tau,0,0)=
\begin{pmatrix}
\cos\dfrac{\dot{\theta}^c}{2}e^{-\frac{i\dot{\varphi}^c}{2}} &
-i\sin\dfrac{\dot{\theta}^c}{2}e^{\frac{i\dot{\varphi}^c}{2}} \\
-i\sin\dfrac{\dot{\theta}^c}{2}e^{-\frac{i\dot{\varphi}^c}{2}} &
\cos\dfrac{\dot{\theta}^c}{2}e^{\frac{i\dot{\varphi}^c}{2}}
\end{pmatrix}}=\nonumber\\
{\renewcommand{\arraystretch}{1.9}
\begin{pmatrix}
e^{\frac{\epsilon-i\varphi}{2}}\left[\cos\dfrac{\theta}{2}\ch\dfrac{\tau}{2}-
i\sin\dfrac{\theta}{2}\sh\dfrac{\tau}{2}\right] &
e^{\frac{-\epsilon+i\varphi}{2}}\left[\cos\dfrac{\theta}{2}\sh\dfrac{\tau}{2}-
i\sin\dfrac{\theta}{2}\ch\dfrac{\tau}{2}\right] \\
e^{\frac{\epsilon-i\varphi}{2}}\left[\cos\dfrac{\theta}{2}\sh\dfrac{\tau}{2}-
i\sin\dfrac{\theta}{2}\ch\dfrac{\tau}{2}\right] &
e^{\frac{-\epsilon+i\varphi}{2}}\left[\cos\dfrac{\theta}{2}\ch\dfrac{\tau}{2}-
i\sin\dfrac{\theta}{2}\sh\dfrac{\tau}{2}\right]
\end{pmatrix}}\nonumber
\end{gather}
and its inverse matrix is
\begin{gather}
\left(\dot{\fg}^T\right)^{-1}=\nonumber\\
{\renewcommand{\arraystretch}{1.9}
\begin{pmatrix}
e^{\frac{-\epsilon+i\varphi}{2}}\left[\cos\dfrac{\theta}{2}\ch\dfrac{\tau}{2}-
i\sin\dfrac{\theta}{2}\sh\dfrac{\tau}{2}\right] &
-e^{\frac{-\epsilon+i\varphi}{2}}\left[\cos\dfrac{\theta}{2}\sh\dfrac{\tau}{2}-
i\sin\dfrac{\theta}{2}\ch\dfrac{\tau}{2}\right] \\
-e^{\frac{\epsilon-i\varphi}{2}}\left[\cos\dfrac{\theta}{2}\sh\dfrac{\tau}{2}-
i\sin\dfrac{\theta}{2}\ch\dfrac{\tau}{2}\right] &
e^{\frac{\epsilon-i\varphi}{2}}\left[\cos\dfrac{\theta}{2}\ch\dfrac{\tau}{2}-
i\sin\dfrac{\theta}{2}\sh\dfrac{\tau}{2}\right]
\end{pmatrix}.}\nonumber
\end{gather}
In accordance with (\ref{DopA}) and (\ref{DopB}) infinitesimal operators
for the conjugate fundamental representation $\dot{\fg}$ are
\begin{gather}
\widetilde{\sA}_1=\frac{i}{2}\begin{bmatrix}
0 & 1\\
1 & 0
\end{bmatrix},\quad
\widetilde{\sA}_2=\frac{1}{2}\begin{bmatrix}
0 & -1\\
1 & 0
\end{bmatrix},\quad
\widetilde{\sA}_3=\frac{1}{2}\begin{bmatrix}
-i & 0\\
0 & i
\end{bmatrix},\nonumber\\
\widetilde{\sB}_1=-\frac{1}{2}\begin{bmatrix}
0 & 1\\
1 & 0
\end{bmatrix},\quad
\widetilde{\sB}_2=\frac{1}{2}\begin{bmatrix}
0 & i\\
-i & 0
\end{bmatrix},\quad
\widetilde{\sB}_3=\frac{1}{2}\begin{bmatrix}
1 & 0\\
0 & -1
\end{bmatrix}.\label{DIF}
\end{gather}
In this case we have
\begin{eqnarray}
\dot{\fg}^T\frac{\partial\left(\dot{\fg}^T\right)^{-1}}{\partial\varphi}&=&
\frac{i}{2}\begin{pmatrix}
\cos\dot{\theta}^c & i\sin\dot{\theta}^c \\
-i\sin\dot{\theta}^c & -\cos\dot{\theta}^c
\end{pmatrix}=-\widetilde{\sA}_3\cos\dot{\theta}^c+
\widetilde{\sA}_2\sin\dot{\theta}^c,\label{PR5}\\
\dot{\fg}^T\frac{\partial\left(\dot{\fg}^T\right)^{-1}}{\partial\epsilon}&=&
\frac{1}{2}\begin{pmatrix}
-\cos\dot{\theta}^c & i\sin\dot{\theta}^c \\
-i\sin\dot{\theta}^c & -\cos\dot{\theta}^c
\end{pmatrix}=-\widetilde{\sB}_3\cos\dot{\theta}^c-
\widetilde{\sB}_2\sin\dot{\theta}^c,\label{PR6}\\
\dot{\fg}^T\frac{\partial\left(\dot{\fg}^T\right)^{-1}}{\partial\theta}&=&
\frac{1}{2}\begin{pmatrix}
0 & i\\
i & 0
\end{pmatrix}=\widetilde{\sA}_1,\label{PR7}\\
\dot{\fg}^T\frac{\partial\left(\dot{\fg}^T\right)^{-1}}{\partial\tau}&=&
\frac{1}{2}\begin{pmatrix}
0 & -1\\
-1 & 0
\end{pmatrix}=\widetilde{\sB}_1.\label{PR8}
\end{eqnarray}
It is easy to verify that relations (\ref{PR1})--(\ref{PR8}) take place
for any representation $\fg\rightarrow\sT_{\fg}$ of the group $\fG_+$.
Substituting these relations into the system (\ref{Complex4}) we obtain
\begin{multline}
\frac{1}{r\sin\theta^c}\sL_1\frac{\partial\boldsymbol{\psi}^\prime}
{\partial\varphi}+
\frac{i}{r\sin\theta^c}\sL_1\frac{\partial\boldsymbol{\psi}^\prime}
{\partial\epsilon}-
\frac{1}{r}\sL_2\frac{\partial\boldsymbol{\psi}^\prime}{\partial\theta}-
\frac{i}{r}\sL_2\frac{\partial\boldsymbol{\psi}^\prime}{\partial\tau}+\\
%\shoveright{
+2\sL_3\frac{\partial\boldsymbol{\psi}^\prime}{\partial r}+
\frac{1}{r}\left[-(\sL_1\sA_2+\sL_2\sA_1)+i(\sL_1\sB_2-\sL_2\sB_1)+\right.
%}
\\
\shoveright{\left.+\ctg\theta^c(i\sL_1\sB_3-\sL_1\sA_3)\right]
\boldsymbol{\psi}^\prime+
\kappa^c\boldsymbol{\psi}=0,}\\
\shoveleft{
\frac{1}{r^\ast\sin\dot{\theta}^c}\sL^\ast_1
\frac{\partial\dot{\boldsymbol{\psi}}^\prime}
{\partial\varphi}-\frac{i}{r^\ast\sin\dot{\theta}^c}\sL^\ast_1
\frac{\partial\dot{\boldsymbol{\psi}}^\prime}{\partial\epsilon}-
\frac{1}{r^\ast}\sL^\ast_2
\frac{\partial\dot{\boldsymbol{\psi}}^\prime}{\partial\theta}+
\frac{i}{r^\ast}\sL^\ast_2
\frac{\partial\dot{\boldsymbol{\psi}}^\prime}{\partial\tau}+}\\
%\shoveright{
2\sL^\ast_3
\frac{\partial\dot{\boldsymbol{\psi}}^\prime}{\partial r^\ast}+
\frac{1}{r^\ast}\left[\sL^\ast_1\widetilde{\sA}_2-
\sL^\ast_2\widetilde{\sA}_1+
i(\sL^\ast_1\widetilde{\sB}_2+\sL^\ast_2\widetilde{\sB}_1)+\right.
%}
\\
%\shoveright{
\left.+
\ctg\dot{\theta}^c(i\sL^\ast_1\widetilde{\sB}_3-
\sL^\ast_1\widetilde{\sA}_3)\right]
\dot{\boldsymbol{\psi}}^\prime+
\dot{\kappa}^c\dot{\boldsymbol{\psi}}=0.
%}
\label{Complex5}
\end{multline}
Now we are in a position that allows to separate variables in 
relativistically invariant system. We decompose the each component
$\psi^k_{lm}$ of the wave function $\boldsymbol{\psi}$ into the series
on the generalized hyperspherical functions. This procedure gives rise
to separation of variables, that is, it reduces the relativistically
invariant system to a system of ordinary differential equations.
Preliminarily, let us calculate elements of the matrices
$V=\sL_1\sB_2-\sL_2\sB_1$, $U=\sL^\ast_1\widetilde{\sA}_2-
\sL^\ast_2\widetilde{\sA}_1$, $G=\sL_1\sA_2+\sL_2\sA_1$,
$W=\sL^\ast_1\widetilde{\sB}_2+\sL^\ast_2\widetilde{\sB}_1$.

First of all, let us find elements of the matrix
$V=\sL_1\sB_2-\sL_2\sB_1$. Using the relations (\ref{BL}) we can write
$V=2i\sL_3+\sB_2\sL_1-\sB_1\sL_2$. Let
\[
V\xi^k_{l,m}=\sum_{l^\prime,m^\prime,k^\prime}
v^{k^\prime k}_{l^\prime l,m^\prime m}\xi^{k^\prime}_{l^\prime,m^\prime}
\]
be an expression of the matrix $V$ for the canonical basis
$\left\{\xi^k_{lm}\right\}$. Taking into account (\ref{L1})--(\ref{L3})
and (\ref{OpB}) we obtain
\begin{multline}
V\xi^k_{l,m}=(2i\sL_3+\sB_2\sL_1-\sB_1\sL_2)\xi^k_{l,m}=\\
=2i\sum_{l^\prime,m^\prime,k^\prime}
c^{k^\prime k}_{l^\prime l,m^\prime m}
\xi^{k^\prime}_{l^\prime,m^\prime}+
\sB_2\sum_{l^\prime,m^\prime,k^\prime}
a^{k^\prime k}_{l^\prime l,m^\prime m}
\xi^{k^\prime}_{l^\prime,m^\prime}-
\sB_1\sum_{l^\prime,m^\prime,k^\prime}
b^{k^\prime k}_{l^\prime l,m^\prime m}
\xi^{k^\prime}_{l^\prime,m^\prime}=\\ 
2i\sum_{l^\prime,m^\prime,k^\prime}
c^{k^\prime k}_{l^\prime l,m^\prime m}
\xi^{k^\prime}_{l^\prime,m^\prime}+
\frac{i}{2}\sum_{l^\prime,m^\prime,k^\prime}
a^{k^\prime k}_{l^\prime l,m^\prime m}
\left(\boldsymbol{\alpha}^{l^\prime}_{m^\prime}
\xi^{k^\prime}_{l^\prime,m^\prime-1}-
\boldsymbol{\alpha}^{l^\prime}_{m^\prime+1}
\xi^{k^\prime}_{l^\prime,m^\prime+1}\right)-\\
-\frac{1}{2}\sum_{l^\prime,m^\prime,k^\prime}
b^{k^\prime k}_{l^\prime l,m^\prime m}
\left(\boldsymbol{\alpha}^{l^\prime}_{m^\prime}
\xi^{k^\prime}_{l^\prime,m^\prime-1}+
\boldsymbol{\alpha}^{l^\prime}_{m^\prime+1}
\xi^{k^\prime}_{l^\prime,m^\prime+1}\right).\nonumber
\end{multline}
Dividing the each of the two latter sums on the two and changing the
summation index in the each four obtained sums, we come to the following
expression
\begin{multline}
V\xi^k_{l,m}=\sum_{l^\prime,m^\prime,k^\prime}\left(
2ic^{k^\prime k}_{l^\prime l,m^\prime m}-
\frac{i}{2}\boldsymbol{\alpha}^{l^\prime}_{m^\prime}
a^{k^\prime k}_{l^\prime l,m^\prime-1,m}+\right.\\
\left.+\frac{i}{2}\boldsymbol{\alpha}^{l^\prime}_{m^\prime+1}
a^{k^\prime k}_{l^\prime l,m^\prime+1,m}-
\frac{1}{2}\boldsymbol{\alpha}^{l^\prime}_{m^\prime}
b^{k^\prime k}_{l^\prime l,m^\prime-1,m}-
\frac{1}{2}\boldsymbol{\alpha}^{l^\prime}_{m^\prime+1}
b^{k^\prime k}_{l^\prime l,m^\prime+1,m}
\right)\xi^{k^\prime}_{l^\prime,m^\prime}.\nonumber
\end{multline}
Therefore, a general element of the matrix $V$ have a form
\begin{multline}
v^{k^\prime k}_{l^\prime l,m^\prime m}=
2ic^{k^\prime k}_{l^\prime l,m^\prime m}-
\frac{i}{2}\boldsymbol{\alpha}^{l^\prime}_{m^\prime}
a^{k^\prime k}_{l^\prime l,m^\prime-1,m}+\\
+\frac{i}{2}\boldsymbol{\alpha}^{l^\prime}_{m^\prime+1}
a^{k^\prime k}_{l^\prime l,m^\prime+1,m}-
\frac{1}{2}\boldsymbol{\alpha}^{l^\prime}_{m^\prime}
b^{k^\prime k}_{l^\prime l,m^\prime-1,m}-
\frac{1}{2}\boldsymbol{\alpha}^{l^\prime}_{m^\prime+1}
b^{k^\prime k}_{l^\prime l,m^\prime+1,m}.\nonumber
\end{multline}
Using the formulae (\ref{L3}), (\ref{L1}) and (\ref{L2}) we find that
\begin{equation}\label{VM}
V:\quad
{\renewcommand{\arraystretch}{1.3}
\left\{\begin{array}{ccc}
v^{k^\prime k}_{l-1,l,m\,m}&=&
ic^{k^\prime k}_{l-1,l}(l+1)\sqrt{l^2-m^2},\\
v^{k^\prime k}_{l\,l,m\,m}&=&
ic^{k^\prime k}_{l\,l}m,\\
v^{k^\prime k}_{l+1,l,m\,m}&=&
-ic^{k^\prime k}_{l+1,l}l\sqrt{(l+1)^2-m^2}.
\end{array}\right.}
\end{equation}
All other elements $v^{k^\prime k}_{l^\prime l,m^\prime m}$ are
equal to zero.

Analogously, using the relations (\ref{DAL}) and the operators
(\ref{DopA}) we find that elements of the matrix
$U=\sL^\ast_1\widetilde{\sA}_2-\sL^\ast_2\widetilde{\sA}_1$ are
\begin{equation}\label{UM}
U:\quad
{\renewcommand{\arraystretch}{1.3}
\left\{\begin{array}{ccc}
u^{\dot{k}^\prime\dot{k}}_{\dot{l}-1,\dot{l},\dot{m}\dot{m}}&=&
-c^{\dot{k}^\prime\dot{k}}_{\dot{l}-1,\dot{l}}(\dot{l}+1)
\sqrt{\dot{l}^2-\dot{m}^2},\\
u^{\dot{k}^\prime\dot{k}}_{\dot{l}\dot{l},\dot{m}\dot{m}}&=&
-c^{\dot{k}^\prime\dot{k}}_{\dot{l}\dot{l}}\dot{m},\\
u^{\dot{k}^\prime\dot{k}}_{\dot{l}+1,\dot{l},\dot{m}\dot{m}}&=&
c^{\dot{k}^\prime\dot{k}}_{\dot{l}+1,\dot{l}}\dot{l}
\sqrt{(\dot{l}+1)^2-\dot{m}^2}.
\end{array}\right.}
\end{equation}
Further, using (\ref{AL}), (\ref{OpA}) and (\ref{DBL}), (\ref{DopB}) it is
easy to verify that all the elements of the matrices
$G=\sL_1\sA_2+\sL_2\sA_1$ and
$W=\sL^\ast_1\widetilde{\sB}_2+\sL^\ast_2\widetilde{\sB}_1$
are equal to zero.

The system (\ref{Complex5}) in the components $\psi^k_{lm}$ is written
as follows
\begin{multline}
\frac{1}{r\sin\theta^c}\sum_{l^\prime,m^\prime,k^\prime}
a^{kk^\prime}_{ll^\prime,mm^\prime}
\frac{\partial\psi^{k^\prime}_{l^\prime m^\prime}}{\partial\varphi}+
\frac{i}{r\sin\theta^c}\sum_{l^\prime,m^\prime,k^\prime}
a^{kk^\prime}_{ll^\prime,mm^\prime}
\frac{\partial\psi^{k^\prime}_{l^\prime m^\prime}}{\partial\epsilon}-\\
-\frac{1}{r}\sum_{l^\prime,m^\prime,k^\prime}
b^{kk^\prime}_{ll^\prime,mm^\prime}
\frac{\partial\psi^{k^\prime}_{l^\prime m^\prime}}{\partial\theta}-
\frac{i}{r}\sum_{l^\prime,m^\prime,k^\prime}
b^{kk^\prime}_{ll^\prime,mm^\prime}
\frac{\partial\psi^{k^\prime}_{l^\prime m^\prime}}{\partial\tau}+
2\sum_{l^\prime,m^\prime,k^\prime}
c^{kk^\prime}_{ll^\prime,mm^\prime}
\frac{\partial\psi^{k^\prime}_{l^\prime m^\prime}}{\partial r}+\\
\shoveright{+\frac{i}{r}\sum_{l^\prime,m^\prime,k^\prime}
v^{kk^\prime}_{ll^\prime,mm^\prime}
\psi^{k^\prime}_{l^\prime m^\prime}+
\frac{2i}{r}\ctg\theta^c\sum_{l^\prime,m^\prime,k^\prime}m^\prime
a^{kk^\prime}_{ll^\prime,mm^\prime}\psi^{k^\prime}_{l^\prime m^\prime}+
\kappa^c\psi^k_{lm}=0,}\\
\shoveleft{
\frac{1}{r^\ast\sin\dot{\theta}^c}
\sum_{\dot{l}^\prime,\dot{m}^\prime,\dot{k}^\prime}
d^{\dot{k}\dot{k}^\prime}_{\dot{l}\dot{l}^\prime,\dot{m}\dot{m}^\prime}
\frac{\partial\psi^{\dot{k}^\prime}_{\dot{l}^\prime\dot{m}^\prime}}
{\partial\varphi}-
\frac{i}{r^\ast\sin\dot{\theta}^c}
\sum_{\dot{l}^\prime,\dot{m}^\prime,\dot{k}^\prime}
d^{\dot{k}\dot{k}^\prime}_{\dot{l}\dot{l}^\prime,\dot{m}\dot{m}^\prime}
\frac{\partial\psi^{\dot{k}^\prime}_{\dot{l}^\prime\dot{m}^\prime}}
{\partial\epsilon}-
}
\\
-\frac{1}{r^\ast}
\sum_{\dot{l}^\prime,\dot{m}^\prime,\dot{k}^\prime}
e^{\dot{k}\dot{k}^\prime}_{\dot{l}\dot{l}^\prime,\dot{m}\dot{m}^\prime}
\frac{\partial\psi^{\dot{k}^\prime}_{\dot{l}^\prime\dot{m}^\prime}}
{\partial\theta}+
\frac{i}{r^\ast}
\sum_{\dot{l}^\prime,\dot{m}^\prime,\dot{k}^\prime}
e^{\dot{k}\dot{k}^\prime}_{\dot{l}\dot{l}^\prime,\dot{m}\dot{m}^\prime}
\frac{\partial\psi^{\dot{k}^\prime}_{\dot{l}^\prime\dot{m}^\prime}}
{\partial\tau}+
2\sum_{\dot{l}^\prime,\dot{m}^\prime,\dot{k}^\prime}
f^{\dot{k}\dot{k}^\prime}_{\dot{l}\dot{l}^\prime,\dot{m}\dot{m}^\prime}
\frac{\partial\psi^{\dot{k}^\prime}_{\dot{l}^\prime\dot{m}^\prime}}
{\partial r^\ast}+\\
%\shoveright{
+\frac{1}{r^\ast}
\sum_{\dot{l}^\prime,\dot{m}^\prime,\dot{k}^\prime}
u^{\dot{k}\dot{k}^\prime}_{\dot{l}\dot{l}^\prime,\dot{m}\dot{m}^\prime}
\psi^{\dot{k}^\prime}_{\dot{l}^\prime\dot{m}^\prime}-
\frac{2i}{r^\ast}\ctg\dot{\theta}^c
\sum_{\dot{l}^\prime,\dot{m}^\prime,\dot{k}^\prime}\dot{m}^\prime
d^{\dot{k}\dot{k}^\prime}_{\dot{l}\dot{l}^\prime,\dot{m}\dot{m}^\prime}
\psi^{\dot{k}^\prime}_{\dot{l}^\prime\dot{m}^\prime}
+\dot{\kappa}^c\psi^{\dot{k}}_{\dot{l}\dot{m}}=0,
%}
\label{Complex6}
\end{multline}
where the coefficients 
$a^{k k^\prime}_{l l^\prime,m m^\prime}$,
$b^{k k^\prime}_{l l^\prime, m m^\prime}$,
$c^{k k^\prime}_{l l^\prime, m m^\prime}$,
$v^{k k^\prime}_{l l^\prime, m m^\prime}$,
$d^{\dot{k}\dot{k}^\prime}_{\dot{l}\dot{l}^\prime,\dot{m}\dot{m}^\prime}$,
$e^{\dot{k}\dot{k}^\prime}_{\dot{l}\dot{l}^\prime,\dot{m}\dot{m}^\prime}$,
$f^{\dot{k}\dot{k}^\prime}_{\dot{l}\dot{l}^\prime,\dot{m}\dot{m}^\prime}$,
$u^{\dot{k}\dot{k}^\prime}_{\dot{l}\dot{l}^\prime,\dot{m}\dot{m}^\prime}$
are defined by the formulae (\ref{L1}), (\ref{L2}), (\ref{L3}), (\ref{VM}),
(\ref{L1'}), (\ref{L2'}), (\ref{L3'}), (\ref{UM}).

With a view to separate the variables in (\ref{Complex6}) let us assume
\begin{equation}\label{F}
\psi^k_{lm}=\boldsymbol{f}^{l_0}_{lmk}(r)
\fM^{l_0}_{mn}(\varphi,\epsilon,\theta,\tau,0,0),
\end{equation}
where $l_0\ge l$, and $-l_0\le m$, $n\le l_0$. Substituting the functions
(\ref{F}) into the system (\ref{Complex6}) and taking into account values
of the coefficients
$a^{k k^\prime}_{l l^\prime,m m^\prime}$,
$b^{k k^\prime}_{l l^\prime, m m^\prime}$,
$c^{k k^\prime}_{l l^\prime, m m^\prime}$,
$v^{k k^\prime}_{l l^\prime, m m^\prime}$,
$d^{\dot{k}\dot{k}^\prime}_{\dot{l}\dot{l}^\prime,\dot{m}\dot{m}^\prime}$,
$e^{\dot{k}\dot{k}^\prime}_{\dot{l}\dot{l}^\prime,\dot{m}\dot{m}^\prime}$,
$f^{\dot{k}\dot{k}^\prime}_{\dot{l}\dot{l}^\prime,\dot{m}\dot{m}^\prime}$,
$u^{\dot{k}\dot{k}^\prime}_{\dot{l}\dot{l}^\prime,\dot{m}\dot{m}^\prime}$
let us collect together the terms with identical radial functions.
In the result we obtain

\begin{multline}
\sum_{k^\prime}c^{kk^\prime}_{l,l-1}\left\{
\left[
2\sqrt{l^2-m^2}\frac{\partial\boldsymbol{f}^{l_0}_{l-1,m,k^\prime}}
{\partial r}-\frac{1}{r}(l+1)\sqrt{l^2-m^2}
\boldsymbol{f}^{l_0}_{l-1,m,k^\prime}\right]
\fM^{l_0}_{mn}(\varphi,\epsilon,\theta,\tau,0,0)+\right.\\
+\frac{1}{2r}\sqrt{(l+m)(l+m-1)}\boldsymbol{f}^{l_0}_{l-1,m-1,k^\prime}
\left[-\frac{1}{\sin\theta^c}\frac{\partial\fM^{l_0}_{m-1,n}}{\partial\varphi}-
\frac{i}{\sin\theta^c}\frac{\partial\fM^{l_0}_{m-1,n}}{\partial\epsilon}
\right.+\\
\shoveright{
\left.+i\frac{\partial\fM^{l_0}_{m-1,n}}{\partial\theta}-
\frac{\partial\fM^{l_0}_{m-1,n}}{\partial\tau}-
\frac{2i(m-1)\cos\theta^c}{\sin\theta^c}\fM^{l_0}_{m-1,n}\right]+}\\
\shoveleft{
+\frac{1}{2r}\sqrt{(l-m)(l-m-1)}\boldsymbol{f}^{l_0}_{l-1,m+1,k^\prime}
\left[\frac{1}{\sin\theta^c}\frac{\partial\fM^{l_0}_{m+1,n}}{\partial\varphi}+
\frac{i}{\sin\theta^c}\frac{\partial\fM^{l_0}_{m+1,n}}{\partial\epsilon}+
\right.}\\
%\shoveright{
\left.
\left.+i\frac{\partial\fM^{l_0}_{m+1,n}}{\partial\theta}-
\frac{\partial\fM^{l_0}_{m+1,n}}{\partial\tau}+
\frac{2i(m+1)\cos\theta^c}{\sin\theta^c}\fM^{l_0}_{m+1,n}\right]\right\}+
%}
\\
\shoveleft{
\sum_{k^\prime}c^{kk^\prime}_{ll}\left\{\left[
2m\frac{\partial\boldsymbol{f}^{l_0}_{l,m,k^\prime}}{\partial r}
-\frac{1}{r}m\boldsymbol{f}^{l_0}_{l,m,k^\prime}\right]
\fM^{l_0}_{m,n}(\varphi,\epsilon,\theta,\tau,0,0)+\right.
}
\\
+\frac{1}{2r}\sqrt{(l+m)(l-m+1)}\boldsymbol{f}^{l_0}_{l,m-1,k^\prime}
\left[\frac{1}{\sin\theta^c}\frac{\partial\fM^{l_0}_{m-1,n}}{\partial\varphi}+
\frac{i}{\sin\theta^c}\frac{\partial\fM^{l_0}_{m-1,n}}{\partial\epsilon}
\right.-\\
\shoveright{
\left.-i\frac{\partial\fM^{l_0}_{m-1,n}}{\partial\theta}+
\frac{\partial\fM^{l_0}_{m-1,n}}{\partial\tau}+
\frac{2i(m-1)\cos\theta^c}{\sin\theta^c}\fM^{l_0}_{m-1,n}\right]+}\\
\shoveleft{
+\frac{1}{2r}\sqrt{(l+m+1)(l-m)}\boldsymbol{f}^{l_0}_{l,m+1,k^\prime}
\left[\frac{1}{\sin\theta^c}\frac{\partial\fM^{l_0}_{m+1,n}}{\partial\varphi}+
\frac{i}{\sin\theta^c}\frac{\partial\fM^{l_0}_{m+1,n}}{\partial\epsilon}+
\right.}\\
%\shoveright{
\left.
\left.+i\frac{\partial\fM^{l_0}_{m+1,n}}{\partial\theta}-
\frac{\partial\fM^{l_0}_{m+1,n}}{\partial\tau}+
\frac{2i(m+1)\cos\theta^c}{\sin\theta^c}\fM^{l_0}_{m+1,n}\right]\right\}+
%}
\\
\sum_{k^\prime}c^{kk^\prime}_{l,l+1}\left\{\left[
2\sqrt{(l+1)^2-m^2}\frac{\partial\boldsymbol{f}^{l_0}_{l+1,m,k^\prime}}
{\partial r}+\frac{1}{r}l\sqrt{(l+1)^2-m^2}
\boldsymbol{f}^{l_0}_{l+1,m,k^\prime}\right]
\fM^{l_0}_{m,n}(\varphi,\epsilon,\theta,\tau,0,0)+\right.\\
+\frac{1}{2r}\sqrt{(l-m+1)(l-m+2)}\boldsymbol{f}^{l_0}_{l+1,m-1,k^\prime}
\left[\frac{1}{\sin\theta^c}\frac{\partial\fM^{l_0}_{m-1,n}}{\partial\varphi}+
\frac{i}{\sin\theta^c}\frac{\partial\fM^{l_0}_{m-1,n}}{\partial\epsilon}
\right.-\\
\shoveright{
\left.-i\frac{\partial\fM^{l_0}_{m-1,n}}{\partial\theta}+
\frac{\partial\fM^{l_0}_{m-1,n}}{\partial\tau}+
\frac{2i(m-1)\cos\theta^c}{\sin\theta^c}\fM^{l_0}_{m-1,n}\right]+}\\
\shoveleft{
+\frac{1}{2r}\sqrt{(l+m+1)(l+m+2)}\boldsymbol{f}^{l_0}_{l+1,m+1,k^\prime}
\left[-\frac{1}{\sin\theta^c}\frac{\partial\fM^{l_0}_{m+1,n}}{\partial\varphi}-
\frac{i}{\sin\theta^c}\frac{\partial\fM^{l_0}_{m+1,n}}{\partial\epsilon}-
\right.}\\
%\shoveright{
\left.
\left.-i\frac{\partial\fM^{l_0}_{m+1,n}}{\partial\theta}+
\frac{\partial\fM^{l_0}_{m+1,n}}{\partial\tau}-
\frac{2i(m+1)\cos\theta^c}{\sin\theta^c}\fM^{l_0}_{m+1,n}\right]\right\}+
%}
\\
+\kappa^c\boldsymbol{f}^{l_0}_{lm}
\fM^{l_0}_{m,n}(\varphi,\epsilon,\theta,\tau,0,0)=0,\nonumber
\end{multline}
\begin{multline}
\sum_{\dot{k}^\prime}c^{\dot{k}\dot{k}^\prime}_{\dot{l},\dot{l}-1}
\left\{\left[
2\sqrt{\dot{l}^2-\dot{m}^2}
\frac{\partial\boldsymbol{f}^{\dot{l}_0}_{\dot{l}-1,\dot{m},\dot{k}^\prime}}
{\partial r^\ast}-\frac{1}{r^\ast}(\dot{l}+1)\sqrt{\dot{l}^2-\dot{m}^2}
\boldsymbol{f}^{\dot{l}_0}_{\dot{l}-1,\dot{m},\dot{k}^\prime}\right]
\fM^{\dot{l}_0}_{\dot{m},\dot{n}}(\varphi,\epsilon,\theta,\tau,0,0)+\right.\\
+\frac{1}{2r^\ast}\sqrt{(\dot{l}+\dot{m})(\dot{l}+\dot{m}-1)}
\boldsymbol{f}^{\dot{l}_0}_{\dot{l}-1,\dot{m}-1,\dot{k}^\prime}\left[
-\frac{1}{\sin\dot{\theta}^c}
\frac{\partial\fM^{\dot{l}_0}_{\dot{m}-1,\dot{n}}}{\partial\varphi}+
\frac{i}{\sin\dot{\theta}^c}
\frac{\partial\fM^{\dot{l}_0}_{\dot{m}-1,\dot{n}}}{\partial\epsilon}+\right.\\
\shoveright{
\left.+i\frac{\partial\fM^{\dot{l}_0}_{\dot{m}-1,\dot{n}}}{\partial\theta}+
\frac{\partial\fM^{\dot{l}_0}_{\dot{m}-1,\dot{n}}}{\partial\tau}+
\frac{2i(\dot{m}-1)\cos\dot{\theta}^c}{\sin\dot{\theta}^c}
\fM^{\dot{l}_0}_{\dot{m}-1,\dot{n}}\right]+
}
\\
\shoveleft{
+\frac{1}{2r^\ast}\sqrt{(\dot{l}-\dot{m})(\dot{l}-\dot{m}-1)}
\boldsymbol{f}^{\dot{l}_0}_{\dot{l}-1,\dot{m}+1,\dot{k}^\prime}\left[
\frac{1}{\sin\dot{\theta}^c}
\frac{\partial\fM^{\dot{l}_0}_{\dot{m}+1,\dot{n}}}{\partial\varphi}-
\frac{i}{\sin\dot{\theta}^c}
\frac{\partial\fM^{\dot{l}_0}_{\dot{m}+1,\dot{n}}}{\partial\epsilon}+\right.}\\
%\shoveright{
\left.
\left.+i\frac{\partial\fM^{\dot{l}_0}_{\dot{m}+1,\dot{n}}}{\partial\theta}+
\frac{\partial\fM^{\dot{l}_0}_{\dot{m}+1,\dot{n}}}{\partial\tau}-
\frac{2i(\dot{m}+1)\cos\dot{\theta}^c}{\sin\dot{\theta}^c}
\fM^{\dot{l}_0}_{\dot{m}+1,\dot{n}}\right]\right\}+
%}
\\
\sum_{\dot{k}^\prime}c^{\dot{k}\dot{k}^\prime}_{\dot{l}\dot{l}}\left\{\left[
2\dot{m}
\frac{\partial\boldsymbol{f}^{\dot{l}_0}_{\dot{l},\dot{m},\dot{k}^\prime}}
{\partial r^\ast}-\frac{1}{r^\ast}\dot{m}
\boldsymbol{f}^{\dot{l}_0}_{\dot{l},\dot{m},\dot{k}^\prime}\right]
\fM^{\dot{l}_0}_{\dot{m},\dot{n}}(\varphi,\epsilon,\theta,\tau,0,0)+\right.\\
+\frac{1}{2r^\ast}\sqrt{(\dot{l}+\dot{m})(\dot{l}-\dot{m}+1)}
\boldsymbol{f}^{\dot{l}_0}_{\dot{l},\dot{m}-1,\dot{k}^\prime}\left[
\frac{1}{\sin\dot{\theta}^c}
\frac{\partial\fM^{\dot{l}_0}_{\dot{m}-1,\dot{n}}}{\partial\varphi}-
\frac{i}{\sin\dot{\theta}^c}
\frac{\partial\fM^{\dot{l}_0}_{\dot{m}-1,\dot{n}}}{\partial\epsilon}-\right.\\
\shoveright{
\left.-i\frac{\partial\fM^{\dot{l}_0}_{\dot{m}-1,\dot{n}}}{\partial\theta}-
\frac{\partial\fM^{\dot{l}_0}_{\dot{m}-1,\dot{n}}}{\partial\tau}-
\frac{2i(\dot{m}-1)\cos\dot{\theta}^c}{\sin\dot{\theta}^c}
\fM^{\dot{l}_0}_{\dot{m}-1,\dot{n}}\right]+
}
\\
\shoveleft{
+\frac{1}{2r^\ast}\sqrt{(\dot{l}+\dot{m}+1)(\dot{l}-\dot{m})}
\boldsymbol{f}^{\dot{l}_0}_{\dot{l},\dot{m}+1,\dot{k}^\prime}\left[
\frac{1}{\sin\dot{\theta}^c}
\frac{\partial\fM^{\dot{l}_0}_{\dot{m}+1,\dot{n}}}{\partial\varphi}-
\frac{i}{\sin\dot{\theta}^c}
\frac{\partial\fM^{\dot{l}_0}_{\dot{m}+1,\dot{n}}}{\partial\epsilon}+\right.}\\
%\shoveright{
\left.
\left.+i\frac{\partial\fM^{\dot{l}_0}_{\dot{m}+1,\dot{n}}}{\partial\theta}+
\frac{\partial\fM^{\dot{l}_0}_{\dot{m}+1,\dot{n}}}{\partial\tau}-
\frac{2i(\dot{m}+1)\cos\dot{\theta}^c}{\sin\dot{\theta}^c}
\fM^{\dot{l}_0}_{\dot{m}+1,\dot{n}}\right]\right\}+
%}
\\
\sum_{\dot{k}^\prime}c^{\dot{k}\dot{k}^\prime}_{\dot{l},\dot{l}+1}\left\{\left[
2\sqrt{(\dot{l}+1)^2-\dot{m}^2}
\frac{\partial\boldsymbol{f}^{\dot{l}_0}_{\dot{l}+1,\dot{m},\dot{k}^\prime}}
{\partial r^\ast}+\frac{1}{r^\ast}\sqrt{(\dot{l}+1)^2-\dot{m}^2}
\boldsymbol{f}^{\dot{l}_0}_{\dot{l}+1,\dot{m},\dot{k}^\prime}\right]
\fM^{\dot{l}_0}_{\dot{m},\dot{n}}(\varphi,\epsilon,\theta,\tau,0,0)+\right.\\
+\frac{1}{2r^\ast}\sqrt{(\dot{l}-\dot{m}+1)(\dot{l}-\dot{m}+2)}
\boldsymbol{f}^{\dot{l}_0}_{\dot{l}+1,\dot{m}-1,\dot{k}^\prime}\left[
\frac{1}{\sin\dot{\theta}^c}
\frac{\partial\fM^{\dot{l}_0}_{\dot{m}-1,\dot{n}}}{\partial\varphi}-
\frac{i}{\sin\dot{\theta}^c}
\frac{\partial\fM^{\dot{l}_0}_{\dot{m}-1,\dot{n}}}{\partial\epsilon}-\right.\\
\shoveright{
\left.-i\frac{\partial\fM^{\dot{l}_0}_{\dot{m}-1,\dot{n}}}{\partial\theta}-
\frac{\partial\fM^{\dot{l}_0}_{\dot{m}-1,\dot{n}}}{\partial\tau}-
\frac{2i(\dot{m}-1)\cos\dot{\theta}^c}{\sin\dot{\theta}^c}
\fM^{\dot{l}_0}_{\dot{m}-1,\dot{n}}\right]+
}
\\
\shoveleft{
+\frac{1}{2r^\ast}\sqrt{(\dot{l}+\dot{m}+1)(\dot{l}+\dot{m}+2)}
\boldsymbol{f}^{\dot{l}_0}_{\dot{l}+1,\dot{m}+1,\dot{k}^\prime}\left[
-\frac{1}{\sin\dot{\theta}^c}
\frac{\partial\fM^{\dot{l}_0}_{\dot{m}+1,\dot{n}}}{\partial\varphi}+
\frac{i}{\sin\dot{\theta}^c}
\frac{\partial\fM^{\dot{l}_0}_{\dot{m}+1,\dot{n}}}{\partial\epsilon}-\right.}\\
%\shoveright{
\left.
\left.-i\frac{\partial\fM^{\dot{l}_0}_{\dot{m}+1,\dot{n}}}{\partial\theta}-
\frac{\partial\fM^{\dot{l}_0}_{\dot{m}+1,\dot{n}}}{\partial\tau}+
\frac{2i(\dot{m}+1)\cos\dot{\theta}^c}{\sin\dot{\theta}^c}
\fM^{\dot{l}_0}_{\dot{m}+1,\dot{n}}\right]\right\}+
%}
\\
+\dot{\kappa}^c\boldsymbol{f}^{\dot{l}_0}_{\dot{l}\dot{m}}
\fM^{\dot{l}_0}_{\dot{m}\dot{n}}(\varphi,\epsilon,\theta,\tau,0,0)=0.
\label{Complex7}
\end{multline}
\begin{sloppypar}\noindent
The each equation of the obtained system contains three generalized
hyperspherical functions $\fM^{l_0}_{mn}$, $\fM^{l_0}_{m-1,n}$,
$\fM^{l_0}_{m+1,n}$ and their conjugate.
We apply the recurrence relations (\ref{SL33})--(\ref{SL34}),
(\ref{SL39})--(\ref{SL40}) to square brackets containing the functions
$\fM^{l_0}_{m-1,n}$ and $\fM^{l_0}_{m+1,n}$. First of all, let us recall
that 
$\fM^{l_0}_{m\pm 1,n}(\varphi,\epsilon,\theta,\tau,0,0)=
e^{-n(\epsilon+i\varphi)}Z^{l_0}_{m\pm 1,n}(\theta,\tau)$ and
$\fM^{\dot{l}_0}_{\dot{m}\pm 1,\dot{n}}(\varphi,\epsilon,\theta,\tau,0,0)=
e^{-\dot{n}(\epsilon-i\varphi)}
Z^{\dot{l}_0}_{\dot{m}\pm 1,\dot{n}}(\theta,\tau)$. Therefore,
$\dfrac{\partial\fM^{l_0}_{m\pm 1,n}}{\partial\varphi}=
-in\fM^{l_0}_{m\pm 1,n}$,
$\dfrac{\partial\fM^{l_0}_{m\pm 1,n}}{\partial\epsilon}=
-n\fM^{l_0}_{m\pm 1,n}$ and
$\dfrac{\partial\fM^{\dot{l}_0}_{\dot{m}\pm 1,\dot{n}}}{\partial\varphi}=
i\dot{n}\fM^{\dot{l}_0}_{\dot{m}\pm 1,\dot{n}}$,
$\dfrac{\partial\fM^{\dot{l}_0}_{\dot{m}\pm 1,\dot{n}}}{\partial\epsilon}=
-\dot{n}\fM^{\dot{l}_0}_{\dot{m}\pm 1,\dot{n}}$.\end{sloppypar}

Thus, in virtue of (\ref{SL34}) the first bracket in (\ref{Complex7})
can be written as follows
\begin{multline}
ie^{-n(\epsilon+i\varphi)}\left[\frac{\partial Z^{l_0}_{m-1,n}}
{\partial\theta}+i\frac{\partial Z^{l_0}_{m-1,n}}{\partial\tau}+\right.\\
\left.
+\frac{2(n-(m-1)\cos\theta^c)}{\sin\theta^c}Z^{l_0}_{m-1,n}\right]=
2i\sqrt{(\dot{l}_0+\dot{m})(\dot{l}_0-\dot{m}+1)}\fM^{l_0}_{m,n}.\label{RR1}
\end{multline}
Further, in virtue of (\ref{SL33}) for the second and fourth brackets
we have
\begin{multline}
ie^{-n(\epsilon+i\varphi)}\left[\frac{\partial Z^{l_0}_{m+1,n}}
{\partial\theta}+i\frac{\partial Z^{l_0}_{m+1,n}}{\partial\tau}-\right.\\
\left.
-\frac{2(n-(m+1)\cos\theta^c)}{\sin\theta^c}Z^{l_0}_{m+1,n}\right]=
2i\sqrt{(\dot{l}_0+\dot{m}+1)(\dot{l}_0-\dot{m})}\fM^{l_0}_{m,n}.\label{RR2}
\end{multline}
Analogously, the third, fifth and sixth brackets can be written as follows
\begin{multline}
-ie^{-n(\epsilon+i\varphi)}\left[\frac{\partial Z^{l_0}_{m-1,n}}
{\partial\theta}+i\frac{\partial Z^{l_0}_{m-1,n}}{\partial\tau}+\right.\\
\shoveright{
\left.
+\frac{2(n-(m-1)\cos\theta^c)}{\sin\theta^c}Z^{l_0}_{m-1,n}\right]=
-2i\sqrt{(\dot{l}_0+\dot{m})(\dot{l}_0-\dot{m}+1)}\fM^{l_0}_{m,n},}\\
\shoveleft{
-ie^{-n(\epsilon+i\varphi)}\left[\frac{\partial Z^{l_0}_{m+1,n}}
{\partial\theta}+i\frac{\partial Z^{l_0}_{m+1,n}}{\partial\tau}-\right.}\\
\left.
-\frac{2(n-(m+1)\cos\theta^c)}{\sin\theta^c}Z^{l_0}_{m+1,n}\right]=
-2i\sqrt{(\dot{l}_0+\dot{m}+1)(\dot{l}_0-\dot{m})}\fM^{l_0}_{m,n}.\label{RR4}
\end{multline}
\begin{sloppypar}
In like manner, the square brackets of the conjugate (dual) part of the
system (\ref{Complex7}) can be rewritten as\end{sloppypar}
\begin{multline}
ie^{-\dot{n}(\epsilon-i\varphi)}\left[
\frac{\partial Z^{\dot{l}_0}_{\dot{m}-1,\dot{n}}}{\partial\theta}-
i\frac{\partial Z^{\dot{l}_0}_{\dot{m}-1,\dot{n}}}{\partial\tau}-\right.\\
\left.
-\frac{2(\dot{n}-(\dot{m}-1)\cos\dot{\theta}^c)}{\sin\dot{\theta}^c}
Z^{\dot{l}_0}_{\dot{m}-1,\dot{n}}\right]=
2i\sqrt{(l_0+m)(l_0-m+1)}\fM^{\dot{l}_0}_{\dot{m},\dot{n}}\label{RR5}
\end{multline}
for the seventh bracket (in virtue of the recurrence relation (\ref{SL40}))
and
\begin{multline}
ie^{-\dot{n}(\epsilon-i\varphi)}\left[
\frac{\partial Z^{\dot{l}_0}_{\dot{m}+1,\dot{n}}}{\partial\theta}-
i\frac{\partial Z^{\dot{l}_0}_{\dot{m}+1,\dot{n}}}{\partial\tau}+\right.\\
\left.
+\frac{2(\dot{n}-(\dot{m}+1)\cos\dot{\theta}^c)}{\sin\dot{\theta}^c}
Z^{\dot{l}_0}_{\dot{m}+1,\dot{n}}\right]=
2i\sqrt{(l_0+m+1)(l_0-m)}\fM^{\dot{l}_0}_{\dot{m},\dot{n}}\label{RR6}
\end{multline}
for the eight and tenth brackets (in virtue of (\ref{SL39})). Finally,
for the ninth, eleventh and twelfth brackets we have
\begin{multline}
-ie^{-\dot{n}(\epsilon-i\varphi)}\left[
\frac{\partial Z^{\dot{l}_0}_{\dot{m}-1,\dot{n}}}{\partial\theta}-
i\frac{\partial Z^{\dot{l}_0}_{\dot{m}-1,\dot{n}}}{\partial\tau}-\right.\\
\shoveright{
\left.
-\frac{2(\dot{n}-(\dot{m}-1)\cos\dot{\theta}^c)}{\sin\dot{\theta}^c}
Z^{\dot{l}_0}_{\dot{m}-1,\dot{n}}\right]=
-2i\sqrt{(l_0+m)(l_0-m+1)}\fM^{\dot{l}_0}_{\dot{m},\dot{n}},}\\
\shoveleft{
-ie^{-\dot{n}(\epsilon-i\varphi)}\left[
\frac{\partial Z^{\dot{l}_0}_{\dot{m}+1,\dot{n}}}{\partial\theta}-
i\frac{\partial Z^{\dot{l}_0}_{\dot{m}+1,\dot{n}}}{\partial\tau}+\right.}\\
\left.
+\frac{2(\dot{n}-(\dot{m}+1)\cos\dot{\theta}^c)}{\sin\dot{\theta}^c}
Z^{\dot{l}_0}_{\dot{m}+1,\dot{n}}\right]=
-2i\sqrt{(l_0+m+1)(l_0-m)}\fM^{\dot{l}_0}_{\dot{m},\dot{n}}.\label{RR8}
\end{multline}
\begin{sloppypar}
Let us replace the square brackets in the system (\ref{Complex7}) via
the relations (\ref{RR1})--(\ref{RR8}) and cancel all the equations by
$\fM^{l_0}_{mn}$ ($\fM^{\dot{l}_0}_{\dot{m}\dot{n}}$).
In such a way, we see that the relativistically invariant system is
reduced to a system of ordinary differential equations:\end{sloppypar}
\begin{multline}
\sum_{k^\prime}c^{kk^\prime}_{l,l-1}\left[
2\sqrt{l^2-m^2}
\frac{d\boldsymbol{f}^{l_0}_{l-1,m,k^\prime}(r)}{d r}+\right.\\
-\frac{1}{r}(l+1)\sqrt{l^2-m^2}\boldsymbol{f}^{l_0}_{l-1,m,k^\prime}(r)+\\
+\frac{i}{r}\sqrt{(l+m)(l+m-1)}
\sqrt{(\dot{l}_0+\dot{m})(\dot{l}_0-\dot{m}+1)}
\boldsymbol{f}^{l_0}_{l-1,m-1,k^\prime}(r)+\\
\shoveright{
\left.+\frac{i}{r}\sqrt{(l-m)(l-m-1)}
\sqrt{(\dot{l}_0+\dot{m}+1)(\dot{l}_0-\dot{m})}
\boldsymbol{f}^{l_0}_{l-1,m+1,k^\prime}(r)\right]+}\\
\shoveleft{
\sum_{k^\prime}c^{kk^\prime}_{ll}\left[
2m\frac{d\boldsymbol{f}^{l_0}_{l,m,k^\prime}(r)}{d r}-
\frac{1}{r}m\boldsymbol{f}^{l_0}_{l,m,k^\prime}(r)-\right.}\\
-\frac{i}{r}\sqrt{(l+m)(l-m+1)}
\sqrt{(\dot{l}_0+\dot{m})(\dot{l}_0-\dot{m}+1)}
\boldsymbol{f}^{l_0}_{l,m-1,k^\prime}(r)+\\
\shoveright{
\left.+\frac{i}{r}\sqrt{(l+m+1)(l-m)}
\sqrt{(\dot{l}_0+\dot{m}+1)(\dot{l}_0-\dot{m})}
\boldsymbol{f}^{l_0}_{l,m+1,k^\prime}(r)\right]+}\\
\shoveleft{
\sum_{k^\prime}c^{kk^\prime}_{l,l+1}\left[
2\sqrt{(l+1)^2-m^2}
\frac{d\boldsymbol{f}^{l_0}_{l+1,m,k^\prime}(r)}{d r}-\right.}\\
+\frac{1}{r}l\sqrt{(l+1)^2-m^2}
\boldsymbol{f}^{l_0}_{l+1,m,k^\prime}(r)-\\
-\frac{i}{r}\sqrt{(l-m+1)(l-m+2)}
\sqrt{(\dot{l}_0+\dot{m})(\dot{l}_0-\dot{m}+1)}
\boldsymbol{f}^{l_0}_{l+1,m-1,k^\prime}(r)-\\
\shoveright{
\left.-\frac{i}{r}\sqrt{(l+m+1)(l+m+2)}
\sqrt{(\dot{l}_0+\dot{m}+1)(\dot{l}_0-\dot{m})}
\boldsymbol{f}^{l_0}_{l+1,m+1,k^\prime}(r)\right]+}\\
+\kappa^c\boldsymbol{f}^{l_0}_{lmk}(r)=0,\nonumber
\end{multline}
\begin{multline}
\sum_{\dot{k}^\prime}c^{\dot{k}\dot{k}^\prime}_{\dot{l},\dot{l}-1}\left[
2\sqrt{\dot{l}^2-\dot{m}^2}
\frac{d\boldsymbol{f}^{\dot{l}_0}_{\dot{l}-1,\dot{m},\dot{k}^\prime}(r)}
{d r^\ast}-\right.\\
-\frac{1}{r^\ast}(\dot{l}+1)\sqrt{\dot{l}^2-\dot{m}^2}
\boldsymbol{f}^{\dot{l}_0}_{\dot{l}-1,\dot{m},\dot{k}^\prime}(r)+\\
+\frac{i}{r^\ast}\sqrt{(\dot{l}+\dot{m})(\dot{l}+\dot{m}-1)}
\sqrt{(l_0+m)(l_0-m+1)}
\boldsymbol{f}^{\dot{l}_0}_{\dot{l}-1,\dot{m}-1,\dot{k}^\prime}(r)+\\
\shoveright{
\left.
+\frac{i}{r^\ast}\sqrt{(\dot{l}-\dot{m})(\dot{l}-\dot{m}-1)}
\sqrt{(l_0+m+1)(l_0-m)}
\boldsymbol{f}^{\dot{l}_0}_{\dot{l}-1,\dot{m}+1,\dot{k}^\prime}(r)\right]+}\\
\shoveleft{
\sum_{\dot{k}^\prime}c^{\dot{k}\dot{k}^\prime}_{\dot{l},\dot{l}}\left[
2\dot{m}
\frac{d\boldsymbol{f}^{\dot{l}_0}_{\dot{l},\dot{m},\dot{k}^\prime}(r)}
{d r^\ast}
-\frac{1}{r^\ast}\dot{m}
\boldsymbol{f}^{\dot{l}_0}_{\dot{l},\dot{m},\dot{k}^\prime}(r)-\right.}\\
-\frac{i}{r^\ast}\sqrt{(\dot{l}+\dot{m})(\dot{l}-\dot{m}-1)}
\sqrt{(l_0+m)(l_0-m+1)}
\boldsymbol{f}^{\dot{l}_0}_{\dot{l},\dot{m}-1,\dot{k}^\prime}(r)+\\
\shoveright{
\left.
+\frac{i}{r^\ast}\sqrt{(\dot{l}+\dot{m}+1)(\dot{l}-\dot{m})}
\sqrt{(l_0+m+1)(l_0-m)}
\boldsymbol{f}^{\dot{l}_0}_{\dot{l},\dot{m}+1,\dot{k}^\prime}(r)\right]+}\\
\shoveleft{
\sum_{\dot{k}^\prime}c^{\dot{k}\dot{k}^\prime}_{\dot{l},\dot{l}+1}\left[
2\sqrt{(\dot{l}+1)^2-\dot{m}^2}
\frac{d\boldsymbol{f}^{\dot{l}_0}_{\dot{l}+1,\dot{m},\dot{k}^\prime}(r)}
{d r^\ast}+\right.}\\
+\frac{1}{r^\ast}\dot{l}\sqrt{(\dot{l}+1)^2-\dot{m}^2}
\boldsymbol{f}^{\dot{l}_0}_{\dot{l}+1,\dot{m},\dot{k}^\prime}(r)-\\
-\frac{i}{r^\ast}\sqrt{(\dot{l}-\dot{m}+1)(\dot{l}-\dot{m}+2)}
\sqrt{(l_0+m)(l_0-m+1)}
\boldsymbol{f}^{\dot{l}_0}_{\dot{l}+1,\dot{m}-1,\dot{k}^\prime}(r)-\\
\shoveright{
\left.
-\frac{i}{r^\ast}\sqrt{(\dot{l}+\dot{m}+1)(\dot{l}+\dot{m}+2)}
\sqrt{(l_0+m+1)(l_0-m)}
\boldsymbol{f}^{\dot{l}_0}_{\dot{l}+1,\dot{m}+1,\dot{k}^\prime}(r)\right]+}\\
+\dot{\kappa}^c
\boldsymbol{f}^{\dot{l}_0}_{\dot{l}\dot{m}\dot{k}^\prime}(r)=0.
\label{RFS}
\end{multline}
\section{Dirac equations}
Let us consider general solutions of the Dirac equations
\begin{equation}\label{Dirac}
i\gamma_\mu\frac{\partial\boldsymbol{\psi}}{\partial x_\mu}-
m\boldsymbol{\psi}=0,
\end{equation}
where $\gamma$--matrices we take in the Weyl basis:
\begin{equation}\label{Gamma}
\gamma_0=\begin{pmatrix}
\sigma_0 & 0\\
0 & -\sigma_0
\end{pmatrix},\;\;
\gamma_1=\begin{pmatrix}
0 & \sigma_1\\
-\sigma_1 & 0
\end{pmatrix},\;\;
\gamma_2=\begin{pmatrix}
0 & \sigma_2\\
-\sigma_2 & 0
\end{pmatrix},\;\;
\gamma_3=\begin{pmatrix}
0 & \sigma_3\\
-\sigma_3 & 0
\end{pmatrix},
\end{equation}
where
\[
\sigma_1=\begin{pmatrix}
0 & 1\\
1 & 0
\end{pmatrix},\quad
\sigma_2=\begin{pmatrix}
0 & -i\\
i & 0
\end{pmatrix},\quad
\sigma_3=\begin{pmatrix}
1 & 0\\
0 & -1
\end{pmatrix}
\]
are the Pauli matrices. 
In accordance with a general Fermi--scheme of interlocking representations
of $\fG_+$ (\ref{Fermi}), 
the equations (\ref{Dirac}) correspond to the following
interlocking scheme
\[
\dgARROWLENGTH=2.5em
\begin{diagram}
\node{\left(\frac{1}{2},0\right)}\arrow{e,<>}
\node{\left(0,\frac{1}{2}\right)}
\end{diagram}
\]
By this reason a Dirac field is represented by a $P$--invariant direct
sum $(1/2,0)\oplus(0,1/2)$ and the equations (\ref{Dirac}) describe
a massive particle with the spin $1/2$.
It is easy to see that $\gamma$--matrices for
the Dirac equations have the structure (\ref{Type1}), where
$\underset{i}{\Delta}{}_2=\sigma_i$, 
$\underset{i}{\overset{\ast}{\Delta}}{}_2=-\sigma_i=\widetilde{\sigma}_i$.
Therefore, in accordance with (\ref{Lambda1}) the
$\sL$--matrices for the Dirac equations in bivector space are
\begin{equation}\label{LM1}
\left.\begin{array}{ccc}
\sL_1=\sigma_1\sigma_0=\begin{pmatrix}
0 & 1\\
1 & 0
\end{pmatrix}, &
\sL_2=\sigma_2\sigma_0=\begin{pmatrix}
0 & -i\\
i & 0
\end{pmatrix}, &
\sL_3=\sigma_3\sigma_0=\begin{pmatrix}
1 & 0\\
0 & -1
\end{pmatrix},\\[0.2cm]
\sL^\ast_1=\widetilde{\sigma}_2\widetilde{\sigma}_3=\begin{pmatrix}
0 & i\\
i & 0
\end{pmatrix}, &
\sL^\ast_2=\widetilde{\sigma}_3\widetilde{\sigma}_1=\begin{pmatrix}
0 & 1\\
-1 & 0
\end{pmatrix}, &
\sL^\ast_3=\widetilde{\sigma}_1\widetilde{\sigma}_2=\begin{pmatrix}
i & 0\\
0 & -i
\end{pmatrix}.
\end{array}\right\}
\end{equation}
It is easy to verify that matrices (\ref{LM1}) satisfy the commutation
relations (\ref{AL})--(\ref{DBL}) with infinitesimal operators
$\sA_i$, $\sB_i$, $\widetilde{\sA}_i$, $\widetilde{\sB}_i$ defined
for the fundamental representation of $\fG_+$ 
(see (\ref{IF}) and (\ref{DIF})).

It is obvious that for the representation $(1/2,0)\oplus(0,1/2)$
we can omit in the system (\ref{RFS}) the index $k$ ($k^\prime$) which
numerates subspaces transforming within one and the same representation of
the group $\fG_+$. Further, assuming that
$c_{\frac{1}{2},-\frac{1}{2}}=c_{\frac{1}{2},\frac{1}{2}}=1$,
$c_{\frac{1}{2},\frac{3}{2}}=0$ we reduce the system (\ref{RFS}) to the
following
\begin{eqnarray}
\frac{d\boldsymbol{f}^l_{\frac{1}{2},\frac{1}{2}}(r)}{dr}\;-\;\;
\frac{1}{2r}\boldsymbol{f}^l_{\frac{1}{2},\frac{1}{2}}(r)\;\,-\;
\frac{i\left(\dot{l}+\frac{1}{2}\right)}{r}
\boldsymbol{f}^l_{\frac{1}{2},-\frac{1}{2}}(r)\,\;+\;
\kappa^c\boldsymbol{f}^l_{\frac{1}{2},\frac{1}{2}}(r)&=&0,\nonumber\\
-\frac{d\boldsymbol{f}^l_{\frac{1}{2},-\frac{1}{2}}(r)}{dr}\;+\;
\frac{1}{2r}\boldsymbol{f}^l_{\frac{1}{2},-\frac{1}{2}}(r)\,+\;
\frac{i\left(\dot{l}+\frac{1}{2}\right)}{r}
\boldsymbol{f}^l_{\frac{1}{2},\frac{1}{2}}(r)\,+\;
\kappa^c\boldsymbol{f}^l_{\frac{1}{2},-\frac{1}{2}}(r)&=&0,\nonumber\\
\frac{d\boldsymbol{f}^{\dot{l}}_{\frac{1}{2},\frac{1}{2}}(r^\ast)}
{dr^\ast}\;-\,
\frac{1}{2r^\ast}\boldsymbol{f}^{\dot{l}}_{\frac{1}{2},\frac{1}{2}}(r^\ast)-
\frac{i\left(l+\frac{1}{2}\right)}{r^\ast}
\boldsymbol{f}^{\dot{l}}_{\frac{1}{2},-\frac{1}{2}}(r^\ast)\;+
\dot{\kappa}^c\boldsymbol{f}^{\dot{l}}_{\frac{1}{2},\frac{1}{2}}
(r^\ast)&=&0,\nonumber\\
-\frac{d\boldsymbol{f}^{\dot{l}}_{\frac{1}{2},-\frac{1}{2}}(r^\ast)}{dr^\ast}+
\frac{1}{2r^\ast}\boldsymbol{f}^{\dot{l}}_{\frac{1}{2},-\frac{1}{2}}(r^\ast)+
\frac{i\left(l+\frac{1}{2}\right)}{r^\ast}
\boldsymbol{f}^{\dot{l}}_{\frac{1}{2},\frac{1}{2}}(r^\ast)+
\dot{\kappa}^c\boldsymbol{f}^{\dot{l}}_{\frac{1}{2},-\frac{1}{2}}
(r^\ast)&=&0,\label{RFD}
\end{eqnarray}
where $\kappa^c=-im$.
If we suppose that 
$\boldsymbol{f}^l_{\frac{1}{2},-\frac{1}{2}}(r)=
-i\boldsymbol{f}^l_{\frac{1}{2},\frac{1}{2}}(r)$ and
$\boldsymbol{f}^{\dot{l}}_{\frac{1}{2},-\frac{1}{2}}(r^\ast)=
-i\boldsymbol{f}^{\dot{l}}_{\frac{1}{2},\frac{1}{2}}(r^\ast)$, then
the system (\ref{RFD}) takes a form
\begin{eqnarray}
\frac{d\boldsymbol{f}^l_{\frac{1}{2},\frac{1}{2}}(r)}{dr}\;-\;
\frac{(\dot{l}+1)}{r}\boldsymbol{f}^l_{\frac{1}{2},\frac{1}{2}}(r)\;+
\kappa^c\boldsymbol{f}^l_{\frac{1}{2},\frac{1}{2}}(r)&=&0,\nonumber\\
\frac{d\boldsymbol{f}^l_{\frac{1}{2},\frac{1}{2}}(r)}{dr}\;\;\;+\;\;\;
\frac{\dot{l}}{r}\boldsymbol{f}^l_{\frac{1}{2},\frac{1}{2}}(r)\;\;\;-\;\;
\kappa^c\boldsymbol{f}^l_{\frac{1}{2},\frac{1}{2}}(r)&=&0,\nonumber\\
\frac{d\boldsymbol{f}^{\dot{l}}_{\frac{1}{2},\frac{1}{2}}(r^\ast)}{dr^\ast}-
\frac{(l+1)}{r^\ast}\boldsymbol{f}^{\dot{l}}_{\frac{1}{2},\frac{1}{2}}(r^\ast)+
\dot{\kappa}^c\boldsymbol{f}^l_{\frac{1}{2},\frac{1}{2}}(r^\ast)
&=&0,\nonumber\\
\frac{d\boldsymbol{f}^{\dot{l}}_{\frac{1}{2},\frac{1}{2}}(r^\ast)}{dr^\ast}\;+\;\,
\frac{l}{r^\ast}\boldsymbol{f}^{\dot{l}}_{\frac{1}{2},\frac{1}{2}}(r^\ast)\;\;-\;\;
\dot{\kappa}^c\boldsymbol{f}^l_{\frac{1}{2},\frac{1}{2}}(r^\ast)
&=&0,\nonumber
\end{eqnarray}
Multiplying the second and fourth equations of the obtained system by 2
and adding the first equation with the second equation and the third
equation with the fourth we come to the following differential equations
\begin{eqnarray}
3\frac{d\boldsymbol{f}^l_{\frac{1}{2},\frac{1}{2}}(r)}{dr}\;+\;
\frac{\dot{l}-1}{r}\boldsymbol{f}^l_{\frac{1}{2},\frac{1}{2}}(r)\;-\;
\kappa^c\boldsymbol{f}^l_{\frac{1}{2},\frac{1}{2}}(r)&=&0,\nonumber\\
3\frac{d\boldsymbol{f}^{\dot{l}}_{\frac{1}{2},\frac{1}{2}}(r^\ast)}{dr^\ast}+
\frac{l-1}{r^\ast}\boldsymbol{f}^{\dot{l}}_{\frac{1}{2},\frac{1}{2}}(r^\ast)-
\dot{\kappa}^c\boldsymbol{f}^{\dot{l}}_{\frac{1}{2},\frac{1}{2}}(r^\ast)
&=&0.\label{RFD2}
\end{eqnarray}
Solutions of the equations (\ref{RFD2}) are expressed via Bessel functions
of the half--integer order:
\begin{eqnarray}
\boldsymbol{f}^l_{\frac{1}{2},\frac{1}{2}}(r)&=&
\sqrt[3]{r}\sum^\infty_{k=0}(-1)^k\left(\frac{2}{\sqrt{3}}\right)^{2k}
\Gamma(\dot{\nu}+k+1)
J_{\dot{\nu}}\left(2\sqrt{\kappa^c}\sqrt[3]{r}\right),\nonumber\\
\boldsymbol{f}^{\dot{l}}_{\frac{1}{2},\frac{1}{2}}(r^\ast)&=&
\sqrt[3]{r^\ast}\sum^\infty_{k=0}(-1)^k\left(\frac{2}{\sqrt{3}}\right)^{2k}
\Gamma(\nu+k+1)
J_{\nu}\left(2\sqrt{\dot{\kappa}^c}\sqrt[3]{r^\ast}\right),\label{SD}
\end{eqnarray}
where $\dot{\nu}=-(\dot{l}-1)$, $\nu=-(l-1)$ and
$\dot{l}=\frac{2\dot{s}+1}{2}$, $l=\frac{2s+1}{2}$,
$s=0,1,2,\ldots$. The Bessel function in (\ref{SD}) has a form
\begin{multline}
J_{\frac{2s+1}{2}}(z)=\sqrt{\frac{2}{\pi z}}\left[
\sin\left(z-\frac{s\pi}{2}\right)\sum^{\frac{s}{2}}_{k=0}
\frac{(-1)^k(s+2k)!}{(2k)!(s-2k)!(2z)^{2k}}+\right.\\
\left.
+\cos\left(z-\frac{s\pi}{2}\right)\sum^{\left[\frac{s-1}{2}\right]}_{k=0}
\frac{(-1)^k(s+2k+1)!}{(2k+1)!(s-2k-1)!(2z)^{2k+1}}\right].\label{Bessel}
\end{multline}
In such a way, solutions of the Dirac equations have the form
\begin{eqnarray}
\psi_1(r,\varphi^c,\theta^c)&=&\boldsymbol{f}^l_{\frac{1}{2},\frac{1}{2}}
(r)\fM^l_{\frac{1}{2},n}(\varphi,\epsilon,\theta,\tau,0,0),\nonumber\\
\psi_2(r,\varphi^c,\theta^c)&=&-i\boldsymbol{f}^l_{\frac{1}{2},\frac{1}{2}}
(r)\fM^l_{-\frac{1}{2},n}(\varphi,\epsilon,\theta,\tau,0,0),\nonumber\\
\dot{\psi}_1(r^\ast,\dot{\varphi}^c,\dot{\theta}^c)&=&
\boldsymbol{f}^{\dot{l}}_{\frac{1}{2},\frac{1}{2}}
(r^\ast)\fM^{\dot{l}}_{\frac{1}{2},\dot{n}}
(\varphi,\epsilon,\theta,\tau,0,0),\nonumber\\
\dot{\psi}_2(r^\ast,\dot{\varphi}^c,\dot{\theta}^c)&=&
-i\boldsymbol{f}^{\dot{l}}_{\frac{1}{2},\frac{1}{2}}
(r^\ast)\fM^l_{-\frac{1}{2},\dot{n}}(\varphi,\epsilon,\theta,\tau,0,0),
\nonumber
\end{eqnarray}
where
\begin{eqnarray}
&&l=\frac{1}{2},\;\frac{3}{2},\;\frac{5}{2};\quad
n=-l,\;-l+1,\;\ldots,\; l;\nonumber\\
&&\dot{l}=\frac{1}{2},\;\frac{3}{2},\;\frac{5}{2};\quad
\dot{n}=-\dot{l},\;-\dot{l}+1,\;\ldots,\; \dot{l},\nonumber
\end{eqnarray}
\[
\fM^l_{\pm\frac{1}{2},n}(\varphi,\epsilon,\theta,\tau,0,0)=
e^{\mp\frac{1}{2}(\epsilon+i\varphi)}Z^l_{\pm\frac{1}{2},n}(\theta,\tau),
\]
\begin{multline}
Z^l_{\pm\frac{1}{2},n}(\theta,\tau)=\cos^{2l}\frac{\theta}{2}
\ch^{2l}\frac{\tau}{2}\sum^l_{k=-l}i^{\pm\frac{1}{2}-k}
\tg^{\pm\frac{1}{2}-k}\frac{\theta}{2}\tnh^{n-k}\frac{\tau}{2}\times\\
\hypergeom{2}{1}{\pm\frac{1}{2}-l+1,1-l-k}{\pm\frac{1}{2}-k+1}
{i^2\tg^2\frac{\theta}{2}}
\hypergeom{2}{1}{n-l+1,1-l-k}{n-k+1}{\tnh^2\frac{\tau}{2}},\nonumber
\end{multline}
\[
\fM^{\dot{l}}_{\pm\frac{1}{2},\dot{n}}(\varphi,\epsilon,\theta,\tau,0,0)=
e^{\mp\frac{1}{2}(\epsilon-i\varphi)}
Z^{\dot{l}}_{\pm\frac{1}{2},\dot{n}}(\theta,\tau),
\]
\begin{multline}
Z^{\dot{l}}_{\pm\frac{1}{2},\dot{n}}(\theta,\tau)=
\cos^{2\dot{l}}\frac{\theta}{2}
\ch^{2\dot{l}}\frac{\tau}{2}
\sum^{\dot{l}}_{\dot{k}=-\dot{l}}i^{\pm\frac{1}{2}-\dot{k}}
\tg^{\pm\frac{1}{2}-\dot{k}}\frac{\theta}{2}
\tnh^{\dot{n}-\dot{k}}\frac{\tau}{2}\times\\
\hypergeom{2}{1}{\pm\frac{1}{2}-\dot{l}+1,1-\dot{l}-\dot{k}}
{\pm\frac{1}{2}-\dot{k}+1}
{i^2\tg^2\frac{\theta}{2}}
\hypergeom{2}{1}{\dot{n}-\dot{l}+1,1-\dot{l}-\dot{k}}
{\dot{n}-\dot{k}+1}{\tnh^2\frac{\tau}{2}}.\nonumber
\end{multline}
\section{Weyl equations}
Let us consider massless Dirac equations
\begin{equation}\label{MD}
i\gamma_\mu\frac{\partial\boldsymbol{\psi}}{\partial x_\mu}=0,
\end{equation}
where $\gamma$--matrices have the form (\ref{Gamma}). Introducing
projection operators $P_\pm=(1\pm\gamma_5)/2$ and following to the
standard procedure (see, for example, \cite{BS93}) we can split the equations
(\ref{MD}) into two (Weyl) equations
\begin{equation}\label{Weyl}
\frac{\partial\boldsymbol{\phi}_1}{\partial x_0}-\boldsymbol{\sigma}
\frac{\partial\boldsymbol{\phi}_1}{\partial\boldsymbol{x}}=0,\quad
\frac{\partial\boldsymbol{\phi}_2}{\partial x_0}+\boldsymbol{\sigma}
\frac{\partial\boldsymbol{\phi}_2}{\partial\boldsymbol{x}}=0,
\end{equation}
where
\[
\boldsymbol{\phi}_1=\frac{1}{2}\begin{pmatrix}
\psi_1-\psi_3\\
\psi_2-\psi_4
\end{pmatrix},\quad
\boldsymbol{\phi}_2=\frac{1}{2}\begin{pmatrix}
\psi_1+\psi_3\\
\psi_2+\psi_4
\end{pmatrix}.
\]
As known, the equations (\ref{Weyl}) describe the massless particle with
the spin $1/2$ (neutrino).

Let us find solutions of the equations (\ref{MD}). In this case,
taking into account (\ref{LM1}) and (\ref{RFS}), we come to a following
system
\begin{eqnarray}
\frac{d\boldsymbol{f}^l_{\frac{1}{2},\frac{1}{2}}(r)}{dr}\;-\;\;
\frac{1}{2r}\boldsymbol{f}^l_{\frac{1}{2},\frac{1}{2}}(r)\;\,-\;
\frac{i\left(\dot{l}+\frac{1}{2}\right)}{r}
\boldsymbol{f}^l_{\frac{1}{2},-\frac{1}{2}}(r)&=&0,\nonumber\\
-\frac{d\boldsymbol{f}^l_{\frac{1}{2},-\frac{1}{2}}(r)}{dr}\;+\;
\frac{1}{2r}\boldsymbol{f}^l_{\frac{1}{2},-\frac{1}{2}}(r)\,+\;
\frac{i\left(\dot{l}+\frac{1}{2}\right)}{r}
\boldsymbol{f}^l_{\frac{1}{2},\frac{1}{2}}(r)&=&0,\nonumber\\
\frac{d\boldsymbol{f}^{\dot{l}}_{\frac{1}{2},\frac{1}{2}}(r^\ast)}
{dr^\ast}\;-\,
\frac{1}{2r^\ast}\boldsymbol{f}^{\dot{l}}_{\frac{1}{2},\frac{1}{2}}(r^\ast)-
\frac{i\left(l+\frac{1}{2}\right)}{r^\ast}
\boldsymbol{f}^{\dot{l}}_{\frac{1}{2},-\frac{1}{2}}(r^\ast)&=&0,\nonumber\\
-\frac{d\boldsymbol{f}^{\dot{l}}_{\frac{1}{2},-\frac{1}{2}}(r^\ast)}{dr^\ast}+
\frac{1}{2r^\ast}\boldsymbol{f}^{\dot{l}}_{\frac{1}{2},-\frac{1}{2}}(r^\ast)+
\frac{i\left(l+\frac{1}{2}\right)}{r^\ast}
\boldsymbol{f}^{\dot{l}}_{\frac{1}{2},\frac{1}{2}}(r^\ast)&=&0.\label{RFW}
\end{eqnarray}
\begin{sloppypar}\noindent
As in the case of the Dirac equations we put
$\boldsymbol{f}^l_{\frac{1}{2},-\frac{1}{2}}(r)=
-i\boldsymbol{f}^l_{\frac{1}{2},\frac{1}{2}}(r)$,
$\boldsymbol{f}^{\dot{l}}_{\frac{1}{2},-\frac{1}{2}}(r^\ast)=
-i\boldsymbol{f}^{\dot{l}}_{\frac{1}{2},\frac{1}{2}}(r^\ast)$ and reduce
the system (\ref{RFW}) to the following equations\end{sloppypar}
\begin{eqnarray}
3\frac{d\boldsymbol{f}^l_{\frac{1}{2},\frac{1}{2}}(r)}{dr}\;\;+\;
\frac{\dot{l}-1}{r}\boldsymbol{f}^l_{\frac{1}{2},\frac{1}{2}}(r)
&=&0,\nonumber\\
3\frac{d\boldsymbol{f}^{\dot{l}}_{\frac{1}{2},\frac{1}{2}}(r^\ast)}{dr^\ast}+
\frac{l-1}{r^\ast}\boldsymbol{f}^{\dot{l}}_{\frac{1}{2},\frac{1}{2}}(r^\ast)
&=&0.\label{RFW2}
\end{eqnarray}
Solutions of the equations (\ref{RFW2}) are
\begin{eqnarray}
\boldsymbol{f}^l_{\frac{1}{2},\frac{1}{2}}(r)&=&
C\sqrt[3]{r}\sum^\infty_{k=0}(-1)^k\left(\frac{2}{\sqrt{3}}\right)^{2k}
\Gamma(k+1)\Gamma(\dot{\nu}+k+1)
J_{\dot{\nu}}\left(2\sqrt[3]{r}\right),\nonumber\\
\boldsymbol{f}^{\dot{l}}_{\frac{1}{2},\frac{1}{2}}(r^\ast)&=&
\dot{C}\sqrt[3]{r^\ast}\sum^\infty_{k=0}(-1)^k
\left(\frac{2}{\sqrt{3}}\right)^{2k}
\Gamma(k+1)\Gamma(\nu+k+1)
J_\nu\left(2\sqrt[3]{r^\ast}\right),\nonumber
\end{eqnarray}
where $\dot{\nu}=-(\dot{l}-1)$, $\nu=-(l-1)$ and the Bessel function
has the form (\ref{Bessel}).

Thus, solutions of the Weyl equations (\ref{Weyl}) are represented by
linear combinations of the solutions of the massless Dirac equations
(\ref{MD}):
\[
\boldsymbol{\phi}_1=\frac{1}{2}\begin{pmatrix}
\psi_1-\dot{\psi}_1\\
\psi_2-\dot{\psi}_2
\end{pmatrix},\quad
\boldsymbol{\phi}_2=\frac{1}{2}\begin{pmatrix}
\psi_1+\dot{\psi}_1\\
\psi_2+\dot{\psi}_2
\end{pmatrix},
\]
where
\begin{eqnarray}
\psi_1(r,\varphi^c,\theta^c)&=&\boldsymbol{f}^l_{\frac{1}{2},\frac{1}{2}}
(r)\fM^l_{\frac{1}{2},n}(\varphi,\epsilon,\theta,\tau,0,0),\nonumber\\
\psi_2(r,\varphi^c,\theta^c)&=&-i\boldsymbol{f}^l_{\frac{1}{2},\frac{1}{2}}
(r)\fM^l_{-\frac{1}{2},n}(\varphi,\epsilon,\theta,\tau,0,0),\nonumber\\
\dot{\psi}_1(r^\ast,\dot{\varphi}^c,\dot{\theta}^c)&=&
\boldsymbol{f}^{\dot{l}}_{\frac{1}{2},\frac{1}{2}}
(r^\ast)\fM^{\dot{l}}_{\frac{1}{2},\dot{n}}
(\varphi,\epsilon,\theta,\tau,0,0),\nonumber\\
\dot{\psi}_2(r^\ast,\dot{\varphi}^c,\dot{\theta}^c)&=&
-i\boldsymbol{f}^{\dot{l}}_{\frac{1}{2},\frac{1}{2}}
(r^\ast)\fM^l_{-\frac{1}{2},\dot{n}}(\varphi,\epsilon,\theta,\tau,0,0),
\nonumber
\end{eqnarray}
and
\begin{eqnarray}
&&l=\frac{1}{2},\;\frac{3}{2},\;\frac{5}{2};\quad
n=-l,\;-l+1,\;\ldots,\; l;\nonumber\\
&&\dot{l}=\frac{1}{2},\;\frac{3}{2},\;\frac{5}{2};\quad
\dot{n}=-\dot{l},\;-\dot{l}+1,\;\ldots,\; \dot{l},\nonumber
\end{eqnarray}
\[
\fM^l_{\pm\frac{1}{2},n}(\varphi,\epsilon,\theta,\tau,0,0)=
e^{\mp\frac{1}{2}(\epsilon+i\varphi)}Z^l_{\pm\frac{1}{2},n}(\theta,\tau),
\]
\begin{multline}
Z^l_{\pm\frac{1}{2},n}(\theta,\tau)=\cos^{2l}\frac{\theta}{2}
\ch^{2l}\frac{\tau}{2}\sum^l_{k=-l}i^{\pm\frac{1}{2}-k}
\tg^{\pm\frac{1}{2}-k}\frac{\theta}{2}\tnh^{n-k}\frac{\tau}{2}\times\\
\hypergeom{2}{1}{\pm\frac{1}{2}-l+1,1-l-k}{\pm\frac{1}{2}-k+1}
{i^2\tg^2\frac{\theta}{2}}
\hypergeom{2}{1}{n-l+1,1-l-k}{n-k+1}{\tnh^2\frac{\tau}{2}},\nonumber
\end{multline}
\[
\fM^{\dot{l}}_{\pm\frac{1}{2},\dot{n}}(\varphi,\epsilon,\theta,\tau,0,0)=
e^{\mp\frac{1}{2}(\epsilon-i\varphi)}
Z^{\dot{l}}_{\pm\frac{1}{2},\dot{n}}(\theta,\tau),
\]
\begin{multline}
Z^{\dot{l}}_{\pm\frac{1}{2},\dot{n}}(\theta,\tau)=
\cos^{2\dot{l}}\frac{\theta}{2}
\ch^{2\dot{l}}\frac{\tau}{2}
\sum^{\dot{l}}_{\dot{k}=-\dot{l}}i^{\pm\frac{1}{2}-\dot{k}}
\tg^{\pm\frac{1}{2}-\dot{k}}\frac{\theta}{2}
\tnh^{\dot{n}-\dot{k}}\frac{\tau}{2}\times\\
\hypergeom{2}{1}{\pm\frac{1}{2}-\dot{l}+1,1-\dot{l}-\dot{k}}
{\pm\frac{1}{2}-\dot{k}+1}
{i^2\tg^2\frac{\theta}{2}}
\hypergeom{2}{1}{\dot{n}-\dot{l}+1,1-\dot{l}-\dot{k}}
{\dot{n}-\dot{k}+1}{\tnh^2\frac{\tau}{2}}.\nonumber
\end{multline}
\section{Maxwell equations}
In accordance with a general Bose--scheme of the interlocking
representations of the group $\fG_+$ (\ref{Bose}), the Maxwell field
$(1,0)\oplus(0,1)$
corresponds to a following interlocking scheme
\[
\dgARROWLENGTH=2.5em
\begin{diagram}
\node{(1,0)}\arrow{e,<>}\node{\left(\frac{1}{2},\frac{1}{2}\right)}
\arrow{e,<>}\node{(0,1)}
\end{diagram}
\]
By this reason and in agreement with a de Broglie theory of fusion
\cite{Bro43} the Maxwell field $(1,0)\oplus(0,1)$ is defined in terms of
tensor products $(1/2,0)\otimes(1/2,0)$ and $(0,1/2)\otimes(0,1/2)$. Indeed,
according to
(\ref{Ten}) the Clifford algebras, corresponded to the Maxwell fields, are
$\C_2\otimes\C_2$ and 
$\overset{\ast}{\C}_2\otimes\overset{\ast}{\C}_2$.
The algebras $\C_2\otimes\C_2$ and 
$\overset{\ast}{\C}_2\otimes\overset{\ast}{\C}_2$ induce spinspaces
$\dS_2\otimes\dS_2$ and $\dot{\dS}_2\otimes\dot{\dS}_2$.
These spinspaces are full representation spaces for tensor products
$\boldsymbol{\tau}_{\frac{1}{2},0}\otimes
\boldsymbol{\tau}_{\frac{1}{2},0}$ and
$\boldsymbol{\tau}_{0,\frac{1}{2}}\otimes
\boldsymbol{\tau}_{0,\frac{1}{2}}$. In accordance with (\ref{Vect}) the
basis `vectors' (spintensors) of the spinspaces $\dS_2\otimes\dS_2$ and
$\dot{\dS}_2\otimes\dot{\dS}_2$ have the form
\begin{eqnarray}
&&\xi^{11}=\xi^1\otimes\xi^1,\;\;\xi^{12}=\xi^1\otimes\xi^2,\;\;
\xi^{21}=\xi^2\otimes\xi^1,\;\;\xi^{22}=\xi^2\otimes\xi^2,\nonumber\\
&&\xi^{\dot{1}\dot{1}}=\xi^{\dot{1}}\otimes\xi^{\dot{1}},\;\;
\xi^{\dot{1}\dot{2}}=\xi^{\dot{1}}\otimes\xi^{\dot{2}},\;\;
\xi^{\dot{2}\dot{1}}=\xi^{\dot{2}}\otimes\xi^{\dot{1}},\;\;
\xi^{\dot{2}\dot{2}}=\xi^{\dot{2}}\otimes\xi^{\dot{2}}.\label{ST}
\end{eqnarray}
Further, the representations 
$\boldsymbol{\tau}_{\frac{1}{2},0}\otimes\boldsymbol{\tau}_{\frac{1}{2},0}$ and
$\boldsymbol{\tau}_{0,\frac{1}{2}}\otimes\boldsymbol{\tau}_{0,\frac{1}{2}}$
are reducible. In virtue of the Clebsh--Gordan formula
\[
\boldsymbol{\tau}_{l_1,\dot{l}_1}\otimes\boldsymbol{\tau}_{l_2,\dot{l}_2}=
\sum_{|l_1-l_2|\leq k\leq l_1+l_2;|\dot{l}_1-\dot{l}_2|\leq\dot{k}\leq
\dot{l}_1+\dot{l}_2}\boldsymbol{\tau}_{k,\dot{k}}
\]
we have
\begin{eqnarray}
\boldsymbol{\tau}_{\frac{1}{2},0}\otimes\boldsymbol{\tau}_{\frac{1}{2},0}
&=&\boldsymbol{\tau}_{0,0}\oplus\boldsymbol{\tau}_{1,0},\nonumber\\
\boldsymbol{\tau}_{0,\frac{1}{2}}\otimes\boldsymbol{\tau}_{0,\frac{1}{2}}
&=&\boldsymbol{\tau}_{0,0}\oplus\boldsymbol{\tau}_{0,1}.\nonumber
\end{eqnarray}
At this point the spinspaces $\dS_2\otimes\dS_2$ and
$\dot{\dS}_2\otimes\dot{\dS}_2$ decompose into direct sums of the
symmetric representation spaces:
\begin{eqnarray}
\dS_2\otimes\dS_2&=&\Sym_{(0,0)}\oplus\Sym_{(2,0)},\nonumber\\
\dot{\dS}_2\otimes\dot{\dS}_2&=&\Sym_{(0,0)}\oplus\Sym_{(0,2)}.\nonumber
\end{eqnarray}
The spintensors $\xi^{11}$, $\xi^{12}=\xi^{21}$, $\xi^{22}$ and
$\xi^{\dot{1}\dot{1}}$, $\xi^{\dot{1}\dot{2}}=\xi^{\dot{2}\dot{1}}$,
$\xi^{\dot{2}\dot{2}}$, obtained after symmetrization from (\ref{ST})
compose the basises of three--dimensional complex spaces $\Sym_{(2,0)}$ and
$\Sym_{(0,2)}$, respectively. Let us introduce independent complex
coordinates $F_1$, $F_2$, $F_3$ and $\overset{\ast}{F}_1$,
$\overset{\ast}{F}_2$, $\overset{\ast}{F}_3$ for the spintensors
$f^{\lambda\mu}$ and $f^{\dot{\lambda}\dot{\mu}}$ (spinor representations
of the electromagnetic tensor), where $F_i=E_i-iB_i$ and
$\overset{\ast}{F}_i=E_i+iB_i$. Explicit expressions of the spinor
representations of the electromagnetic tensor are
\begin{gather}
\boldsymbol{\tau}_{1,0}\quad\left\{\begin{array}{ccl}
f^{11}&\sim&4(F_1+iF_2),\\
f^{12}&\sim&4F_3,\\
f^{22}&\sim&4(F_1-iF_2);
\end{array}\right.\nonumber\\
\boldsymbol{\tau}_{0,1}\quad\left\{\begin{array}{ccl}
f^{\dot{1}\dot{1}}&\sim&4(\overset{\ast}{F}_1+i\overset{\ast}{F}_2),\\
f^{\dot{1}\dot{2}}&\sim&4\overset{\ast}{F}_3,\\
f^{\dot{2}\dot{2}}&\sim&4(\overset{\ast}{F}_1-i\overset{\ast}{F}_2).
\end{array}\right.\nonumber
\end{gather}
In such a way, we see that complex linear combinations $\bF=\bE-i\bB$ and
$\overset{\ast}{F}=\bE+i\bB$, transformed within
$\boldsymbol{\tau}_{1,0}$ and $\boldsymbol{\tau}_{0,1}$ representations,
are coincide with the Majorana--Oppenheimer wave functions. 
At the beginning of thirties of the last century
Majorana \cite{Maj} and Oppenheimer \cite{Opp31} proposed to consider
the Maxwell theory of electromagnetism as the wave mechanics of the photon.
They introduced a wave function of the form
\begin{equation}\label{MO1}
\boldsymbol{\psi}=\bE-i\bB,
\end{equation}
where $\bE$ and $\bB$ are electric and magnetic fields. In virtue of this
the standard Maxwell equations can be rewritten
\begin{equation}\label{1}
\diver\boldsymbol{\psi}=\rho,\quad i\rot\boldsymbol{\psi}=\bj+
\frac{\partial\boldsymbol{\psi}}{\partial t},
\end{equation}
where $\boldsymbol{\psi}=(\psi_1,\psi_2,\psi_3)$, $\psi_k=E_k-iB_k$
($k=1,2,3$). In accordance with correspondence principle
$\left(-i\partial/\partial x_i\rightarrow p_i; +i\partial/\partial t
\rightarrow W\right)$ and in absence of electric charges and currents
the equations (\ref{1}) take a Dirac--like form
\begin{equation}\label{2}
(W-\boldsymbol{\alpha}\cdot\boldsymbol{p})\boldsymbol{\psi}=0
\end{equation}
with transversality condition
\[
\boldsymbol{p}\cdot\boldsymbol{\psi}=0.
\]
At this point three matrices
\begin{equation}\label{3}
\alpha^1=\begin{pmatrix}
0 & 0 & 0\\
0 & 0 & i\\
0 &-i & 0
\end{pmatrix},\quad\alpha^2=\begin{pmatrix}
0 & 0 & -i\\
0 & 0 & 0\\
i & 0 & 0
\end{pmatrix},\quad\alpha^3=\begin{pmatrix}
0 & i & 0\\
-i& 0 & 0\\
0 & 0 & 0
\end{pmatrix}
\end{equation}
satisfy the angular--momentum commutation rules
\[
\ld\alpha_i,\alpha_k\rd=-i\varepsilon_{ikl}\alpha_l\quad
(i,k,l=1,2,3).
\]
Further, for the complex conjugate wave function
\begin{equation}\label{MO2}
\boldsymbol{\psi}^\ast=\bE+i\bB
\end{equation}
there are the analogous Dirac--like equations
\begin{equation}\label{4}
(W+\boldsymbol{\alpha}\cdot\boldsymbol{p})\boldsymbol{\psi}^\ast=0.
\end{equation}
In such a way,
$\boldsymbol{\psi}$ ($\boldsymbol{\psi}^\ast$) may be considered as
a wave function of the photon satisfying the massless Dirac--like equations
(Maxwell equations in the Dirac form are considered also in \cite{Mos58,Mos59}).
In contrast to the Gupta--Bleuler phenomenology, where the 
non--observable four--potential $A_\mu$ is quantized, the main
advantage of the Majorana--Oppenheimer formulation of electrodynamics
lies in the fact that it deals directly with observable quantities,
such as the electric and magnetic fields
\cite{Esp98,Gia85,Rec90,Var01b,Var023}.

It is easy to see that Maxwell equations in the Majorana--Oppenheimer form
can be mapped into the bivector space $\R^6$. At this point, for the
$\sL$--matrices we have
\begin{equation}\label{LM2}
\left.\begin{array}{ccc}
\sL_1=\begin{pmatrix}
0 & 0 & 0\\
0 & 0 & i\\
0 &-i & 0
\end{pmatrix}, &
\sL_2=\begin{pmatrix}
0 & 0 &-i\\
0 & 0 & 0\\
i & 0 & 0
\end{pmatrix}, &
\sL_3=\begin{pmatrix}
0 & i & 0\\
-i& 0 & 0\\
0 & 0 & 0
\end{pmatrix},\\ \vspace{0.4cm}
\sL^\ast_1=\begin{pmatrix}
0 & 0 & 0\\
0 & 0 &-1\\
0 & 1 & 0
\end{pmatrix}, &
\sL^\ast_2=\begin{pmatrix}
0 & 0 & 1\\
0 & 0 & 0\\
-1& 0 & 0
\end{pmatrix}, &
\sL^\ast_3=\begin{pmatrix}
0 &-1 & 0\\
1 & 0 & 0\\
0 & 0 & 0
\end{pmatrix}.
\end{array}\right\}
\end{equation}
Infinitesimal operators in the representation spaces $\Sym_{(2,0)}$ and
$\Sym_{(0,2)}$ take a form
\begin{equation}\label{IF2}
\left.\begin{array}{ccc}
\sA_1=\begin{bmatrix}
0 & 0 & 0\\
0 & 0 &-1\\
0 & 1 & 0
\end{bmatrix}, &
\sA_2=\begin{bmatrix}
0 & 0 & 1\\
0 & 0 & 0\\
-1& 0 & 0
\end{bmatrix}, &
\sA_3=\begin{bmatrix}
0 &-1 & 0\\
1 & 0 & 0\\
0 & 0 & 0
\end{bmatrix}, \\[0.4cm]
\sB_1=\begin{bmatrix}
0 & 0 & 0\\
0 & 0 &-i\\
0 & i & 0
\end{bmatrix}, &
\sB_2=\begin{bmatrix}
0 & 0 & i\\
0 & 0 & 0\\
-i& 0 & 0
\end{bmatrix}, &
\sB_3=\begin{bmatrix}
0 &-i & 0\\
i & 0 & 0\\
0 & 0 & 0
\end{bmatrix}, \\[0.4cm]
\widetilde{\sA}_1=\begin{bmatrix}
0 & 0 & 0\\
0 & 0 & 1\\
0 &-1 & 0
\end{bmatrix}, &
\widetilde{\sA}_2=\begin{bmatrix}
0 & 0 &-1\\
0 & 0 & 0\\
1 & 0 & 0
\end{bmatrix}, &
\widetilde{\sA}_3=\begin{bmatrix}
0 & 1 & 0\\
-1& 0 & 0\\
0 & 0 & 0
\end{bmatrix}, \\[0.4cm]
\widetilde{\sB}_1=\begin{bmatrix}
0 & 0 & 0\\
0 & 0 & i\\
0 &-i & 0
\end{bmatrix}, &
\widetilde{\sB}_2=\begin{bmatrix}
0 & 0 &-i\\
0 & 0 & 0\\
i & 0 & 0
\end{bmatrix}, &
\widetilde{\sB}_3=\begin{bmatrix}
0 & i & 0\\
-i& 0 & 0\\
0 & 0 & 0
\end{bmatrix}.
\end{array}\right\}
\end{equation}
It is easy to verify that the matrices (\ref{LM2}) and operators (\ref{IF2})
satisfy the commutation relations (\ref{AL})--(\ref{DBL}).

In the case of spin 1 the system (\ref{RFS}) takes a form
\begin{eqnarray}
2\frac{d\boldsymbol{f}^l_{1,1}(r)}{dr}\,\;-\;\,\frac{1}{r}\boldsymbol{f}^l_{1,1}(r)\;\,-\;\;
\frac{i\sqrt{2\dot{l}(\dot{l}+1)}}{r}\boldsymbol{f}^l_{1,0}(r)\;\;+\;\,\;
\frac{i\sqrt{2\dot{l}(\dot{l}+1)}}{r}\boldsymbol{f}^l_{0,0}(r)&=&0,\nonumber\\
2\frac{d\boldsymbol{f}^l_{0,0}(r)}{dr}\;-\;\frac{2}{r}\boldsymbol{f}^l_{0,0}(r)\;\;-\;\;
\frac{i\sqrt{2\dot{l}(\dot{l}+1)}}{r}\boldsymbol{f}^l_{1,-1}(r)\,\;+\;\;
\frac{i\sqrt{2\dot{l}(\dot{l}+1)}}{r}\boldsymbol{f}^l_{1,1}(r)&=&0,\nonumber\\
-2\frac{d\boldsymbol{f}^l_{1,-1}(r)}{dr}\;\,+\;\frac{1}{r}\boldsymbol{f}^l_{1,-1}(r)\;+\;\;
\frac{i\sqrt{2\dot{l}(\dot{l}+1)}}{r}\boldsymbol{f}^l_{1,0}(r)\;+\,\,
\frac{i\sqrt{2\dot{l}(\dot{l}+1)}}{r}\boldsymbol{f}^l_{0,0}(r)&=&0,\nonumber\\
2\frac{d\boldsymbol{f}^{\dot{l}}_{1,1}(r^\ast)}{dr^\ast}\;-\;\frac{1}{r^\ast}
\boldsymbol{f}^{\dot{l}}_{1,1}(r^\ast)\;-\;\frac{i\sqrt{2l(l+1)}}{r^\ast}
\boldsymbol{f}^{\dot{l}}_{1,0}(r^\ast)\,+\frac{i\sqrt{2l(l+1)}}{r^\ast}
\boldsymbol{f}^{\dot{l}}_{0,0}(r^\ast)&=&0,\nonumber\\
2\frac{d\boldsymbol{f}^{\dot{l}}_{0,0}(r^\ast)}{dr^\ast}\,\,-\;\frac{2}{r^\ast}
\boldsymbol{f}^{\dot{l}}_{0,0}(r^\ast)\,-\frac{i\sqrt{2l(l+1)}}{r^\ast}
\boldsymbol{f}^{\dot{l}}_{1,-1}(r^\ast)+\frac{i\sqrt{2l(l+1)}}{r^\ast}
\boldsymbol{f}^{\dot{l}}_{1,1}(r^\ast)&=&0,\nonumber\\
-2\frac{d\boldsymbol{f}^{\dot{l}}_{1,-1}(r^\ast)}{dr^\ast}+\frac{1}{r^\ast}
\boldsymbol{f}^{\dot{l}}_{1,-1}(r^\ast)+\frac{i\sqrt{2l(l+1)}}{r^\ast}
\boldsymbol{f}^{\dot{l}}_{1,0}(r^\ast)+\frac{i\sqrt{2l(l+1)}}{r^\ast}
\boldsymbol{f}^{\dot{l}}_{0,0}(r^\ast)&=&0,\label{RFM}
\end{eqnarray}
It is easy to see that in the system (\ref{RFM}) there are two
superfluous components $\boldsymbol{f}^l_{0,0}(r)$ and
$\boldsymbol{f}^{\dot{l}}_{0,0}(r^\ast)$. The requirement
$\boldsymbol{f}^l_{0,0}(r)=\boldsymbol{f}^{\dot{l}}_{0,0}(r^\ast)=0$
gives rise to $\boldsymbol{f}^l_{1,-1}(r)=\boldsymbol{f}^l_{1,1}(r)$
and $\boldsymbol{f}^{\dot{l}}_{1,-1}(r^\ast)=
\boldsymbol{f}^{\dot{l}}_{1,1}(r^\ast)$ (it follows from the second and
fourth equations). Taking into account these relations we can rewrite
the system (\ref{RFM}) as follows
\begin{eqnarray}
2\frac{d\boldsymbol{f}^l_{1,1}(r)}{dr}\;\;\;-\;\;
\frac{1}{r}\boldsymbol{f}^l_{1,1}(r)\;\;-\;\;
\frac{i\sqrt{2\dot{l}(\dot{l}+1)}}{r}\boldsymbol{f}^l_{1,0}(r)
&=&0,\nonumber\\
-2\frac{d\boldsymbol{f}^l_{1,-1}(r)}{dr}\;\;+\,\,
\frac{1}{r}\boldsymbol{f}^l_{1,-1}(r)\;+\;
\frac{i\sqrt{2\dot{l}(\dot{l}+1)}}{r}\boldsymbol{f}^l_{1,0}(r)
&=&0,\nonumber\\
2\frac{d\boldsymbol{f}^{\dot{l}}_{1,1}(r^\ast)}{dr^\ast}\,-\;\frac{1}{r^\ast}
\boldsymbol{f}^{\dot{l}}_{1,1}(r^\ast)\;-\;\frac{i\sqrt{2l(l+1)}}{r^\ast}
\boldsymbol{f}^{\dot{l}}_{1,0}(r^\ast)&=&0,\nonumber\\
-2\frac{d\boldsymbol{f}^{\dot{l}}_{1,-1}(r^\ast)}{dr^\ast}+\frac{1}{r^\ast}
\boldsymbol{f}^{\dot{l}}_{1,-1}(r^\ast)+\frac{i\sqrt{2l(l+1)}}{r^\ast}
\boldsymbol{f}^{\dot{l}}_{1,0}(r^\ast)&=&0,\label{RFM2}
\end{eqnarray}
Solutions of the system (\ref{RFM2}) also can be expressed via the Bessel
functions of integer order:
\begin{eqnarray}
\boldsymbol{f}^l_{1,1}(r)&=&\boldsymbol{f}^l_{1,-1}(r)=
c_1\sum^\infty_{k=0}(-1)^k2^{2k}\Gamma(k+1)\Gamma(-\dot{l}+k+1)
\left(2\sqrt{r}\right)^{\dot{l}-1}J_{-\dot{l}}\left(2\sqrt{r}\right),
\nonumber\\
\boldsymbol{f}^l_{1,0}(r)&=&\frac{c_2}{\sqrt{2\dot{l}(\dot{l}+1)}}
\sum^\infty_{k=0}(-1)^k(2k-3)2^{2k}\Gamma(k+1)\Gamma(-\dot{l}+k+1)
\left(2\sqrt{r}\right)^{\dot{l}}J_{-\dot{l}}\left(2\sqrt{r}\right),
\nonumber\\
\boldsymbol{f}^{\dot{l}}_{1,1}(r^\ast)&=&
\boldsymbol{f}^{\dot{l}}_{1,-1}(r^\ast)=
\dot{c}_1\sum^\infty_{k=0}(-1)^k2^{2k}\Gamma(k+1)\Gamma(-l+k+1)
\left(2\sqrt{r^\ast}\right)^{l-1}J_{-l}\left(2\sqrt{r^\ast}\right),
\nonumber\\
\boldsymbol{f}^{\dot{l}}_{1,0}(r^\ast)&=&
\frac{\dot{c}_2}{\sqrt{2l(l+1)}}
\sum^\infty_{k=0}(-1)^k(2k-3)2^{2k}\Gamma(k+1)\Gamma(-l+k+1)
\left(2\sqrt{r^\ast}\right)^{l}J_{-l}\left(2\sqrt{r^\ast}\right),
\nonumber
\end{eqnarray}
where
\[
J_{-\nu}(z)=\sum^\infty_{k=0}\frac{(-1)^k}{\Gamma(k+1)\Gamma(-\nu+k+1)}
\left(\frac{z}{2}\right)^{-\nu+2k}.
\]
In such a way, solutions of the Maxwell equations in vacuum are
\begin{eqnarray}
\psi_1(r,\varphi^c,\theta^c)&=&\boldsymbol{f}^l_{1,1}(r)
\fM^l_{1,n}(\varphi,\epsilon,\theta,\tau,0,0),\nonumber\\
\psi_2(r,\varphi^c,\theta^c)&=&\boldsymbol{f}^l_{1,0}(r)
\fM^l_{0,n}(\varphi,\epsilon,\theta,\tau,0,0),\nonumber\\
\psi_3(r,\varphi^c,\theta^c)&=&\boldsymbol{f}^l_{1,-1}(r)
\fM^l_{-1,n}(\varphi,\epsilon,\theta,\tau,0,0),\nonumber\\
\dot{\psi}_1(r^\ast,\dot{\varphi}^c,\dot{\theta}^c)&=&
\boldsymbol{f}^{\dot{l}}_{1,1}(r^\ast)
\fM^{\dot{l}}_{1,\dot{n}}(\varphi,\epsilon,\theta,\tau,0,0),\nonumber\\
\dot{\psi}_2(r^\ast,\dot{\varphi}^c,\dot{\theta}^c)&=&
\boldsymbol{f}^{\dot{l}}_{1,0}(r^\ast)
\fM^{\dot{l}}_{0,\dot{n}}(\varphi,\epsilon,\theta,\tau,0,0),\nonumber\\
\dot{\psi}_3(r^\ast,\dot{\varphi}^c,\dot{\theta}^c)&=&
\boldsymbol{f}^{\dot{l}}_{1,-1}(r^\ast)
\fM^{\dot{l}}_{-1,\dot{n}}(\varphi,\epsilon,\theta,\tau,0,0),\nonumber
\end{eqnarray}
where
\begin{eqnarray}
&&l=1,\;2,\;3;\quad
n=-l,\;-l+1,\;\ldots,\; l;\nonumber\\
&&\dot{l}=1,\;2,\;3;\quad
\dot{n}=-\dot{l},\;-\dot{l}+1,\;\ldots,\; \dot{l},\nonumber
\end{eqnarray}
\[
\fM^l_{\pm 1,n}(\varphi,\epsilon,\theta,\tau,0,0)=
e^{\mp(\epsilon+i\varphi)}Z^l_{\pm 1,n}(\theta,\tau),
\]
\begin{multline}
Z^l_{\pm 1,n}(\theta,\tau)=\cos^{2l}\frac{\theta}{2}
\ch^{2l}\frac{\tau}{2}\sum^l_{k=-l}i^{\pm 1-k}
\tg^{\pm 1-k}\frac{\theta}{2}\tnh^{n-k}\frac{\tau}{2}\times\\
\hypergeom{2}{1}{\pm 1-l+1,1-l-k}{\pm 1-k+1}
{i^2\tg^2\frac{\theta}{2}}
\hypergeom{2}{1}{n-l+1,1-l-k}{n-k+1}{\tnh^2\frac{\tau}{2}},\nonumber
\end{multline}
\[
\fM^l_{0,n}(\varphi,\epsilon,\theta,\tau,0,0)=
Z^l_{0,n}(\theta,\tau),
\]
\begin{multline}
Z^l_{0,n}(\theta,\tau)=\cos^{2l}\frac{\theta}{2}
\ch^{2l}\frac{\tau}{2}\sum^l_{k=-l}i^{-k}
\tg^{-k}\frac{\theta}{2}\tnh^{n-k}\frac{\tau}{2}\times\\
\hypergeom{2}{1}{-l+1,1-l-k}{-k+1}
{i^2\tg^2\frac{\theta}{2}}
\hypergeom{2}{1}{n-l+1,1-l-k}{n-k+1}{\tnh^2\frac{\tau}{2}},\nonumber
\end{multline}
\[
\fM^{\dot{l}}_{\pm 1,\dot{n}}(\varphi,\epsilon,\theta,\tau,0,0)=
e^{\mp(\epsilon-i\varphi)}
Z^{\dot{l}}_{\pm 1,\dot{n}}(\theta,\tau),
\]
\begin{multline}
Z^{\dot{l}}_{\pm 1,\dot{n}}(\theta,\tau)=
\cos^{2\dot{l}}\frac{\theta}{2}
\ch^{2\dot{l}}\frac{\tau}{2}
\sum^{\dot{l}}_{\dot{k}=-\dot{l}}i^{\pm 1-\dot{k}}
\tg^{\pm 1-\dot{k}}\frac{\theta}{2}
\tnh^{\dot{n}-\dot{k}}\frac{\tau}{2}\times\\
\hypergeom{2}{1}{\pm 1-\dot{l}+1,1-\dot{l}-\dot{k}}
{\pm 1-\dot{k}+1}
{i^2\tg^2\frac{\theta}{2}}
\hypergeom{2}{1}{\dot{n}-\dot{l}+1,1-\dot{l}-\dot{k}}
{\dot{n}-\dot{k}+1}{\tnh^2\frac{\tau}{2}},\nonumber
\end{multline}
\[
\fM^{\dot{l}}_{0,\dot{n}}(\varphi,\epsilon,\theta,\tau,0,0)=
Z^{\dot{l}}_{0,\dot{n}}(\theta,\tau),
\]
\begin{multline}
Z^{\dot{l}}_{0,\dot{n}}(\theta,\tau)=
\cos^{2\dot{l}}\frac{\theta}{2}
\ch^{2\dot{l}}\frac{\tau}{2}
\sum^{\dot{l}}_{\dot{k}=-\dot{l}}i^{-\dot{k}}
\tg^{-\dot{k}}\frac{\theta}{2}
\tnh^{\dot{n}-\dot{k}}\frac{\tau}{2}\times\\
\hypergeom{2}{1}{-\dot{l}+1,1-\dot{l}-\dot{k}}
{-\dot{k}+1}
{i^2\tg^2\frac{\theta}{2}}
\hypergeom{2}{1}{\dot{n}-\dot{l}+1,1-\dot{l}-\dot{k}}
{\dot{n}-\dot{k}+1}{\tnh^2\frac{\tau}{2}}.\nonumber
\end{multline}
\begin{sloppypar}
Moving further on
the general Bose-- and Fermi--schemes of the interlocking representations
(\ref{Bose})--(\ref{Fermi}) we can obtain in the same way solutions
of all other higher--spin equations in terms of the functions on the
Lorentz group.
\end{sloppypar}

\bigskip
\begin{flushleft}
{\small
Department of Mathematics\\
Siberia State University of Industry\\
Kirova 42, Novokuznetsk 654007\\
Russia}
\end{flushleft}
\end{document}